\documentclass[useAMS,usenatbib]{mn2e}
\usepackage{epsfig}
\usepackage{graphicx}
\usepackage{times}
\usepackage{color}

\def\cm3{cm$^{-3}$}
\def\kms{km~s$^{-1}$}

\def\lsun{L$_{\odot}$}
\def\rsun{R$_{\odot}$}

\def\msun{M$_{\odot}$}

\let\ts=\thinspace

\def\one{{\,\sc i}}
\def\two{{\,\sc ii}}

\def\four{{\,\sc iv}}

\def\sev{{\sc vii}}
\def\beq{\begin{equation}}
\def\eeq{\end{equation}}

\def\lesssim{\mathrel{\hbox{\rlap{\hbox{\lower4pt\hbox{$\sim$}}}\hbox{$<$}}}}
\def\gtrsim{\mathrel{\hbox{\rlap{\hbox{\lower4pt\hbox{$\sim$}}}\hbox{$>$}}}}

\def\kepler{{\sc kepler}}
\def\cmfgen{{\sc cmfgen}}
\def\mesa{{\sc mesa star}}
\def\v1d{{\sc v1d}}

\newcommand{\iso}[2]{\ensuremath{^{#1}\rm{#2}}}

\def\aj{AJ}
\def\pasp{PASP}
\def\apj{ApJ}
\def\apjs{ApJS}
\def\apjl{ApJL}
\def\aap{A\&A}
\def\araa{ARA\&A}

\def\apss{Ap\&SS}
\def\mnras{MNRAS}

\def\physrep{Phys.~Rep.}
\def\ssr{Space Science Reviews}

\def\gray{$\gamma$-ray}
\def\grays{$\gamma$-rays}

\title[SNe II-P and their RSG progenitors]{Type II-Plateau supernova radiation:
dependencies on progenitor and explosion properties}
  \author[Luc Dessart et al.]{\vspace{0.3cm}
  Luc Dessart,$^{1}$\thanks{email: Luc.Dessart@oamp.fr}
  D. John Hillier,$^{2}$ Roni Waldman,$^{3}$ and Eli Livne.$^3$  \\
  $^1$:  Aix Marseille Universit\'e, CNRS, LAM (Laboratoire d'Astrophysique de Marseille),
  UMR\,7326, 13388, Marseille, France \\
  $^2$ Department of Physics and Astronomy \& Pittsburgh Particle Physics,
  Astrophysics, and Cosmology Center (PITT PACC), University of Pittsburgh, \\
  Pittsburgh, PA 15260, USA \\
  $^3$: Racah Institute of Physics, The Hebrew University, Jerusalem 91904, Israel}

\date{Accepted 2013 May 13. Received 2013 May 13; in original form 2013 March 13}

\pagerange{\pageref{firstpage}--\pageref{lastpage}} \pubyear{2013}

\begin{document}

\maketitle

\label{firstpage}

\begin{abstract}
We explore the properties of Type II-Plateau (II-P) supernovae (SNe) together with their red-supergiant
(RSG) star progenitors.
Using \mesa, we modulate the parameters (e.g., mixing length, overshoot, rotation, metallicity)
that control the evolution of a 15\,\msun\ main-sequence star to produce a variety of physical
pre-SN models and SN II-P ejecta.
We extend previous modeling of SN II-P radiation to include photospheric and nebular phases, as well as
multi-band light curves and spectra. Our treatment does not assume local thermodynamic
equilibrium, is time dependent, treats explicitly the effects of line blanketing, and incorporates
non-thermal processes.
We find that the color properties of SNe II-P require large model atoms for Fe\one\ and Fe\two,
much larger than adopted in \citet{DH11}.
The color properties also imply RSG progenitors of limited extent ($\sim$\,500\,\rsun) --- larger progenitor
stars produce a SN II-P radiation that remains too blue for too long. This finding calls for a reduction of
RSG radii, perhaps through a strengthening of convective energy transport in RSG envelopes.
Increased overshoot and rotation reduce the ratio of ejecta to helium-core mass, similarly to
an increase in main-sequence mass, and thus complicate the inference of progenitor masses.
In contrast to the great sensitivity on progenitor radius,  SN II-P color evolution appears insensitive to
variations in explosion energy.
Finally, we document the numerous SN II-P signatures that vary with progenitor metallicity,
revealing their potential for metallicity determinations  in the nearby and distant Universe.
\end{abstract}


\section{Introduction}

   Since the 1970s, red-supergiant (RSG) stars have been recognized as the potential
   progenitors of Type II-Plateau (II-P) supernovae (SNe; \citealt{grassberg_etal_71,falk_arnett_77}).
   The association of RGS with Type II-P SNe has been confirmed by the detection of
   the progenitor star in pre-explosion images (e.g., \citealt{smartt_09}).
    In current surveys, they represent about 50\% of all core-collapse SNe
    \citep{arcavi_etal_10,smith_etal_11}.
    The early radiation-hydrodynamics simulations of SN II-P light curves
   have been upgraded in recent years
   with improved transport (in particular with the code {\sc stella} of \citealt{blinnikov_etal_98})
   and with dedicated studies of well-observed SNe II-P, although generally limited to grey
   radiation transport \citep{utrobin_07,DLW10a,DLW10b,bersten_etal_11,pumo_zampieri_11}.
   An alternate approach is to
   model the dynamical phase of the SN with grey transport and switch to more sophisticated
   radiative-transfer schemes when the ejecta is in homologous expansion. \citet{KW09}
   followed this approach by combining {\sc kepler}
   hydrodynamical inputs of RSG explosions  \citep{weaver_etal_78}
   with {\sc sedona} for the subsequent radiative transfer modeling to study SN II-P light
   curves. However, little information was extracted from
   the spectra to assess the properties of the progenitor star. This aspect is difficult to
   address because it requires a treatment of the gas in full
   non-local thermodynamic equilibrium (non-LTE; \citealt{DH11}), a treatment of time-dependent
   ionization \citep{UC05,DH08}, and a treatment of line blanketing
   due to metal species \citep{hoeflich_88,eastman_etal_94,DH10,li_etal_12}.
   Focusing primarily on the early-time light curves, these studies do not exploit the information
   from nebular epochs, apart from the inference of the original mass of \iso{56}Ni. It is only recently
   that dedicated studies of the nebular phase of SNe II-P have sought information on
   the progenitor helium core \citep{DH11,DLW10b,maguire_etal_12,jerkstrand_etal_12}.
   Inferring the {\it ejecta} mass requires not just the inference of the H-rich envelope mass
   through the modeling of the photospheric-phase  light curve and spectra,
   but also of the helium-core mass, from nebular-phase studies.

   In \citet{DH11}, we presented exploratory simulations of SN II-P radiation from 10\,d
   until 3\,yr after explosion, thus covering from photospheric to nebular conditions.
   Over this time span, the spectrum formation region scans the entire ejecta, from the
   surface of the progenitor at early times down to the inner ejecta at late times. While the non-LTE
   time-dependent treatment of radiation transport produced a promising match to the
   fundamental SN II-P properties, our models overestimated the bolometric luminosity and were
   too blue. Two progenitor masses were used (15 and 25\,\msun\
   on the main sequence), but no parameter variation was allowed for in either the pre-SN evolution
   or the explosion.

   There is much debate today about the progenitor masses of SNe II-P
   \citep{smartt_09,utrobin_chugai_09,DLW10b,jerkstrand_etal_12} but this is just one of
   numerous issues about SN II-P progenitors that need study. Being end-points of stellar evolution, their
   properties are controlled by nuclear reactions, opacities, energy transport, fluid instabilities
   in stellar interiors, as well as variations associated with initial rotation or metallicity \citep{WHW02}.
   The treatment of convection by means of the mixing-length theory (MLT), or core
   overshooting at the edge of convective regions, is far from satisfactory \citep{meakin_arnett_07}.

  One of our goals is to use SN II-P radiation to help constrain the processes that control
  stellar evolution, i.e., the associated parameters that appear in the stellar structure
  and evolution equations.
  We want to determine the influence of these parameters on the final properties of
  massive stars at death and how these connect to SN II-P radiation properties.
  In this study, we undertake a limited investigation of the influence of the MLT parameter,
  rotation, mass loss, core-overshooting, and metallicity for a 15\,\msun\ progenitor star.
  To complement this, we also
  investigate the effect of varying the SN II-P explosion energy. This is in the spirit of
  \citet{utrobin_07}, but now the diversity of progenitor/ejecta is produced by modeling the diverse
  evolutionary paths followed by a massive star from the main sequence until death.
  In other words, rather than
  crafting a pre-SN star in hydrostatic equilibrium, we adopt different parameters controlling
  stellar evolution and compute physical models of a massive star as it evolves from the main
  sequence until core collapse. With this approach, we can connect the radiation properties
  of SN II-P model ejecta to the pre-SN star and infer the parameters that influenced its evolution.

In SNe II-P, the light curve is primarily determined by the influence of the shock on the progenitor envelope
and the resulting temperature and density structure of the ejecta. Thus having a physical progenitor model
is crucial for understanding and predicting the light curve. The non-LTE effects, and the radiative transfer
are crucial for spectral formation, and for obtaining accurate colors.

  In this paper, we start off by reviewing the successes and failures of \citet{DH11}. We present the physical
  improvements that are necessary to obtain a better agreement with observations (Section~\ref{sect_s15}).
  We explore in particular the color problem of our previous SN II-P simulations with \cmfgen\
  when confronted to observations of SN\,1999em.
  To find the origin of this problem, we use \mesa\ \citep{paxton_etal_11,paxton_etal_13}
  to generate a grid of 15\,\msun\ models evolved
  from the main sequence until iron-core collapse. This grid is selected to cover parameters known
  to impact the evolution of massive stars and in particular their properties at core collapse.
  In Section~\ref{sect_mesa}, we present the properties of these simulations at the onset of collapse.
  We also discuss the properties of the ejecta produced through a piston-driven explosion with  \v1d\
  \citep{livne_93,DLW10a,DLW10b}, with allowance  for explosive nucleosynthesis.
  In Section~\ref{sect_rt}, we describe the \cmfgen\ light curve and spectra for the whole grid of
  models evolved until core collapse with \mesa, and subsequently exploded with \v1d.
  In Section~\ref{sect_comp_mesa_kepler}, we first present a comparison of SN II-P radiation
  properties computed with \cmfgen\ \citep{HM98_lb,DH05_qs_SN, DH08_time,HD12}
  for two 15\,\msun\ stars that share similar properties at death but were evolved
  with \kepler\ (model s15e12 of \citealt{DH11}) and \mesa.
  We then discuss the dependencies of this radiation on progenitor and explosion properties, and specifically
  the impact of progenitor radius (Section~\ref{sect_rad}), core overshooting (Section~\ref{sect_os}),
  metallicity (Section~\ref{sect_z}), and explosion energy (Section~\ref{sect_ekin}).
  We conclude and discuss the implications of our results in Section~\ref{sect_disc}.
  Throughout this work, we compare our results to a well-observed SN II-P representative of the II-P
  class of objects. We choose SN 1999em \citep{hamuy_etal_01,leonard_etal_02a}, and adopt the distance
  of 11.5\,Mpc and reddening $E(B-V)=$\,0.1\,mag
  inferred by \citet{DH06_SN1999em}. We delay a thorough comparison to the broad class of SNe II-P to
  a subsequent study.

 \section{Comparison with \citet{DH11}}
\label{sect_s15}

\subsection{Successes/failures of SN II-P models presented in \citet{DH11}}

 \citet{DH11} presented SN II-P time-dependent non-LTE radiative transfer simulations,
covering the photospheric phase, the transition to the nebular phase, and the nebular phase
until three years after the explosion
of a RSG star. The progenitor evolution was computed until core collapse with
\kepler\ for two stars, 15 and 25\,\msun\ on the main sequence, without rotation and at solar metallicity.
With such inputs, \citet{DH11} reproduced the basic morphology of SN II-P light curves,
the length and brightness of the plateau, the typical spectral evolution from optically-thick to
optically-thin conditions.
Balmer lines appear strong throughout the recombination phase, a feature associated with a
time-dependent effect on the ionization \citep{UC05,DH08_time}. This effect influences all lines
to some extent. The ejecta kinetic energy of 1.2\,B, the ejecta mass of $\sim$\,11\,\msun\  and the \iso{56}Ni
mass of $\sim$\,0.08\,\msun\ in model s15e12 (now referred to as model s15O for s15 ``old";
see Table~\ref{tab_progprop}
for ejecta and progenitor parameters) appear roughly compatible with the SN\,1999em
characteristics at all times, including the pseudo-contiuum due to Fe\two\ and the fine structure lines
associated with O\one\ and Ca\two\ at about one year  after explosion.

At the quantitative level, a number of discrepancies are apparent.
First, the representative plateau luminosity (taken at 50\,d after explosion) is
$\sim$\,6$\times$10$^8$\,\lsun\ for the s15O (formerly named s15e12) model
and 10$^9$\,\lsun\ for the s25e12 model, both greater by a factor of 2-3 than inferred from standard
SNe II-P \citep{bersten_hamuy_09}. SN II-P simulations of \citet{KW09}, which are
based on very similar \kepler\ ejecta inputs have comparable plateau luminosities
(within 10\% of our values), and thus share the same discrepancy.
Second, our s15O model exhibits a bell-shape $V$-band light curve
(the $R$ and $I$ band light curves have a similar morphology),
instead of the plateau generally observed. Probably related, our predicted $U$-band
light curve fades too slowly, even exhibiting a rise or a plateau at
early times in model s25e12. This was observed by \citet{gezari_etal_08}, but for about a week only.
Even in those events detected soon after explosion, SNe II-P exhibit a $U$-band fading
after discovery  \citep{quimby_etal_07,brown_etal_07,brown_etal_09}. The $B$-band light curve
may show a short rise at first, but within a week initiates a fading. Hence, our simulations
are too luminous and tend to remain blue for too long.
Third, the nebular phase spectra of \citet{DH11} systematically predict no H$\alpha$ emission
although a strong line is observed (in, e.g., SN\,1999em; \citealt{leonard_etal_02a}).
In the \citet{DH11} SN II-P ejecta models, significant mixing is applied, making hydrogen present
in the inner ejecta, so the discrepancy is unlikely caused by insufficient mixing.
Finally, the transition from thick to thin conditions at the end of the plateau comes
in \citet{DH11} with a sudden disappearance of continuum-like flux --- the spectrum
suddenly becomes a pure emission spectrum (as in a genuine nebula) while SNe II-P exhibit
a pseudo-continuum for months after the end of the plateau.
This transition phase between thick and thin conditions is not modelled in detail
by any other code in the community so we cannot compare our synthetic spectra
with alternate works.

\subsection{Improvements since \citet{DH11}}
\label{sect_s15N1}

Before invoking inadequate progenitor or explosion properties, we must first investigate
shortcomings potentially affecting the radiative-transfer solution. We identify three limitations
to the \citet{DH11} approach:

\begin{enumerate}
\item Insufficient opacity: This is a major concern for any radiative-transfer modeling of astrophysical
plasmas, in particular in association with metal line blanketing. \citet{DH11}
discussed the important role of sub-dominant species like Sc\two, Ti\two, or
Cr\two\ during the recombination phase. \citet{li_etal_12} emphasized
the importance of Fe\one\ to correct (at least in part) the overestimate in the $U$-band flux
for SN 1987A models \citep{DH10}. In \citet{DH11}, we employed a modest model atom for iron,
with 136 levels for Fe\one\ and 115 levels for Fe\two.
The models also lacked the neutral state for many species, including Ne, Mg, Si, S, Ar, K, and Ca.
\item Neglect of non-thermal processes: As demonstrated by \citet{lucy_91,KF92,KF98a,KF98b,
li_etal_12,dessart_etal_12}, \grays\ from radioactive decay give rise to high-energy electrons
that can non-thermally excite and ionize the gas, exacerbating the departures from LTE.
Such non-thermal processes thrive in low-ionization conditions, which prevail in SNe II-P during the
recombination phase and beyond.
The energy channeled into excitation and ionization reduces the fraction going into heat
and reduces the gas temperature. Hence, the simulations of \citet{DH11} do not model the non-LTE
properties of the gas adequately, particularly during the nebular phase.
\item Assumption of local energy deposition: While the \gray\ mean free path remains small
compared to the scale of the ejecta, preventing \gray\ escape, \grays\ may leak out of the core
and deposit their energy within the overlying H-rich envelope. Local energy deposition underestimates
the contribution of the envelope to the emergent spectrum \citep{maguire_etal_12}.
\end{enumerate}

In a new simulation, named s15N (for s15 ``new"),
we re-run the model s15 with a large model atom for Fe\one\ (1142 levels; 77420 transitions) and  Fe\two\
(827 levels; 44831 transitions).
We include additional neutral species, and specifically Ne\one, Mg\one, Si\one, S\one, Ar\one, Ar\two, K\one,
Ca\one.
We treat non-thermal processes as explained and tested in \citet{li_etal_12,dessart_etal_12}.
We also allow for non-local energy deposition by solving the \gray\ transport problem at each time
step \citep{HD12}. However, we keep the same hydrodynamical input (i.e., s15e12) for
s15N as previously used for s15O.

During a model sequence, we adjust the size of the model atoms to have the most complete description
of opacity sources.
At early times, the wide range in ionization forces us to include Fe\one\ to Fe\sev\ while late in the plateau
phase, only Fe\one\ to Fe\four\ are needed. When we reduce the number of ions to treat, we increase the
number of levels for the dominant ions.
As a result, the total number of levels and transitions varies along any given sequence. For model s15N,
we treat a total of 8308 levels and 197499 bound-bound transitions up to day 60,
7103 levels and 220993 transitions from day 60 to day 130, and
5281 levels and 188457 transitions beyond 130\,d.

 In our approach, we typically include all the metal line transitions with a $gf$ value greater than 0.002 at early
times. When metal line blanketing becomes visible in the spectrum, we lower this value to 0.0001.
However, this cut only applies to elements whose atomic weight is greater than 20,
does not apply to the lowest $n$ levels
($n$ is typically 9), and a transition is omitted only when there is at least $m$ ($m$ is typically 9) stronger
downward transitions from the level. Thus, this procedures does not cut important transitions to ground levels,
and forbidden and semi-forbidden transitions among low-lying states.
With this procedure, we include lines in a very different way from that used
in some SN Ia studies  \citep{kasen_etal_08}, which use a $gf$ cut alone for the selection.
We have performed numerous tests and we find that the large model atoms we employ in this work
for metal species yield converged results
(see also \citealt{DH11,li_etal_12}), i.e., increasing further the size of model atoms does not influence
noticeably the emergent spectra and colors. The much larger number of lines
quoted in SNe Ia studies also stems from their stronger impact in H-deficient ejecta, where,
unlike in SNe II, metal line blanketing is the primary source of opacity.

\begin{figure}
\epsfig{file=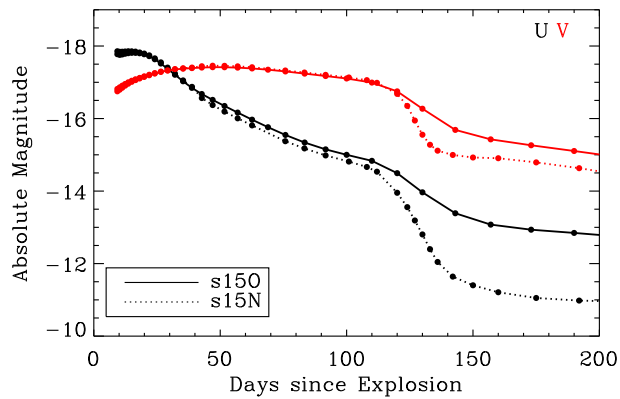,width=8.5cm,bbllx=20,bblly=10,bburx=510,bbury=340,clip=}
\caption{Comparison between the absolute $U$-band (black) and $V$-band (red) synthetic light curves
for models s15O (solid; i.e., model s15 ``old") and s15N (dotted; i.e., model s15 ``new")
as a function of days since explosion. The differences, most
visible at nebular times, stem from the use of bigger Fe\one\ and Fe\two\ model atoms and the
treatment of non-thermal processes and non-local energy deposition in model s15N.
\label{fig_col_s15_old_new}
}
\end{figure}

\begin{figure}
\epsfig{file=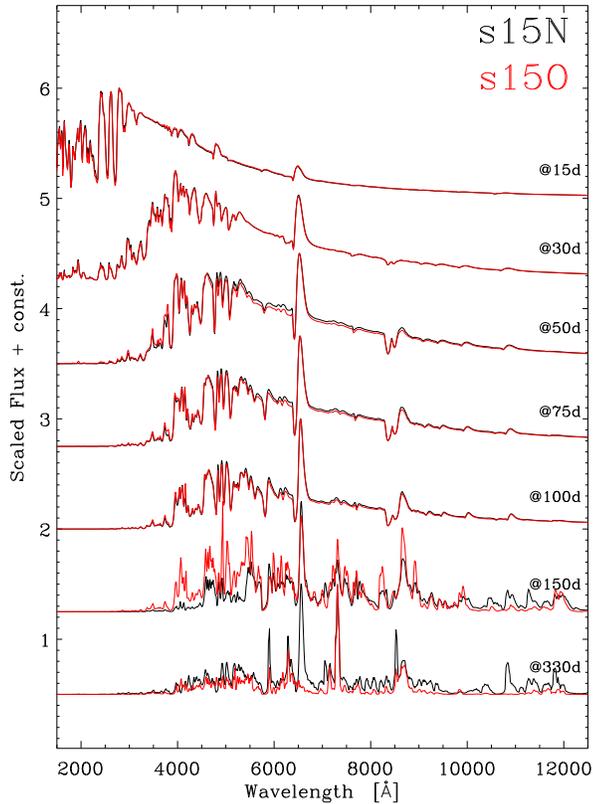,width=8.5cm,bbllx=20,bblly=60,bburx=566,bbury=793,clip=}
\caption{Comparison of the spectral evolution at photospheric and nebular times for models
s15N (black) and s15O (red). Notice the modest differences during the phtospheric phase,
the  stronger blanketing in the s15N model at early nebular times, and the prediction of strong
H$\alpha$ throughout the nebular phase in model s15N. We present line identifications at
multiple epochs in Figs.~A1---A4.
\label{fig_spec_s15_old_new}
}
\end{figure}

\subsection{New results}
\label{sect_s15N2}

During the photospheric phase, we obtain little difference in color (Fig.~\ref{fig_col_s15_old_new}) and
spectral properties (Fig.~\ref{fig_spec_s15_old_new}) between models s15N and s15O.
However, when the ejecta becomes nebular, line blanketing from both Fe\one\ and Fe\two\
weaken the flux shortward of 6000\,\AA\  and increase it in the red, primarily in the form of line emission.
The impact on the  $V$ band light curve is noticeable although modest, while the effect in the $U$ band
is huge.
At the times shown, model s15N is systematically 2 magnitudes fainter than model s15O in the $U$ band.
The luminosity of the SN model is the same; the enhanced blanketing redistributes the flux and
changes the lines that act as coolant for the ejecta. In between strong emission lines, there
is also additional flux from overlapping weak lines (associated especially with Fe\one\ and Fe\two),
whose cumulative effect mimics a pseudo-continuum.

Recently, \citet{li_etal_12} have discussed the influence of non-thermal
processes in the Type II-pec SN\,1987A, and in particular their decisive role
for enhancing or maintaining the Balmer line strength/width when the SN
becomes nebular. Non-thermal processes have a direct impact on level
excitation and species ionization. Because they increase the population
of excited levels, they can also favor indirectly the photoionization rates.
As in SN Ib/c where non-thermal processes are key for the production of
He\one\ lines, we find that they are essential in SNe II-P to produce strong H$\alpha$
and He\one\,10830\,\AA. In model s15N, both lines are present when no equivalent
feature is seen in s15O (Fig.~\ref{fig_spec_s15_old_new}).
We also predict He\one\,7065\,\AA\ but it is overestimated compared to observations of
SN 1999em (the same problem occurs in a SN II-pec model when compared to
SN 1987A \citet{li_etal_12}).
The nebular models of \citet{jerkstrand_etal_12} do not predict an observable He\one\,7065\,\AA.
However the assumed structure of their models during the nebular phase is very different
from that assumed here.
The hydrogen (helium) ionization is higher by two (ten) orders of magnitude in the
inner ejecta in model s15N compared to model s15O at 300\,d after explosion.
However, hydrogen and helium remain primarily neutral so the associated dominant ion
are H\one\ and He\one\ in both s15N and s15O.

Prior to $\sim$\,300\,d after explosion, the decay energy is trapped within the ejecta in all
our simulations. However, as early as 100-200\,d, some \grays\ can escape the core and
deposit their energy in the inner H-rich ejecta.
This causes a temperature reduction of $\sim$\,1000\,K in ejecta shells with $v\lesssim$\,2000\,\kms\
at 300\,d between models s15N and s15O, aggravated by the channeling of that energy into non-thermal
excitation/ionization. In contrast, the overlying H-rich shells of the s15N ejecta are
typically 2000-3000\,K hotter than those of model s15O where local-energy deposition
is assumed.

The inclusion of additional neutral species gives rise to lines not predicted in \citet{DH11}.
Because of their low ionization potentials, these are present at nebular times, i.e., beyond
the end of the plateau and/or later. In this set, we have the K\one\ resonance line at
7600\,\AA,\footnote{This line, blended with O\one\,7777\,\AA, is mentioned by
\citet{chornock_etal_10a} to explain
the observed 7770\,\AA\ feature in SN\,2006ov. We find that in general, past the end of the plateau
phase, the broad feature is primarily due to K\one.}
Mg\one\ lines (primarily in the near-IR), while Si\one, S\one\ and Ca\one\ lines, which are
predicted in pair-instability SN nebular spectra \citep{dessart_etal_13}, remain too weak to be visible here.
We find lines that form preferentially in what used to be the progenitor core
(e.g., the doublet [Ca\two]\,7300\,\AA\ and He\,\one\,7065\,\AA),
or preferentially in the H-rich ejecta shells (H$\alpha$), and lines that form in both
(e.g., [O\one]\,6300\,\AA\ doublet, He\,\one\,10830\,\AA,
Fe\two\,5169\,\AA, and the Ca\two\ triplet at 8500\,\AA).
Na\one\,D hardly changes between models s15N and s15O. This is expected since
it is a resonance line; it scatters whatever overlapping background flux is emitted from deeper
ejecta layers. Some illustrations of line identifications are provided in the appendix, in
Figs.~A1--A4.

\begin{figure}
\epsfig{file=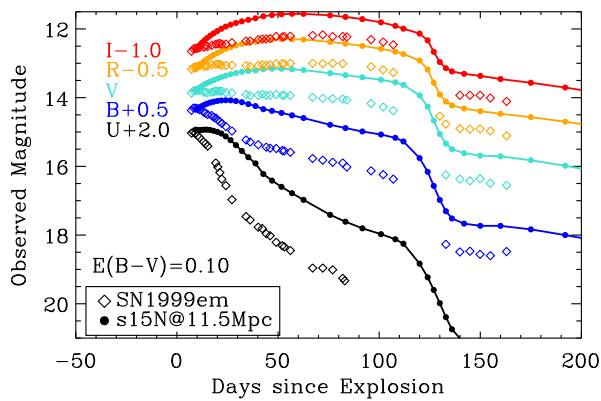,width=8.5cm,bbllx=20,bblly=10,bburx=510,bbury=340,clip=}
\caption{Comparison between the s15N model (reddened with $E(B-V)=$\,0.10\,mag and the
extinction law of \citealt{CCM88_ISE})  and the multi-band light curves of
SN\,1999em, scaled to the SN distance of 11.5\,Mpc \citep{DH06_SN1999em}.
\label{fig_mag_s15new_vs_99em}
}
\end{figure}

\begin{figure*}
\epsfig{file=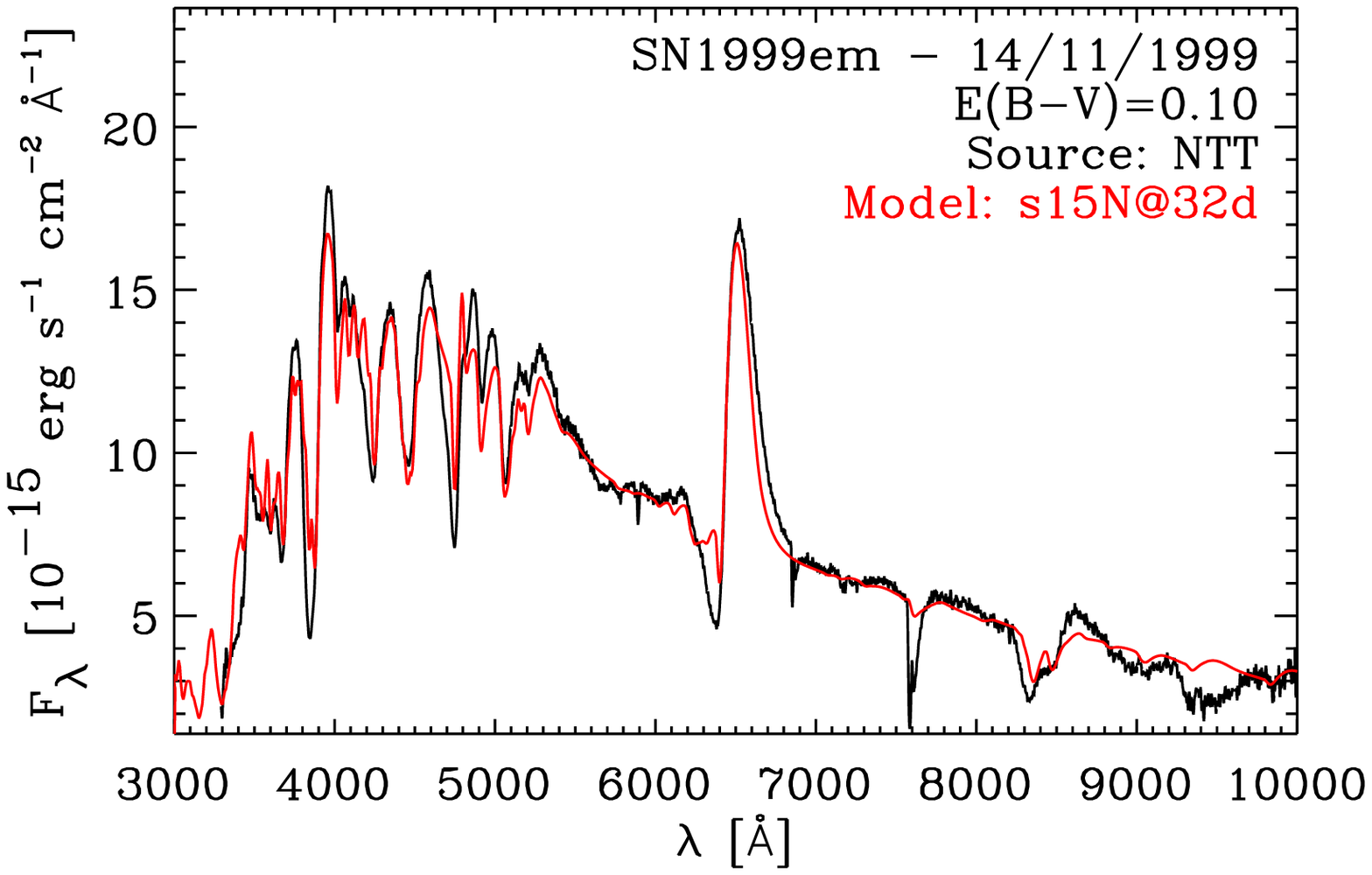,width=8.5cm}
\epsfig{file=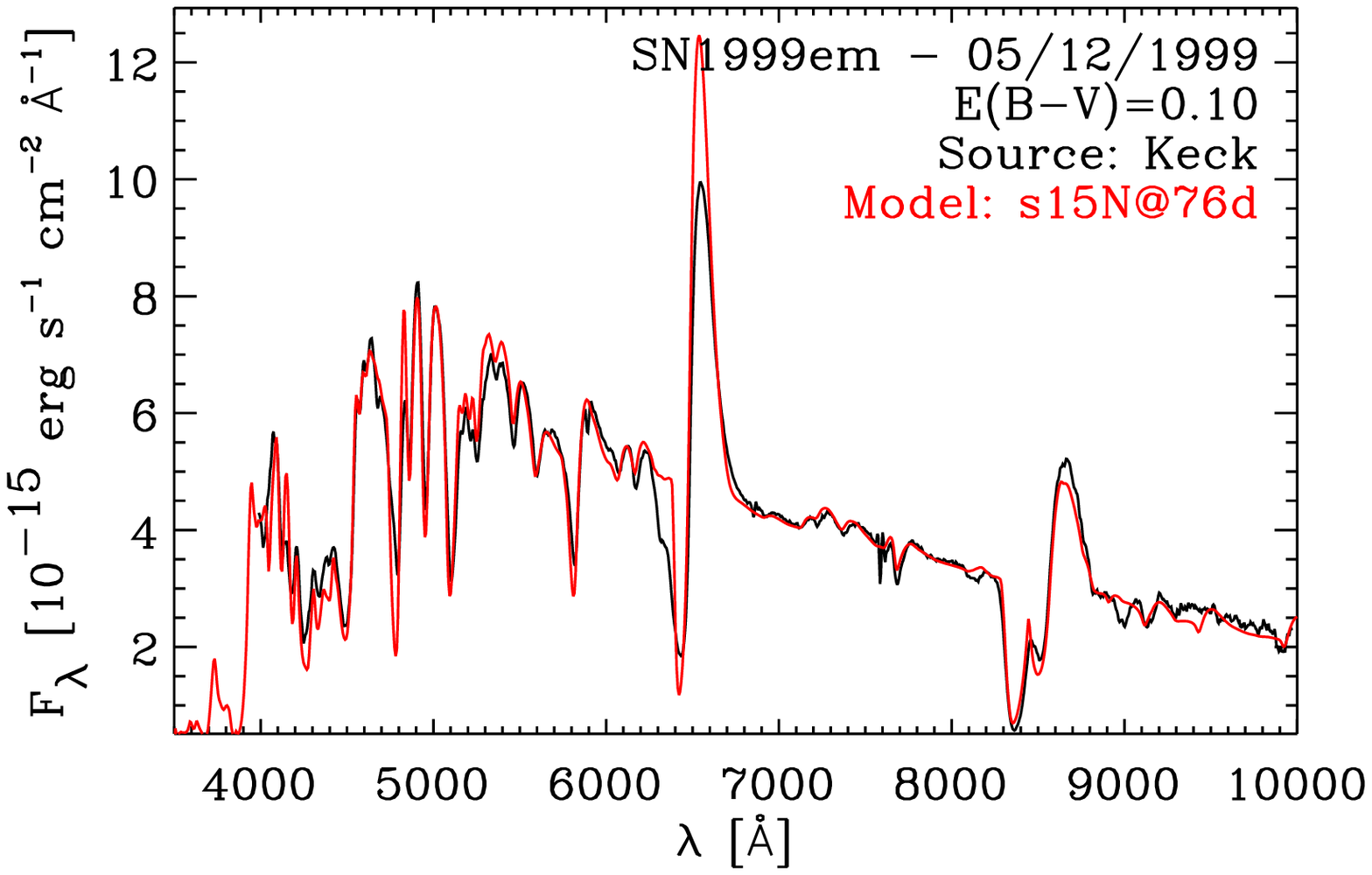,width=8.5cm}
\epsfig{file=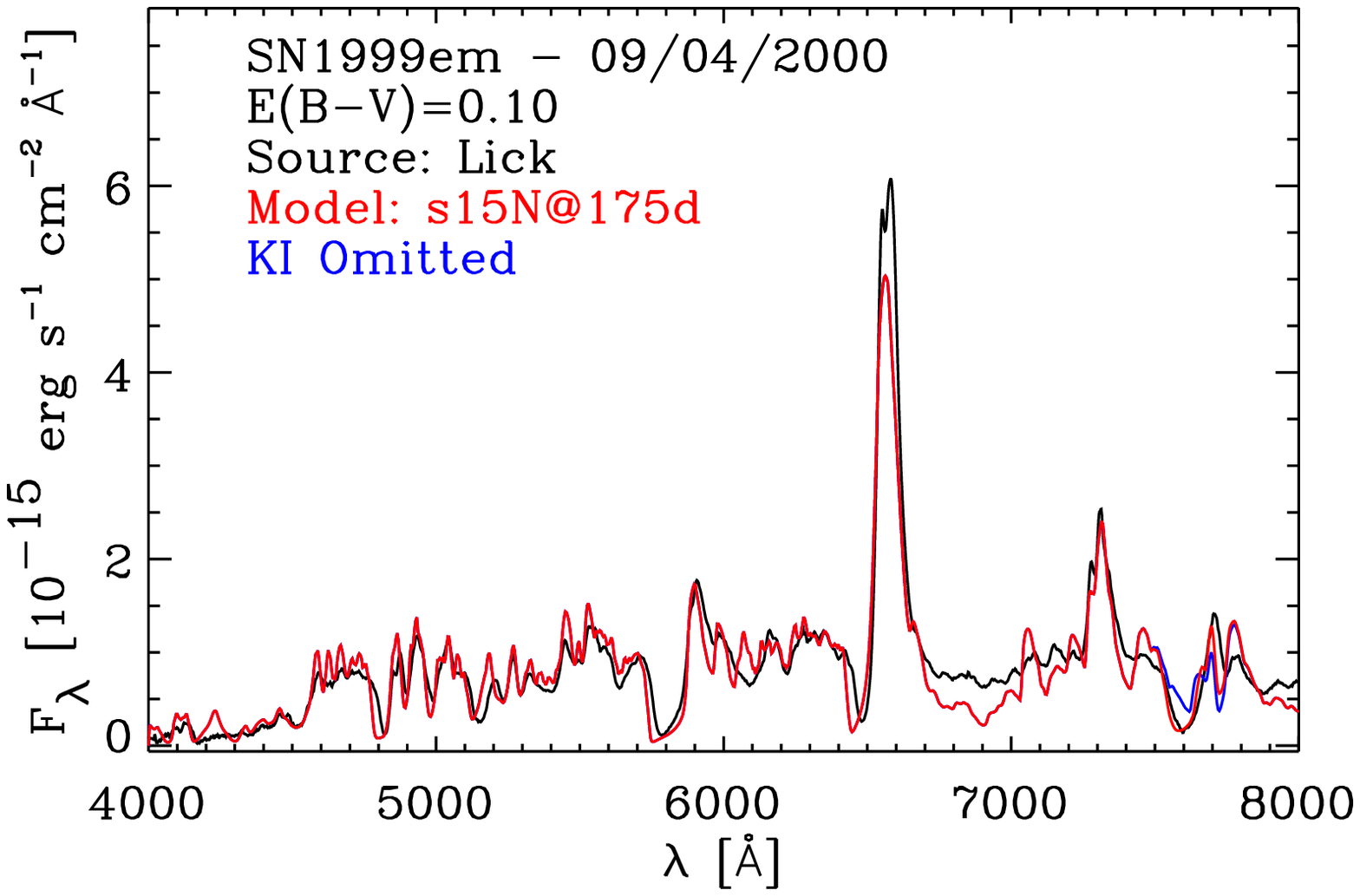,width=8.5cm}
\epsfig{file=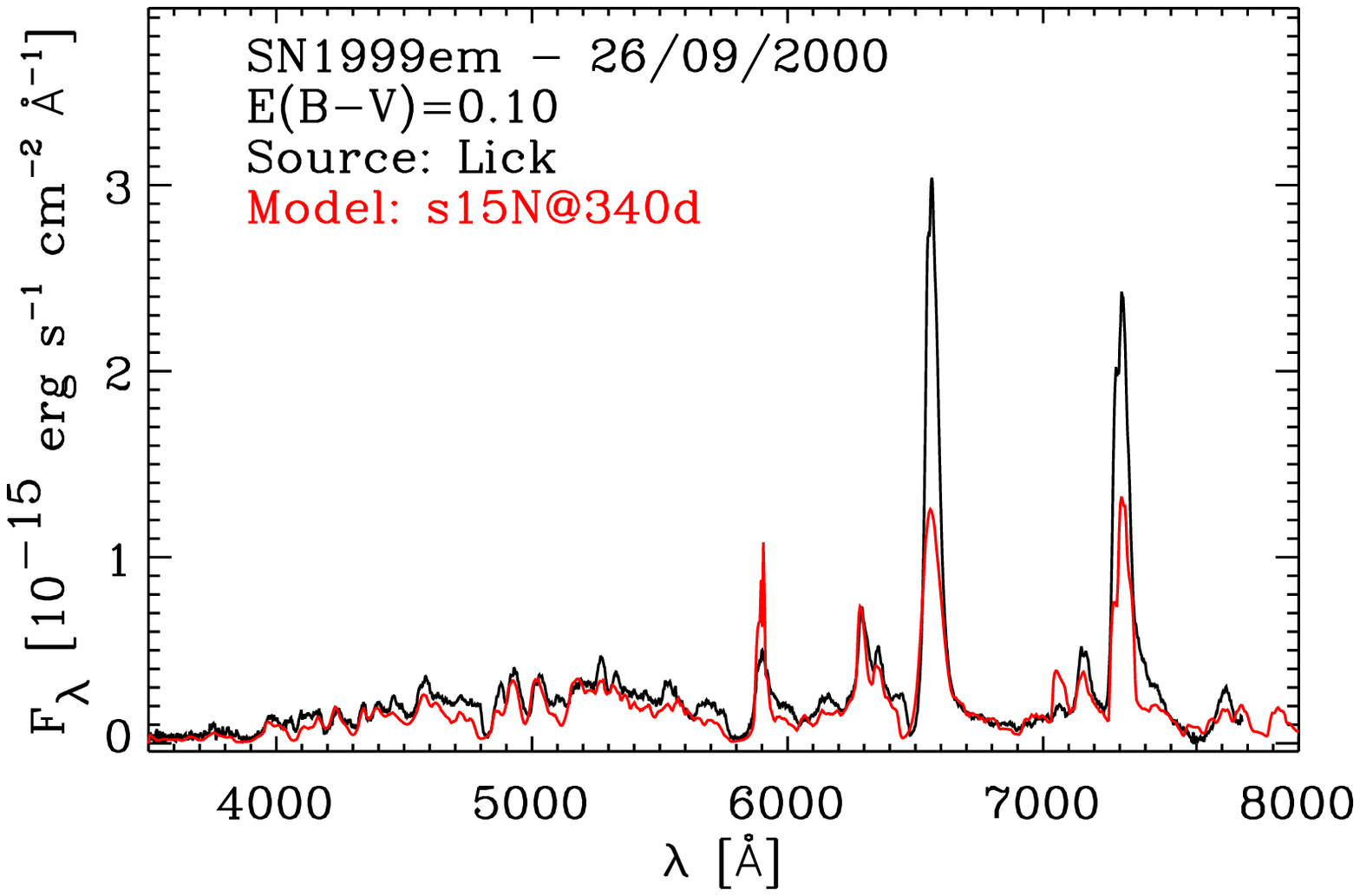,width=8.5cm}
\caption{Spectral comparison between model s15N (reddened with $E(B-V)=$\,0.10\,mag and the Cardelli
extinction law; \citealt{CCM88_ISE})  and multi-epoch observations of
SN\,1999em, scaled to the SN distance of 11.5\,Mpc \citep{DH06_SN1999em}.
For each epoch, a slight time shift is allowed for to correct for
the mismatch in color shown in Fig.~\ref{fig_mag_s15new_vs_99em}.
In the bottom left panel, the blue curve corresponds to the synthetic flux when K\one\ bound-bound
transitions are taken out --- this alters only one feature near 7600\,\AA\ associated with the resonance
transition of K\one.
\label{fig_spec_s15new_vs_99em}
}
\end{figure*}

Overall, the agreement between model s15N and SN\,1999em is much better than with s15O
(Fig.~\ref{fig_mag_s15new_vs_99em} and Fig.~\ref{fig_spec_s15new_vs_99em}), which
implies that employing a large model atom, in particular for iron, and treating non-thermal
processes are both important for the modeling of SNe II-P (non-local energy deposition is
less influential within the first year of a SN II-P). However, some problems remain.
Our model plateau luminosity for SN 1999em is still too high, the $V/R/I$-band light curves still exhibit a
peculiar bell shape, and the $U$-band flux is still too slowly decreasing with time compared to observations.

Given that all our attempts to resolve this remaining color problem have failed, the mismatch may after all be
physical. In particular, it does not seem to be related to an opacity issue.
The solution may be found in alternate progenitor and/or explosion models.
Intuitively, it could be related to the energy budget of the ejecta, with the slower-than-observed cooling
of the photosphere (fading of the UV, $U$, and $B$-band regions) associated with a RSG progenitor radius,
as produced by stellar evolution, that is generally too large.

\section{Grid of \mesa\ computations for a 15\,\msun\ main-sequence star}
\label{sect_mesa}

  The previous sections suggest that the \cmfgen\ simulation s15N based on the \kepler\ model s15e12
  \citep{DH11} reproduces
  most of the generic signatures of SNe II-P but fails to reproduce their color evolution. Assuming that this shift
  is physical and that it does not stem from inadequate explosion properties (as may be argued for
  the shift in plateau luminosity), we now investigate if different progenitor properties could resolve this
  persistent problem. To limit the parameter space, we focus
  on one main-sequence mass only and vary some of the parameters that control its evolution until
  core collapse.

\begin{figure}
\epsfig{file=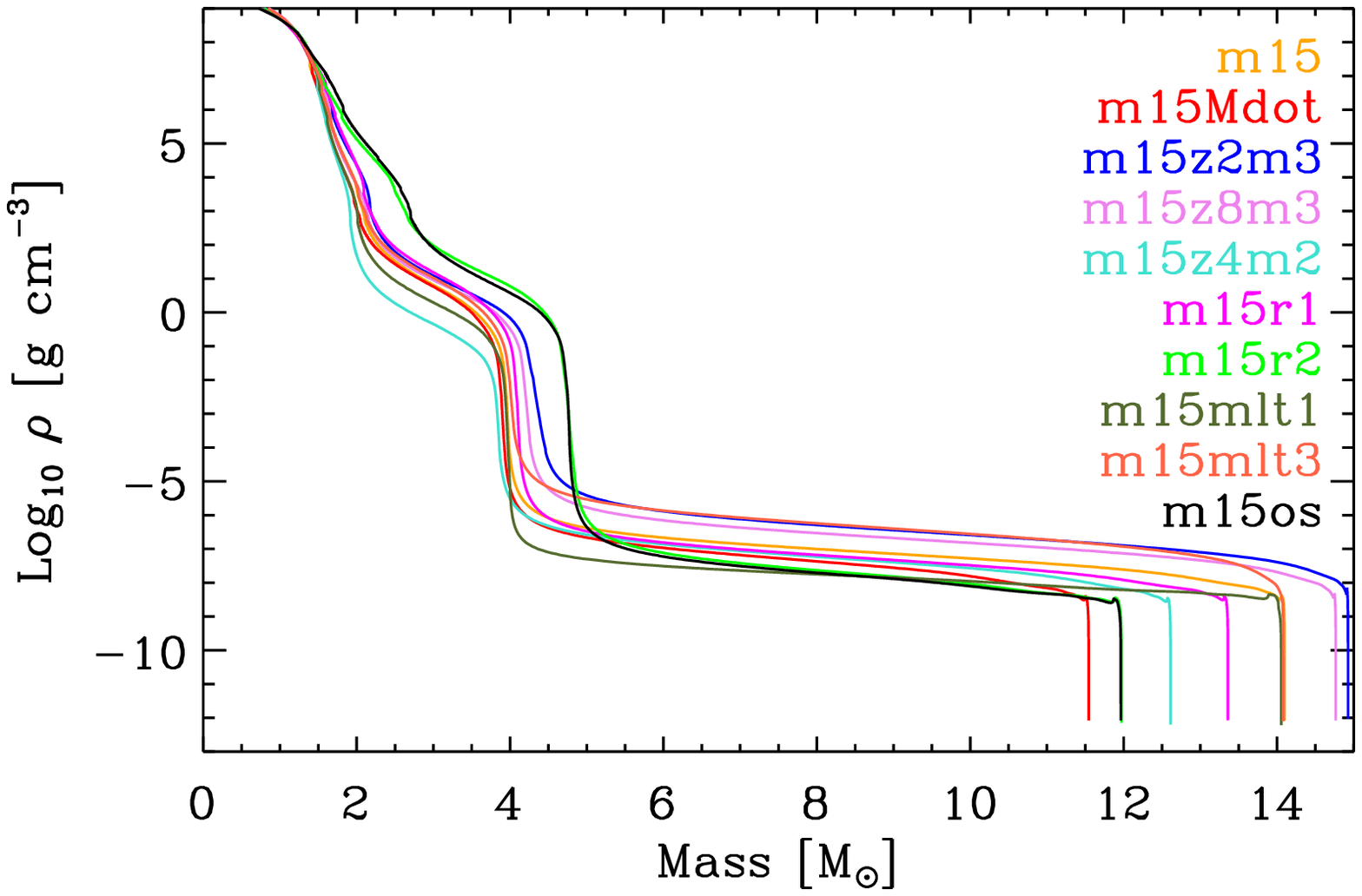,width=8.5cm}
\epsfig{file=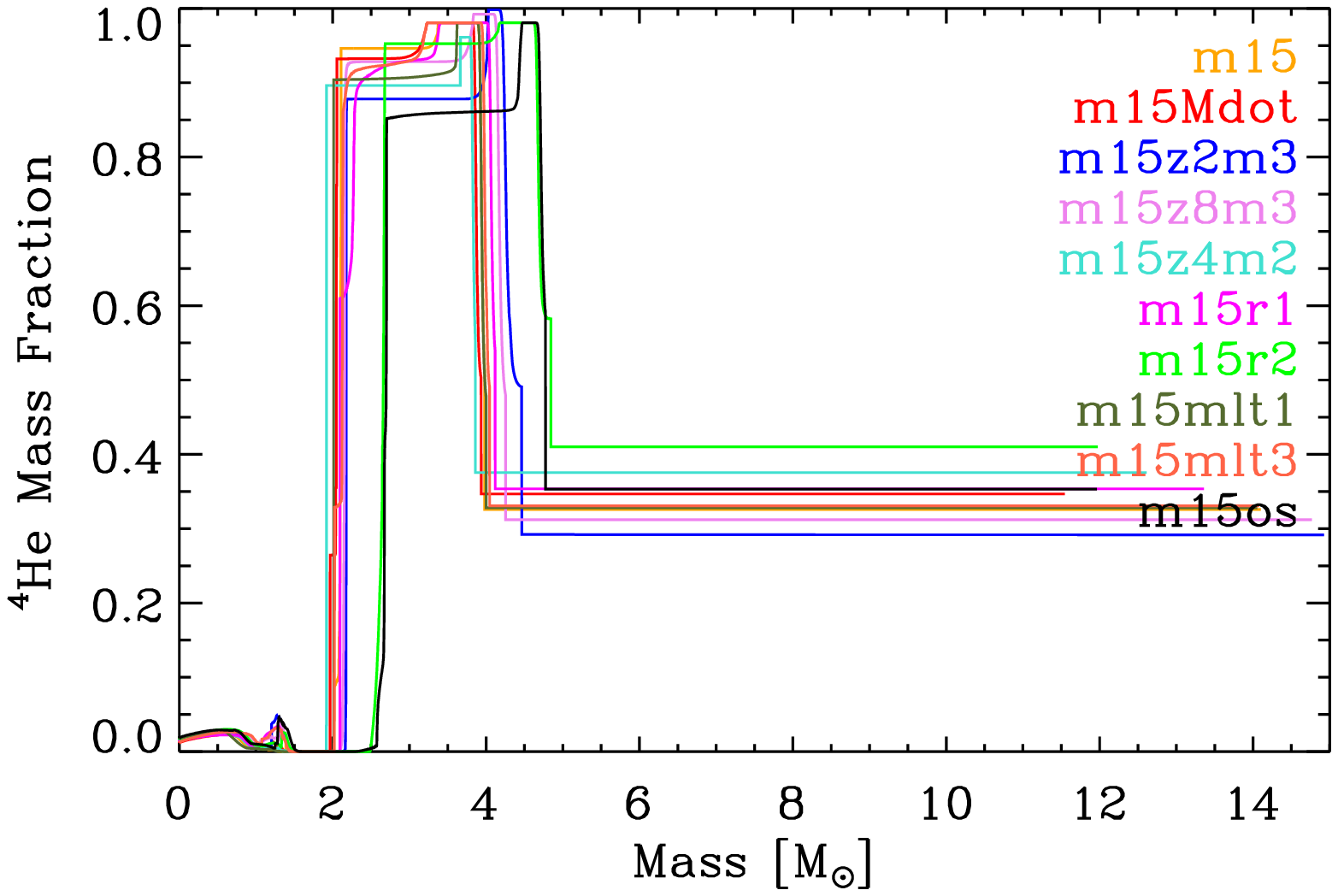,width=8.5cm}
\epsfig{file=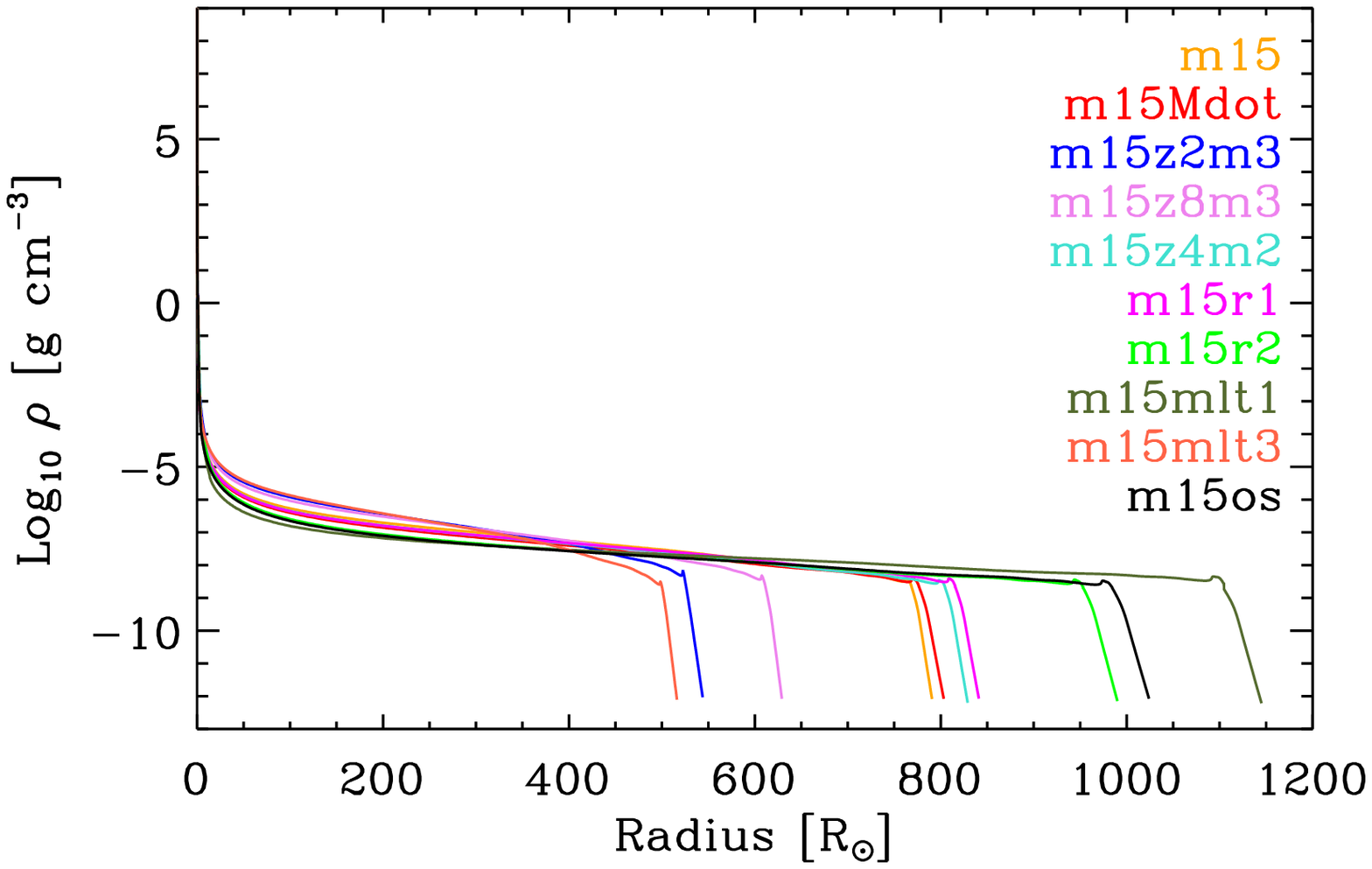,width=8.5cm}
\caption{Slices through the \mesa\ progenitors at the onset of core collapse showing
the mass density (top row) and the \iso{4}He mass fraction (middle row) versus Lagrangian
mass, and the density versus radius (bottom row).
For the model nomenclature, see Section~\ref{sect_naming}.
\label{fig_mesa_slices}
}
\end{figure}

  Using \mesa, we generate a grid of models starting with the same main sequence mass of 15\,\msun.
The parameters used for our reference \mesa\ simulation named m15 are solar metallicity (we take $Z=0.02$),
zero rotation, a mixing-length parameter $\alpha=1.6$, a standard resolution
({\sc mesh\_delta\_coeff}=1), no core-overshooting, the mass loss rate recipes dubbed ``Dutch''
with a scaling of 0.8. Using this m15 reference model parameters, we compute a wide variety of
additional models in which specific parameters are modified, covering different mixing-length
parameters ($\alpha=$\,1.1 and 3 for models m15mlt1 and mt15mlt3),
enhanced mass loss ({\sc dutch\_wind\_eta} = 2 for model m15Mdot),
an exponential overshoot parameter of 0.016 (model m15os; see Section 5.2
of \citealt{paxton_etal_11} for a description on how overshooting is implemented),
models with an equatorial velocity of 100 and 200\,\kms\ on the zero-age main sequence
(models m15r1 and m15r2),
and finally models evolved at metallicities of 0.002, 0.008, 0.04 (named m15z2m3, m15z8m3, and m15z4m2).
In all simulations, we adopt the Schwarschild criterion for convection.
In Table~\ref{tab_progprop}, we summarize the pre-SN properties for each model and show various
slices through the stellar envelope at that time in Fig.~\ref{fig_mesa_slices}.

At the onset of core collapse (central density of $\sim$10$^{9}$\,g\,cm$^{-3}$), the reference model
m15 (also called m15e1p3 when compared to models with different kinetic energies, or m15z2m2 when
compared to models of different metallicity), is a 14.09\,\msun\ RSG star with a luminosity of 63141\,\lsun,
a radius of
768\,\rsun, an effective temperature of 3303\,K. It possesses an H-rich envelope of 10.21\,\msun, an
helium core of 3.88\,\msun\ (set by the inner edge of the H-rich envelope), while the outer edge of
the iron core is at 1.6\,\msun\ (taken here at the location when the electron fraction drops below 0.49 ---
at this point the \iso{54}Fe mass fraction rises suddenly from 7$\times$10$^{-5}$ in the envelope to 0.05).

\subsection*{Influence of the mixing-length parameter}

In the reference model, and in fact in all RSG star models, the entire H envelope is convective. Varying the
mixing-length parameter thus does not alter the fraction of the H-rich envelope that is convectively unstable,
but merely modifies the
efficiency of convective energy transport. We tested what stellar radius our RSG would have if energy transport
in the envelope was done by radiative diffusion alone, which can be determined by reducing $\alpha$ to
something very small (we tested with $\alpha=$\,0.1). In this case, the radius increases enormously
when the star becomes a RSG, becoming as large as $\sim$\,2500\,\rsun\ at the onset of collapse
in our \mesa\ calculation.
More realistically, we choose $\alpha=$\,1.1, 1.6, and 3 which produces RSG surface radii of 1107\,\rsun,
768\,\rsun, and 501\,\rsun\ (models m15mlt1, m15mlt2, and m15mlt3).
Importantly,  such non-rotating solar-metallicity 15\,\msun\ stars
 reach the RSG phase when hydrogen core/shell burning is completed. Hence, the strong
 convection that affects their envelopes late in their life has only a small influence on the helium core properties ---
 uncertainties in RSG mass loss rates, because they impact the star late in its life, have little impact on
 the He core as well.
 Consequently, models m15mlt1, m15mlt2, and m15mlt3 differ primarily in surface radius (or H-rich envelope
 density). We use these models to gauge the impact of the RSG radius on SN II-P radiation
 (Section~\ref{sect_rad}).

\subsection*{Influence of rotation}

Compared to the non-rotating reference model m15, rotation produces higher mass helium cores, higher
luminosities that foster a stronger mass loss rate, and more extended RSG stars at death. Enhanced
initial rotation rate also leads to a smaller $M_{\rm total}/M_{\rm He \ts core}$ at core collapse.
Here, we employ modest initial rotational
velocities of 100 and 200\,\kms\ at the equator and on the main sequence (model m15r1 and m15r2).

\subsection*{Influence of core overshooting}

Introducing core overshooting has a similar impact as increasing rotation (model m15os).
In recent simulations of massive stars for the analysis of VLT-FLAMES survey observations,
\citet{brott_etal_11} motivate the use of a (step-function) overshoot of 0.335.
This extends the convective zone by 0.335 times the local pressure scale height with a constant
diffusion coefficient set to the value at the edge of the original convective zone.
In model m15os, we employ an exponential overshoot of 0.016, implying an exponential
decrease of the diffusion coefficient according to Eq.~2 of \citet{paxton_etal_11}.
In practice, the effect of a 0.1 step-function overshoot is equivalent to an 0.008 exponential overshoot.

Using the step-function overshoot in \mesa, we explore the impact of this process on the evolution
of a 15\,\msun\ star and its properties at core collapse.
We find that increasing the overshoot parameter from 0 to 0.5 leads to a reduction of the final star mass from
13.57 to 9.40\,\msun\ and an increase
in helium core mass from 3.96 to 6.61\,\msun. The ratio $M_{\rm total}/M_{\rm He \ts core}$ varies
from 3.42 down to 1.42, in the same order.  These modulations are non trivial.
The variations in pre-SN structure that result from enhanced core overshooting (or rotation) parallel
those obtained for enhanced main-sequence mass, which are known to influence SN ejecta
kinematics and SN II-P radiation properties \citep{DLW10b}.

\begin{table*}
\caption{
Summary of model properties used as initial conditions for \cmfgen\ simulations.
The quantities tabulated are in general self explanatory. Following
the columns containing the cumulative ejecta masses in H, He, and O, we give the
Lagrangian mass corresponding to specific locations in the pre-SN star:
$M_{r,{\rm Y_e}}$ corresponds to the edge of the iron core, i.e.,  where the electron fraction rises
to 0.49 as we progress outward from the star center (this is also the piston location for the explosion);
$M_{r,S_4}$ stands for the Lagrangian mass where the entropy above the iron core rises
to 4\,k$_{\rm B}$ per baryon,
$M_{r,\rm He}$ to the base of the Helium-rich shell, and
$M_{r,\rm H}$ to the base of the H-rich envelope.
The last three columns give some ejecta properties that result from the imposed explosion parameters.
The quoted  \iso{56}Ni mass corresponds to that originally produced in the explosion.
Each model was started from a 15\,\msun\ main-sequence star, but evolved
with different codes and under different conditions.
For the model nomenclature, see end of Section~\ref{sect_naming}.
}
\label{tab_progprop}
\begin{tabular}{l@{\hspace{2mm}}c@{\hspace{2mm}}c@{\hspace{2mm}}c@{\hspace{2mm}}c@{\hspace{2mm}}c@{\hspace{2mm}}
c@{\hspace{2mm}}c@{\hspace{2mm}}c@{\hspace{2mm}}c@{\hspace{2mm}}c@{\hspace{2mm}}c@{\hspace{2mm}}c@{\hspace{2mm}}
c@{\hspace{2mm}}c@{\hspace{2mm}}c@{\hspace{2mm}}c@{\hspace{2mm}}c@{\hspace{2mm}}}
\hline
   model   &    $M_{\rm final}$ & Age &  $T_{\rm eff}$  &    $R_{\star}$ &$L_{\star}$   &       $Z$     &  $M_{\rm H}$
    &         $M_{\rm He}$  &   $M_{\rm O}$   &    $M_{r,{\rm Y_e}}$ &    $M_{r,S_4}$ &  $M_{r,\rm He}$  &  $M_{r,\rm H}$ &  $M_{\rm H, env}$ &
          $M_{\rm ej}$   &  $E_{\rm kin}$  &  $M_{\iso{56}Ni}$  \\
           &     [\msun]  & [Myr] &        [K]  &    [\rsun] &    [\lsun]   &     &   [\msun]   &  [\msun]  &    [\msun]
           &     [\msun]        &  [\msun]          &   [\msun]          &  [\msun]    & [\msun] & [B]  & [\msun]    \\
\hline
  s15O     &     12.79   &  13.24   &    3300   &   810  &     70235  &  0.020  &   5.490 &      4.000  &     0.816     &   \dots & 1.72  &      2.93    &    4.13  &      8.66  &    10.93  &   1.20  & 0.087 \\
  s15N     &     12.79   &  13.24   &    3300   &   810  &     70235  &  0.020  &   5.490 &      4.000  &     0.816     &   \dots  & 1.72  &      2.93    &    4.13  &      8.66  &    10.93  &   1.20  & 0.087 \\
     m15   &     14.09   &  12.39   &    3303   &   768  &     63141  &  0.020  &   6.630 &      5.105  &     0.325     &    1.62 & 1.62  &      2.00    &    3.88  &     10.21  &    12.48  &   1.27  &   0.050  \\
\hline
 m15Mdot   &     11.53   &  12.65   &    3249   &   776  &     60426  &  0.020  &   4.839 &      4.444  &     0.316      &  1.50 &   1.61  &      1.95    &    3.81  &      7.72  &    10.01  &   1.28  &   0.081  \\
  s15N     &     12.79   &  13.24   &    3300   &   810  &     70235  &  0.020  &   5.490 &      4.000  &     0.816           &  \dots   & 1.72  &      2.93    &    4.13  &      8.66  &    10.93  &   1.20  & 0.087 \\
\hline
 m15mlt1        &     14.01   &  12.36   &    3318   &  1107  &    106958  &  0.020  &   6.516 &      5.167  &     0.354  &    1.36 &    1.61  &      2.00    &    3.89  &     10.13  &    12.57  &   1.24  &   0.121   \\
m15mlt2   &     14.09   &  12.39   &    3303   &   768  &     63141  &  0.020  &   6.630 &      5.105  &     0.325     &   1.62  &  1.62  &      2.00    &    3.88  &     10.21  &    12.48  &   1.27  &   0.050  \\
 m15mlt3   &     14.08   &  12.41   &    4106   &   501  &     64218  &  0.020  &   6.542 &      5.173  &     0.383          &   1.54 & 1.55  &      2.02    &    3.92  &     10.16  &    12.52  &   1.34  &   0.086  \\
 \hline
     m15   &     14.09   &  12.39   &    3303   &   768  &     63141  &  0.020  &   6.630 &      5.105  &     0.325      &   1.62  & 1.62  &      2.00    &    3.88  &     10.21  &    12.48  &   1.27  &   0.050  \\
   m15os   &     11.93   &  13.21   &    3175   &   984  &     88612  &  0.020  &   4.510 &      4.385  &     0.762     &   1.63      &  1.76  &      2.28    &    4.64  &      7.29  &    10.28  &   1.40  &   0.15   \\
\hline
   m15r0   &     14.09   &  12.39   &    3303   &   768  &     63141  &  0.020  &   6.630 &      5.105  &     0.325     &  1.62  &   1.62  &      2.00    &    3.88  &     10.21  &    12.48  &   1.27  &   0.050  \\
   m15r1   &     13.34   &  12.95   &    3277   &   815  &     68932  &  0.020  &   5.800 &      5.126  &     0.404     &   1.60 &  1.65  &      2.09    &    4.00  &      9.34  &    11.73  &   1.35  &   0.10   \\
   m15r2   &     11.94   &  14.88   &    3221   &   953  &     87895  &  0.020  &   4.109 &      5.001  &     0.512     &   1.52 &   1.80  &      2.46    &    4.61  &      7.34  &    10.39  &   1.34  &   0.19   \\
 \hline
 m15z2m3   &     14.92   &  13.57   &    4144   &   524  &     72890  &  0.002  &   7.483 &      5.048  &     0.507     &   1.61 &   1.65  &      2.13    &    4.15  &     10.77  &    13.29  &   1.35  &   0.081  \\
 m15z8m3   &     14.76   &  13.34   &    3813   &   611  &     71052  &  0.008  &   7.183 &      5.252  &     0.428     &    1.63 &  1.65  &      1.90    &    4.09  &     10.67  &    13.12  &   1.27  &   0.036  \\
 m15z2m2   &     14.09   &  12.39   &    3303   &   768  &     63141  &  0.020  &   6.630 &      5.105  &     0.325     &  1.62 &   1.62  &      2.00    &    3.88  &     10.21  &    12.48  &   1.27  &   0.050  \\
 m15z4m2   &     12.60   &  10.88   &    3137   &   804  &     56412  &  0.040  &   5.119 &      5.042  &     0.387     &  1.40  &  1.49  &      1.91    &    3.77  &      8.83  &    11.12  &   1.24  &   0.095  \\
\hline
  m15e0p6  &     14.09   &  12.39   &    3303   &   768  &     63141  &  0.020  &   6.630 &      5.105  &     0.325     &   1.62 &  1.62  &      2.00    &    3.88  &     10.21  &    12.46  &   0.63  &   0.046  \\
  m15e1p3   &     14.09   &  12.39   &    3303   &   768  &     63141  &  0.020  &   6.630 &      5.105  &     0.325     &  1.62 &  1.62  &      2.00    &    3.88  &     10.21  &    12.48  &   1.27  &   0.050  \\
  m15e2p9  &     14.09   &  12.39   &    3303   &   768  &     63141  &  0.020  &   6.630 &      5.105  &     0.325     &    1.62 &  1.62  &      2.00    &    3.88  &     10.21  &    12.48  &   2.91  &   0.052  \\
\hline
\end{tabular}
\end{table*}

\subsection*{Influence of metallicity}

Because of the adopted metallicity dependence of RSG mass loss rates, our \mesa\ models
m15z2m3, m15z8m3, m15 (also called model m15z2m2 in this context),
and m15z4m2 have final H-rich envelope masses of 10.77--8.83\,\msun.
The efficiency of convective energy transport being set in all four
simulations through a mixing-length parameter of 1.6, the variation in
metallicity, which changes the opacity in the envelope, alters the stellar
radius (since the energy flux to transport from the edge of the core to the stellar surface is
essentially the same between these 4 models). Consequently, for smaller metallicities (opacities), we
obtain smaller RSG radii. Our RSG models are thus both more massive and more compact at
lower metallicities.

\subsection*{Piston-driven explosions with \v1d}

Each \mesa\ model in our grid, once it has reached the onset of core collapse,
is exploded with \v1d\ \citep{livne_93,DLW10a,DLW10b} by driving
a piston at the inner boundary in order to yield an ejecta kinetic energy of
$\sim$\,1.2\,B. For simplicity
and to avoid biases between models,  we choose the edge of the iron core to position the piston
in all cases (corresponding to the $M_{r, {\rm Y_e}}$ in Table~\ref{tab_progprop}).
Depending on the core properties, the combustion generates various amounts of \iso{56}Ni.
We make no attempt at adjusting the piston properties to yield the same  \iso{56}Ni mass,
but we note that the explosive nucleosynthesis and fallback are quite sensitive to the exact
location of the piston and its adopted trajectory.
To gauge the influence of the explosion energy on the SN II-P characteristics, we also run two additional
models designed to have  ejecta kinetic energies of 0.6 and 2.9\,B (models m15e0p6 and m15e2p9) --- when
these models are discussed, the reference model, which is characterized by a 1.27\,B ejecta
kinetic energy, is named m15e1p3.

All ejecta are mixed using a boxcar algorithm with a width of 0.4\,\msun. This
mixes the He-core material efficiently but leads to modest mixing of
\iso{56}Ni into the H-rich envelope. In practice, the sequence of m15
simulations with $\sim$\,1.2\,B ejecta kinetic energies have little \iso{56}Ni
beyond $\sim$\,2000\,\kms.
A final property of our ejecta models is that, suffering mild fallback, their inner velocity is often
very low. This poses a challenge for \cmfgen\ since the velocity space to
cover is much larger (say from 50 to 20000\,\kms). Inevitably, this
causes some numerical diffusion, whereby strong gradients are softened as we
remap the grid at each time step. Furthermore, the slow inner ejecta shells expand
less and remain dense out to late times, inhibiting forbidden-line emission.
Our artificial piston-driven explosions are thus limited in physical
consistency, potentially affecting SN observables when radiation arises from
the inner ejecta, as it does at nebular times (this issue is relevant for all
piston-driven explosions). Because of these limitations, we postpone
a detailed investigation of the influence of mixing in our SN II-P simulations.

An important property of all RSG stars is the strong compactness of the helium core
(which typically fits within 1\,\rsun) and the very extended H-rich envelope. This holds for
all models shown in Fig.~\ref{fig_mesa_slices}, but also for RSG models of different
main-sequence masses \citep{WH07,DLW10a}. The H-rich envelope always has a small
binding energy while the helium core is highly bound. Owing much to this property, the plateau
phase of SNe II-P corresponds to epochs when the photosphere is within the (originally extended)
H-rich envelope and as such the plateau phase can only serve as a diagnostic of the H-rich envelope
mass \citep{DH11}. Hence, it does not (and it cannot) provide a reliable constraint on the helium core mass,
nor on the ejecta mass. We stress this property because the so-called mass discrepancy emphasized by
\citet{utrobin_07,utrobin_chugai_09} seems to stem from the large inferred helium-core mass,
although photospheric-phase light curve modeling constrains the H-rich envelope mass only. Furthermore,
their adopted density structure seems peculiar for a RSG since only the inner 2\,\msun\ of the progenitor
model is highly bound, even for the 29\,\msun\ main-sequence star employed for
SN\,2004et \citep{utrobin_chugai_09}.

\subsection*{Summary of model nomenclature}
\label{sect_naming}

Model s15O is based on the s15e12 ejecta model and is described in detail in \citet{DH11}.
Model s15N is identical except for improvements in the radiative transfer, which are discussed
in Section~\ref{sect_s15}. In contrast, all models started with ``m15'' were computed with \mesa\ and
exploded with \v1d.
We divide Table~\ref{tab_progprop} in groups in which a single parameter has been varied.
These distinct groups include models computed
with a different core-overshooting parameter (m15 and m15os),
with different initial rotation rates (m15r0, m15r1, m15r2),
at different metallicities (m15z2m3, m15z8m3, m15z2m2, m15z4m2),
and exploded with different explosion energies (m15e0p6, m15e1p3, m15e2p9).
In each sub group, the reference model m15 is used, but for clarity its name is edited
to reflect the quantity that was altered. Hence, model m15 becomes m15z2m2 when compared
to models computed at different metallicity etc. Similarly, we use m15r0, m15mlt2, m15e1p3.

In the following sections, we illustrate the results from simulations started from \mesa\ and
\v1d\ inputs, evolved with \cmfgen, and using the same setup (radiative-transfer technique,
model atoms etc) as for the simulation s15N described in Sections~\ref{sect_s15N1}--\ref{sect_s15N2}.

\section{Diversity of SN II-P radiation properties for a 15\,\msun\ main sequence star}
\label{sect_rt}

\begin{figure}
\epsfig{file=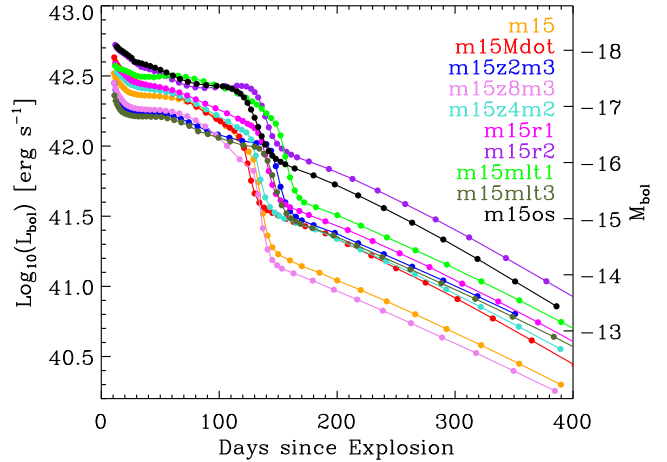,width=8.5cm}
\caption{
Bolometric light curves computed with \cmfgen\ and based on \mesa\ simulations
for a 15\,\msun\ main-sequence star evolved under different assumptions, and exploded the same
way with \v1d\ using a piston speed of 30000\,\kms. All models in this figure are within 10\% of a
reference ejecta kinetic energy of 1.2\,B. Models with different ejecta kinetic energies are
presented in Fig.~\ref{fig_m15_ekin_lc}.
[See Section~\ref{sect_mesa} and Table~\ref{tab_progprop} for details.]
\label{fig_lbol_m15_grid}
}
\end{figure}

   One important goal of SN II-P research is to determine the progenitor mass, both at the time
of explosion and on the main sequence. In our exploration, although we use the same main-sequence
mass, the different evolutionary paths associated with different metallicities, rotation rates,
mixing-length parameters or mass loss rates, lead to a range of pre-SN properties (primarily final mass,
He-core mass, H-rich envelope mass, radius). In turn, the resulting stellar explosions display
a range of SN II-P bolometric light curves (Fig.~\ref{fig_lbol_m15_grid}), multi-band light curves
(see Figs.~B1---B4 for a full illustration),  
as well as spectral evolution.

  For models characterized by the same $\sim$\,1.2\,B ejecta kinetic energy, the plateau  brightness
  varies by about a factor of 2; the plateau phase lasts between 120 and 150\,d; the nebular luminosity
  spans a factor of $\sim$\,7. The plateau brightness increases with progenitor radius following
  the reduced cooling from expansion. The plateau length is reduced for a smaller ejecta mass and/or
  a smaller progenitor radii (see also \citealt{LN83,popov_93,Y04}).
  The plateau phase starts when hydrogen-recombination sets in, which occurs
  when the gas temperature at the photosphere is $\sim$\,5500\,K. Subsequently,  the recession of the
  photosphere to deeper ejecta layers occurs at a rate comparable to the expansion rate of the corresponding
  layers, which combined to the fixed photospheric temperature, leads to a constant luminosity.
  The ``plateau" radii in these simulations are typically of $\sim$\,2$\times$\,10$^{15}$\,cm.
  In some models with a large \iso{56}Ni mass (e.g., s15N or m15r2), the bolometric
  light curve shows a mild slant upward at the end of the ``plateau".

  The nebular flux directly scales with the mass of \iso{56}Ni synthesized in the explosion, which in our
  approach, increases for progenitors with a higher mass helium core. It is maximum for the m15r2
  and m15os models whose helium core mass is $\sim$\,4.6\,\msun. In some models with the same
  helium core mass, we obtain a range of \iso{56}Ni masses because of differences in inner core structures and
  the Lagrangian mass adopted for the piston (see Table~\ref{tab_progprop}).
  All these results are as expected, but they highlight the difficulty of inferring the progenitor
  mass, because it is necessarily subject to uncertainties from stellar evolution
  (e.g., affecting $R_{\star}$) or unknowns about the progenitor star environment (e.g., $Z$).

  In the following sections, we study in more detail the impact on SN II-P properties arising from
  variations in progenitor radius (Section~\ref{sect_rad}), core overshooting (Section~\ref{sect_os}),
  metallicity (Section~\ref{sect_z}), and explosion energy (Section~\ref{sect_ekin}).
  We first present a comparison of SN II-P radiation properties for models computed with \kepler\
  and \mesa\ \& \v1d.

\section{Comparison between \kepler\ and \mesa}
\label{sect_comp_mesa_kepler}

   \mesa\ is a public code and thus offers a great potential for the community to study stellar evolution
   and, in our field of research, produce new and diverse pre-SN massive star models, resulting from
   either single or binary-star evolution. A lot of benchmarking has been done in \mesa, some of which
   is described in \citet{paxton_etal_11,paxton_etal_13}. Here, we investigate what \mesa\ parameters
   are needed to make a 15\,\msun\ main-sequence star produce a SN II-P explosion comparable to
   s15N, whose pre-SN evolution (and explosion) was computed with \kepler\ \citep{WH07}.

   Of all light curves and ejecta/envelope properties presented in Fig.~\ref{fig_lbol_m15_grid} and
   Table~\ref{tab_progprop}, progenitor/explosion model m15Mdot is a near twin of s15N, with
   $R_{\star}$/$M_{\rm H \ts env}$/$E_{\rm kin}$/$M_{\iso{56}Ni}$ of
   776\,\rsun/7.72\,\msun/1.28\,B/0.081\,\msun\ compared to
   810\,\rsun/8.66\,\msun/1.20\,B/0.087\,\msun\ in s15N.
   The mixing imposed in model m15Mdot (which is done with the same procedure as in other m15
   simulations) is weaker than in model s15N, i.e., the \iso{56}Ni remains more centrally
   concentrated in model m15Mdot. For H and He, models m15Mdot and s15N differ
   in cumulative mass at the 10\% level, while for subdominant species whose mass-fraction distribution
   tends to peak in the core, the differences are up to a factor of 2 (for example, the total oxygen mass
   is 0.78\,\msun\ in model
   s15N but only 0.27\,\msun\ in model m15Mdot). The progenitor core properties differ between the
   two simulations, in part because of the lack of core overshooting in model m15Mdot, and a much
   coarser nuclear network in
   \mesa\ that includes 21 isotopes only while model s15N contains the yields computed for
   over 2000 isotopes.
  For intermediate-mass and iron-group elements (respectively IMEs and IGEs)
  with known blanketing effects (e.g., Sc\two, Cr\two\ etc.; see
  \citealt{DH11}) but not accounted for in the \mesa\ calculation, we set their abundance to
  the solor-metallicity value \citep{GS98}.
  Hence, in practice, the m15Mdot model has a similar H-rich envelope composition to s15N but
  neglects any depth variation in IME induced by advanced nuclear-burning stages (e.g., Na is
  produced through oxygen burning and its mass fraction raised in the O-rich shell by a factor of
  $\sim$\,10 over the solar value in model s15N while in model m15Mdot the sodium mass fraction
  is 3.4$\times$\,10$^{-5}$ throughout the ejecta). For IMEs like
  Ne, Mg, Si, S, or Ca, which are treated explicitly in the approx-21 nuclear network used
  here \citep{timmes_99},
  the distribution with depth of their mass fraction is comparable for models m15Mdot and s15N.

\begin{figure}
\epsfig{file=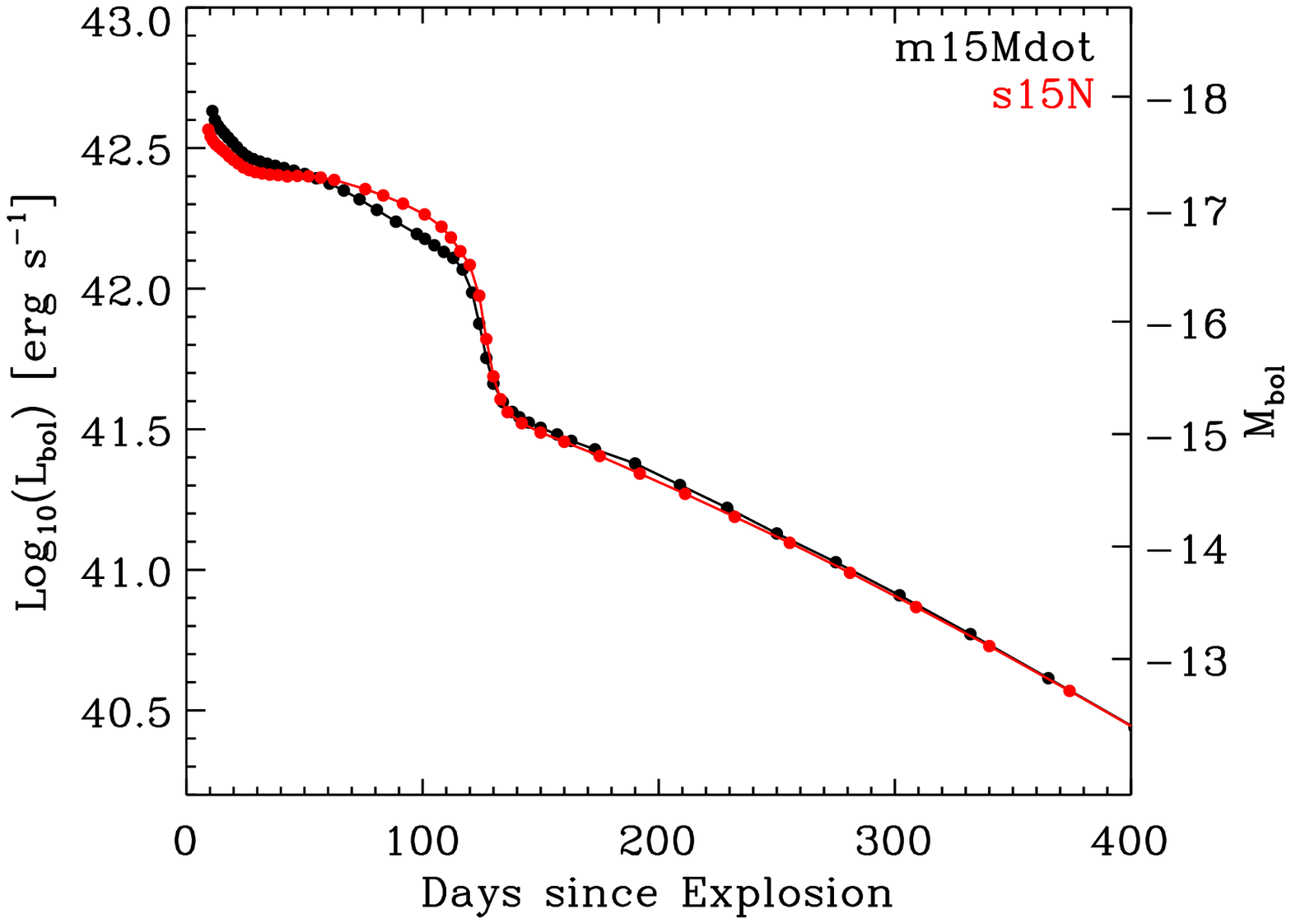,width=8.cm,bbllx=-10,bblly=10,bburx=510,bbury=370,clip=}
\epsfig{file=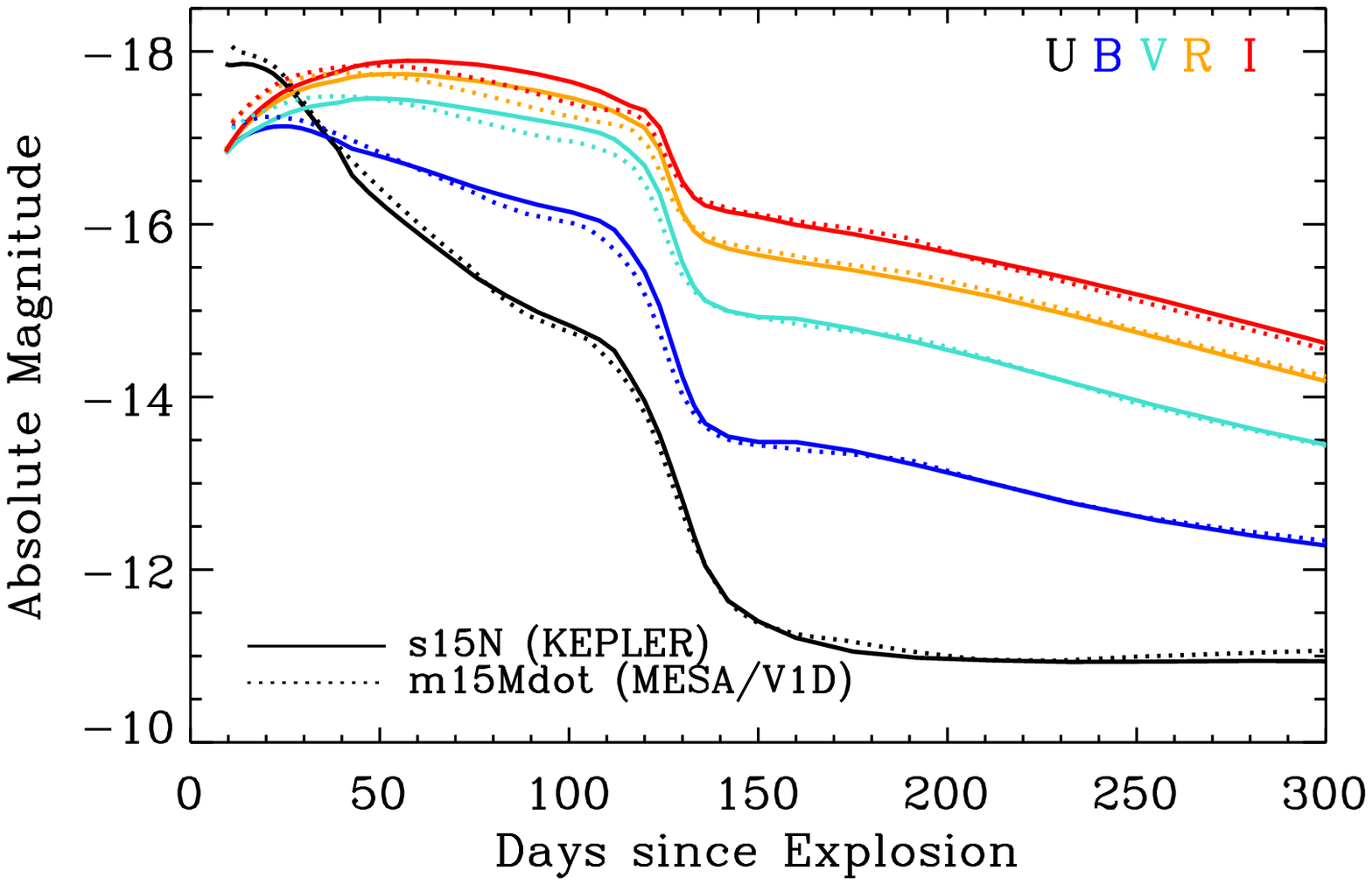,width=8.cm,bbllx=10,bblly=15,bburx=510,bbury=330,clip=}
\epsfig{file=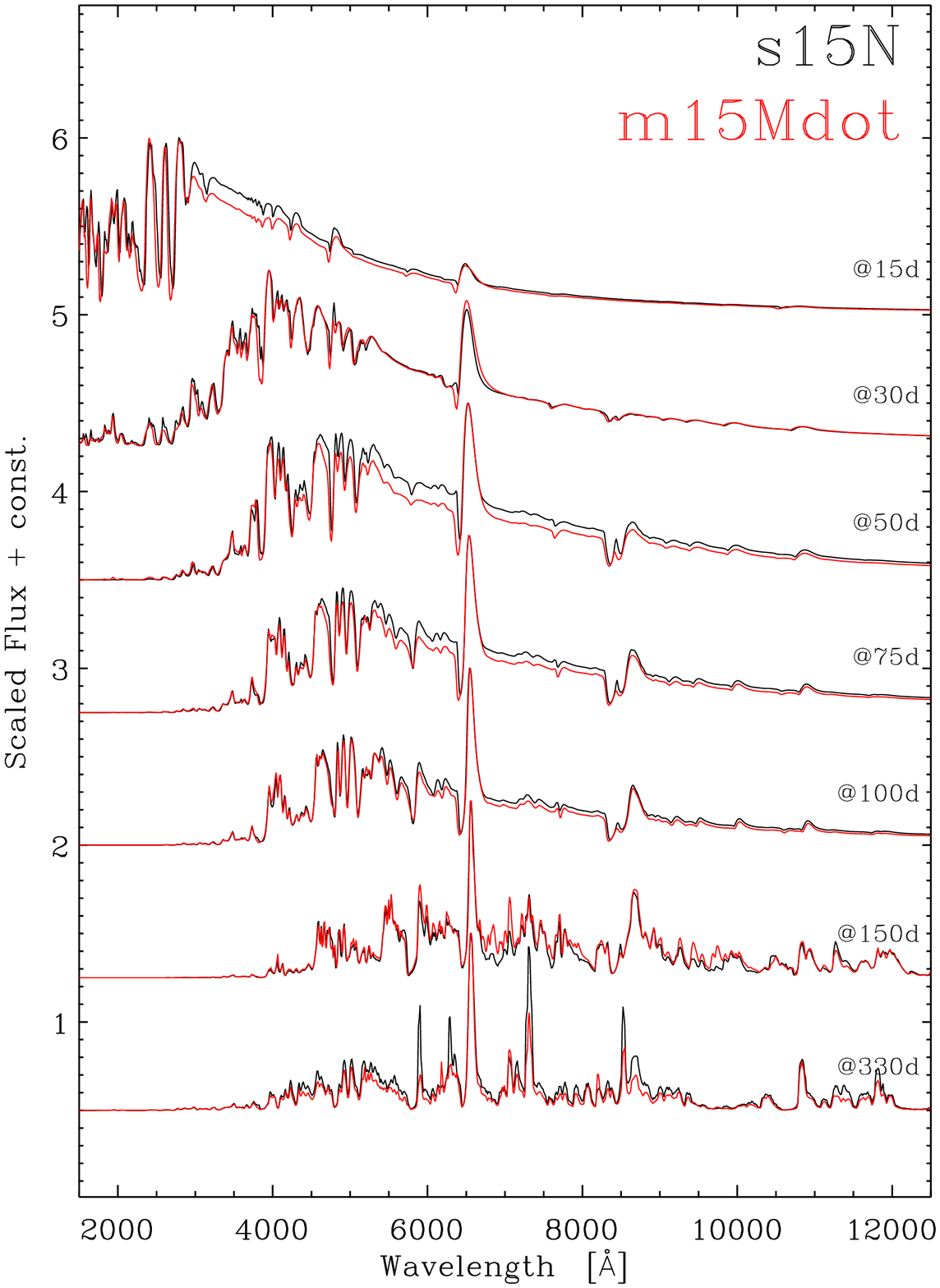,width=8.cm,bbllx=10,bblly=60,bburx=566,bbury=793,clip=}
\caption{Comparison of  \cmfgen\ results based on a pre-SN and explosion
model computed with \kepler\ (model s15N) or based on a pre-SN evolution computed with \mesa\
and exploded with \v1d\ (model m15Mdot). Both models have the same main-sequence mass of 15\,\msun,
but also have similar pre-SN star and ejecta properties (Table~\ref{tab_progprop}).
We show the bolometric light curves (top), multi-band light curves (middle), and spectral evolution (bottom;
spectra are normalized at the redmost wavelength and stacked vertically for clarity).
\label{fig_s15_m15}
}
\end{figure}

  Turning to the SN II-P radiation properties, we find that
  the bolometric light curves for model s15N and m15Mdot are analogous (Fig.~\ref{fig_s15_m15}),
  overlapping closely at early photospheric times and  nebular epochs. We have a 25\% difference in bolometric
  flux during the second half of the plateau. Inspecting the photospheric conditions at 90\,d after explosion,
  we find that models m15Mdot and s15N have at that location the same temperature of 5200\,K but  radii
  of 2.334$\times$\,10$^{15}$\,cm and 2.589$\times$\,10$^{15}$\,cm, which translates into the 23\% offset
  in bolometric flux. At corresponding ejecta velocities of $\sim$\,3000\,\kms, the s15N ejecta remains more
  optically thick, which likely arises from the $\sim$\,10\% larger
  \iso{56}Ni mass and the stronger mixing into the H-rich envelope.
  The length of the plateaus coincides. More striking is the comparable color evolution followed by each
  simulation. As we argue in the next section, this largely results from the similar progenitor radii.
  Finally, the spectral evolution is nearly identical for each model, revealing the same features, with the
  same strengths and widths at most times. The larger difference is seen at nebular times. The decay energy
  injection being the same, the spectral differences arise at such late times from the smaller mass of oxygen
  and sodium (weaker O\one\ and Na\one\,D), sometimes combined with a different species distribution.
  This is the case for Ca which is more centrally concentrated in model m15Mdot. This is likely the cause
  of the weaker Ca\two\ triplet and [Ca\two] 7300\,\AA\ doublet in model m15Mdot.

   This comparison suggests that with the \mesa\ parameters employed to calculate the evolution of
   model m15Mdot, the final RSG structure and composition are qualitatively and quantitatively similar
   to the \kepler\ model s15 \citep{WH07}. This also validates the combined approach with \mesa\ and \v1d\
   for feeding SN ejecta into \cmfgen.
   Accounting for sub-dominant species by setting their abundance to the environmental mass fraction also
   proves to be a suitable approach. The modest contrast we find also shows  how degenerate the SN II-P
   observables are with respect to the pre-SN conditions --- the sensitivity to abundance variations is weak
   for many species. In the next sections, we discuss in detail some dependencies of SN II-P radiation to
   changes in progenitor and explosions properties.

\section{Dependency on progenitor radius}
\label{sect_rad}

   To assess the influence of the progenitor radius on SN II-P
   radiation, we selected models m15mlt1 ($R_{\star}=$\,1107\,\rsun) and m15mlt3
   ($R_{\star}=$\,501\,\rsun). Models evolved at different metallicity or with enhanced
   rotation also cover a range of pre-SN star radii too, but they differ in other ways, while models
   m15mlt1 and m15mlt3 have essentially the same He-core and H-rich envelope mass.
   Below, we first discuss the \cmfgen\ predictions for these two models, then compare
   to SN\,1999em observations, before drawing the implications of these results.

\begin{figure}
\epsfig{file=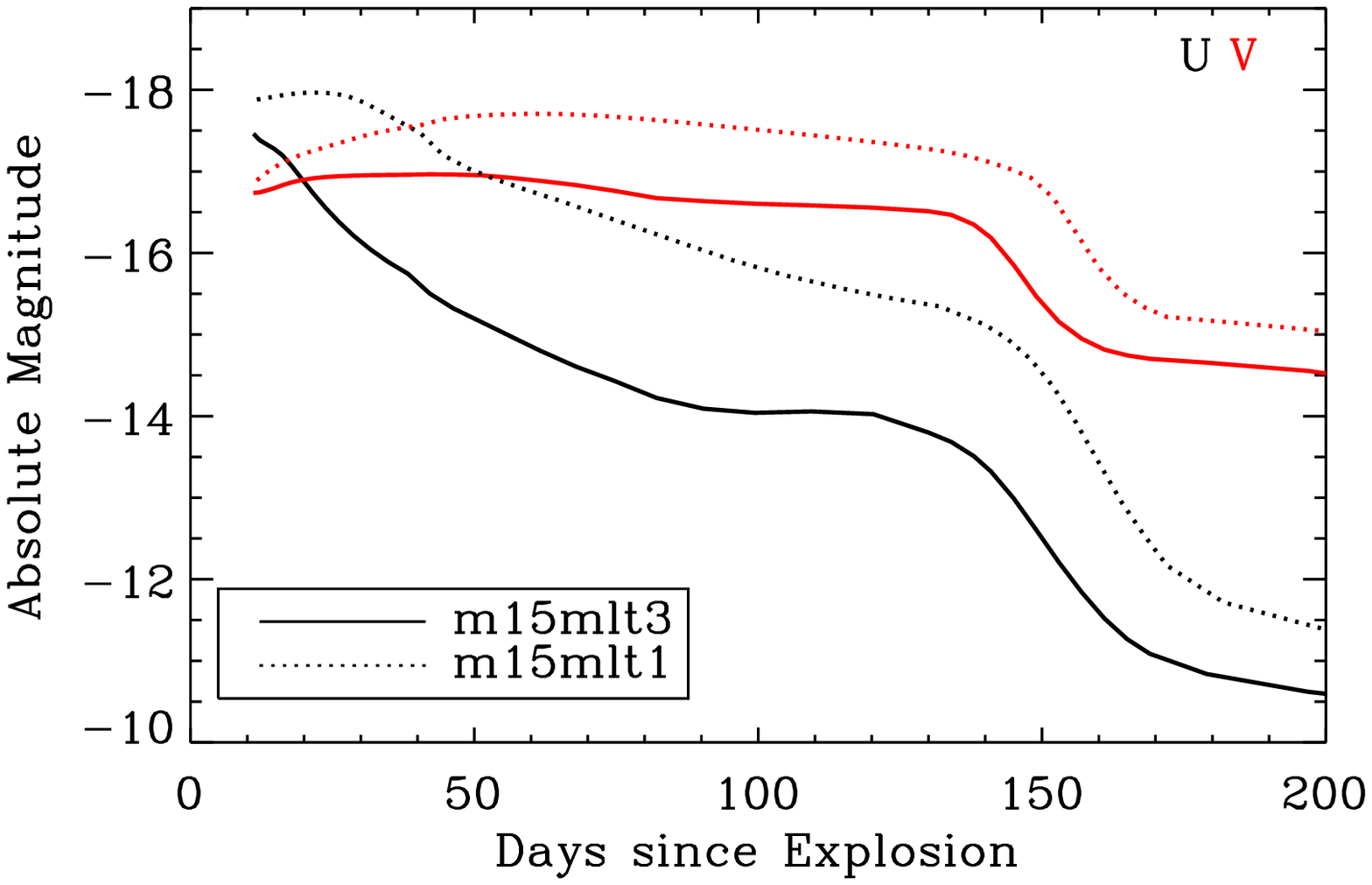,width=8.cm}
\epsfig{file=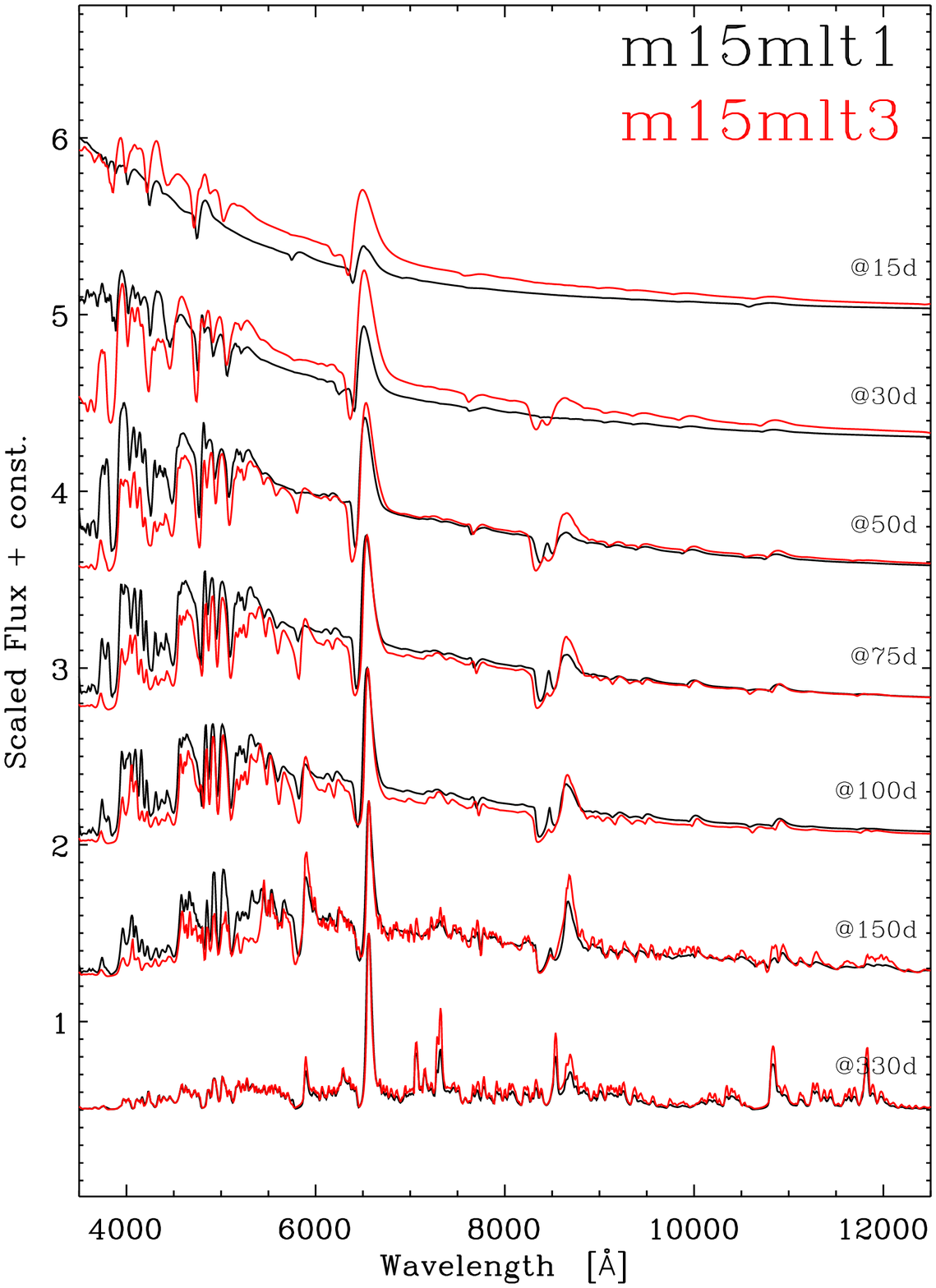,width=8cm}
\caption{
Comparison of the photometric (top) and spectral evolution (bottom) for models m15mlt1
and m15mlt3. Model m15mlt1 corresponds to the explosion of a star with a larger progenitor radius
($R_{\star}=$\,1110\,\rsun), which results
from the smaller adopted mixing-length parameter. Consequently, it has a brighter and bluer plateau, evidenced
spectroscopically through the delayed photosphere recombination compared to model m15mlt3
($R_{\star}=$\,500\,\rsun), which arose from a more compact progenitor.
Having similar cores, they exhibit similar nebular-phase spectra (an offset of $\sim$\,40\%
in \iso{56}Ni mass leads to a comparable shift in nebular flux level).
\label{comp_mlt}
}
\end{figure}

   \subsection{Results for the luminosity, color, and spectral evolution}

   Owing to its more compact progenitor star, model m15mlt3 is subject to a stronger
   cooling through expansion. We thus expect to see a smaller plateau luminosity and
   a more rapid transition to the recombination phase compared to
   model m15mlt1 (the contrast with model m15mlt2 is present albeit weaker).

   Model m15mlt3 indeed produces a genuine plateau in the $V$-band light curve
   (one magnitude fainter than the representative brightness in m15mlt1),
   and a systematic fading in the $U$-band (fainter by about 2 magnitudes than in model m15mlt1)
   from the onset of the simulation at 10\,d after explosion (top panel of Fig.~\ref{comp_mlt}).
   The onset of the plateau phase (i.e., when $T_{\rm phot}$ drops down to $\sim$\,5500\,K, or when
   the $V$-band starts to flatten after the explosion) is at $\sim$\,20\,d after explosion in m15mlt3
   but as much as $\sim$\,45\,d in model m15mlt1.

   Hence, a factor of two change in progenitor radius produces an interesting morphological change
   in a SN II-P multi-band light curve, reflecting modestly the stronger trend observed in
   Type II SN colors arising from compact BSG-star explosions (e.g., SN\,1987A, for which \cmfgen\ does reproduce
   the $U$-band fading satisfactorily when the model atoms are complete enough;
   \citealt{DH10,li_etal_12}) and from more extended RSG-star explosions (see, e.g., \citealt{brown_etal_09}).

   The impact of the progenitor radius on SN II-P radiation is better revealed by inspecting
   the spectral evolution, from the photospheric to the nebular phase (bottom panel of Fig.~\ref{comp_mlt}).
   Confirming the color evolution discussed above, model m15mlt3 systematically shows a cooler
   spectrum during the plateau phase, transitioning to the recombination phase
   about 20 \,d earlier.
   This is noticeable through the earlier appearance of Ca\two\ and Fe\two\ lines as well as the weakening of
   the UV flux, which results from enhanced photospheric cooling and line blanketing.
   The contrast with model m15mlt1 ebbs as time progresses. Leaving aside the 40\% difference in
   absolute flux level between the two (which stems from a 40\% difference in \iso{56}Ni mass), the nebular
   phase spectra are essentially identical, i.e., they have the same color, identical lines of a similar width
   but with a strength that differs by few tens of percent.

   A change in progenitor radius therefore affects primarily the plateau phase. The correlations obtained
   by \citet{LN85} suggest a 0.5\,mag brighter plateau for our m15mlt1 model parameters compared
   to m15mlt3, while we obtain a larger contrast of $\sim$\,1\,mag. Our computation of the full spectrum
   at all times in non-LTE and with large
   model atoms reveal an important change in color evolution and light-curve morphology. These changes
   suggest that a smaller progenitor radius may in fact cure the persistent color problem we encounter with models
   s15O/s15N (Section~\ref{sect_s15}) when comparing to SN II-P light-curve properties.

\begin{figure*}
\epsfig{file=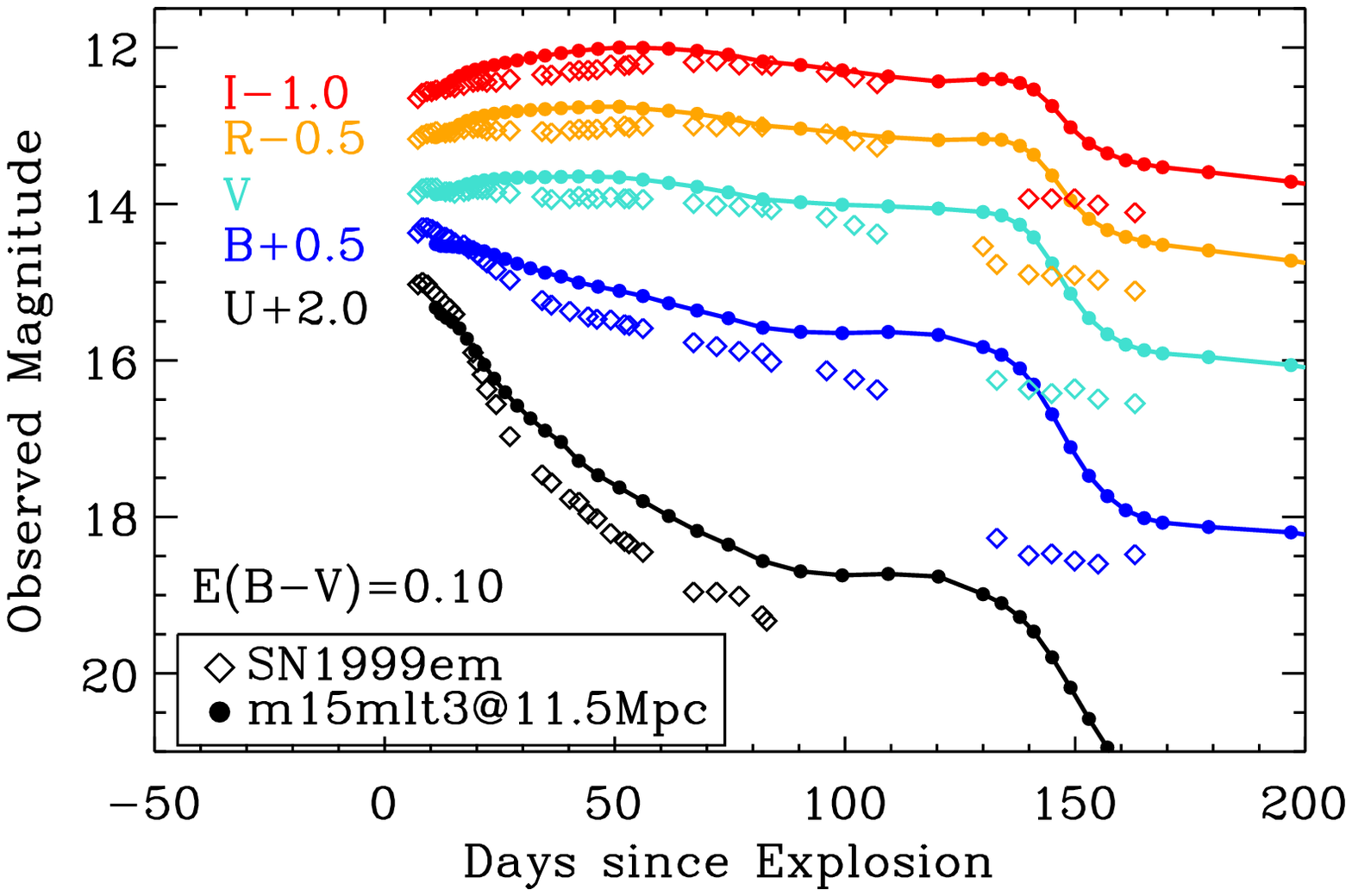,width=8.5cm}
\epsfig{file=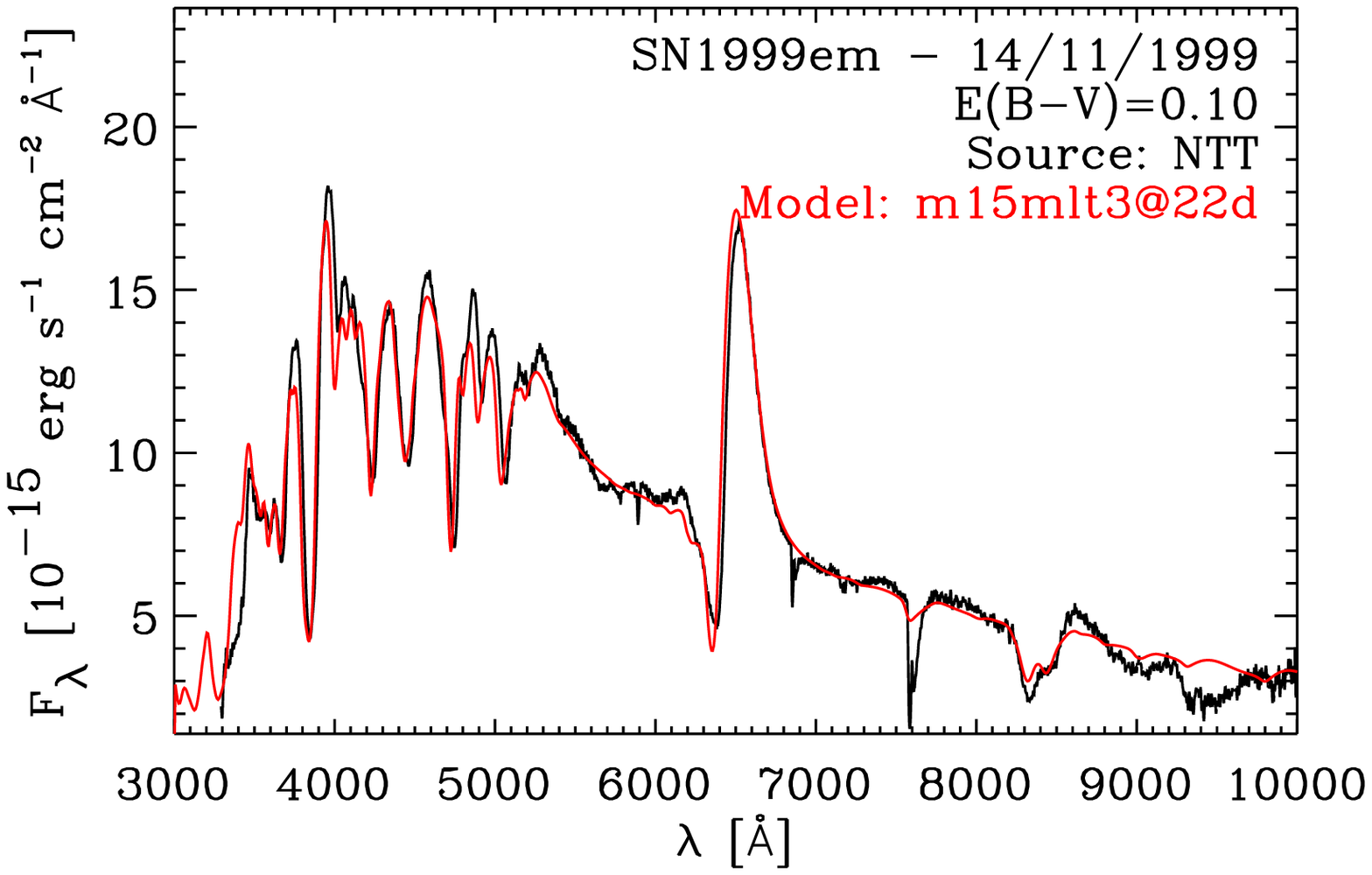,width=8.5cm}
\epsfig{file=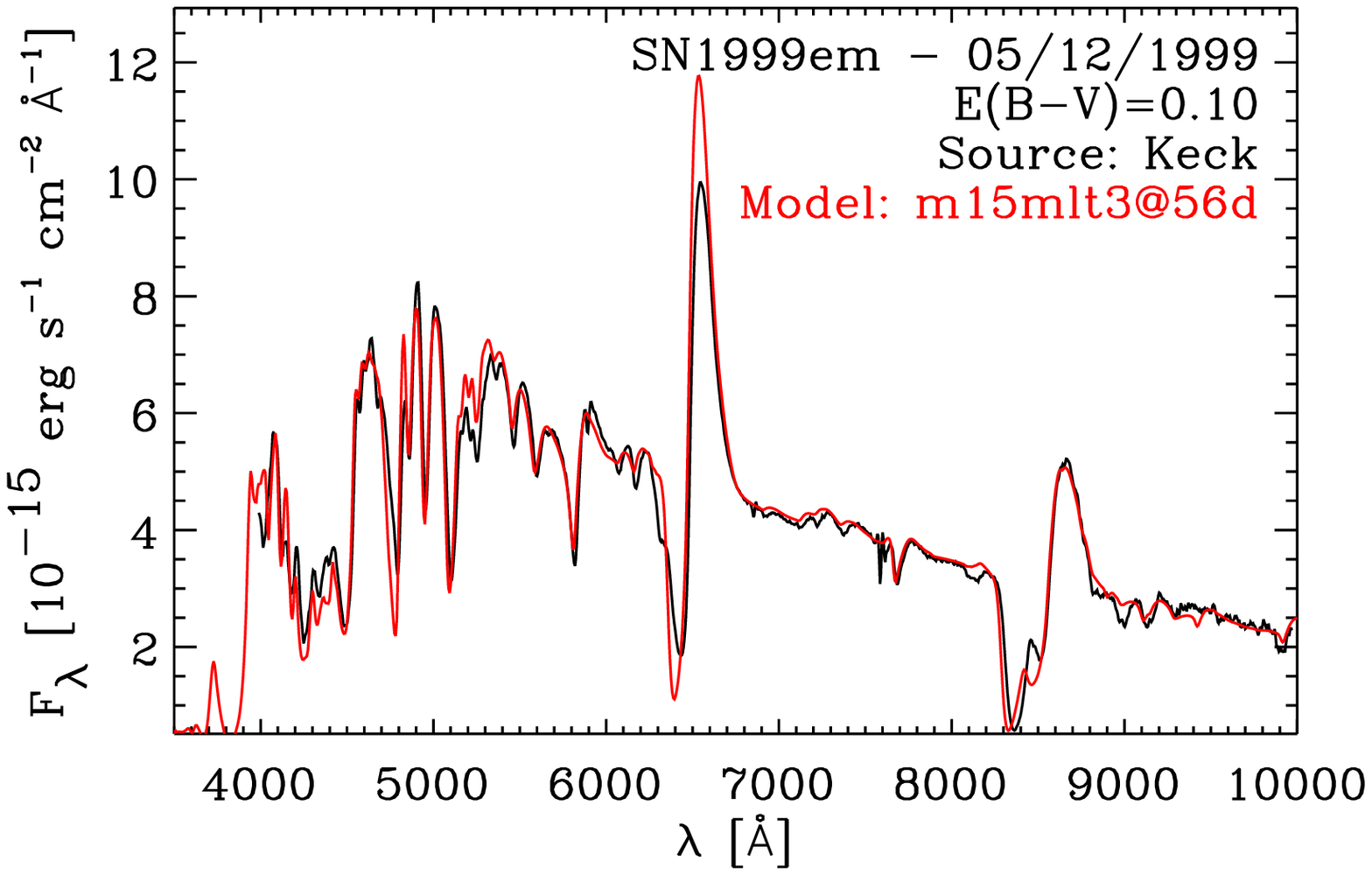,width=8.5cm}
\epsfig{file=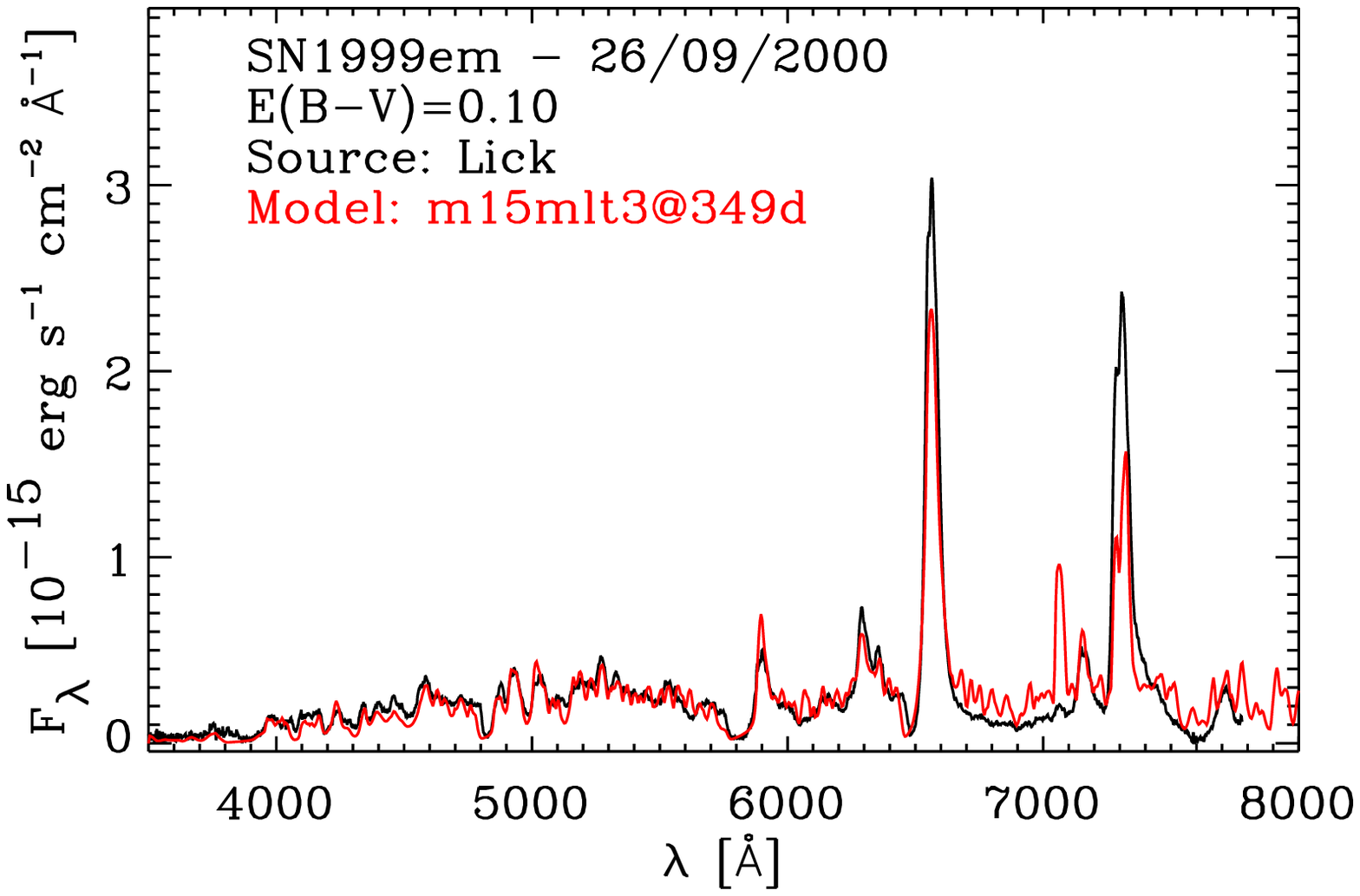,width=8.5cm}
\caption{Comparison between the observations of SN\,1999em and  model m15mlt3
(these results should be compared with those for model s15N shown in
Figs.~\ref{fig_mag_s15new_vs_99em}--\ref{fig_spec_s15new_vs_99em}),  which arises from a
more compact RSG star ($R_{\star}=$\,500\,\rsun) than model s15N ($R_{\star}=$\,810\,\rsun).
The larger H-rich envelope mass produces a longer plateau than observed (something easily corrected
for if we invoke a smaller mass RSG or stronger mass loss) while other synthetic signatures match the
SN\,1999em multi-band light curves and spectra simultaneously.
\label{fig_spec_mlt_99em}
}
\end{figure*}

\subsection{Comparison to observations}

  When comparing model m15mlt3 (model with the smaller progenitor radius and faster cooling/recombination)
  with the observations of SN\,1999em, we find that the agreement is much improved compared
 to what we obtained previously with model s15N (top-left panel of Fig.~\ref{fig_spec_mlt_99em}).
 The multi-band light curves agree rather well, albeit
 a slight offset in $U$ and $B$ bands. We have a genuine plateau as observed, although we find that
 the plateau length in model m15mlt3 is too long. This stems from its large H-rich envelope mass of
 10.16\,\msun, which is 1.5\,\msun\ larger than in models s15N or m15Mdot whose plateau length is
 on the order of 120\,d (in rough agreement with SN\,1999em), and from the somewhat larger
 \iso{56}Ni mass of the model (0.086\,\msun in m15mlt3 compared to either the 0.036\,\msun\ inferred
 by \citet{utrobin_07} or to the 0.056\,\msun\ inferred by \citet{bersten_etal_11}).

  Spectroscopically, the agreement with observations is better than obtained with model s15N
  (modulo a time shift at the end of the photospheric phase since the plateau is $\sim$\,30\,d longer
  in model m15mlt3 than observed in SN\,1999em). The comparison at 22\,d is remarkable
  (top-right panel of Fig.~\ref{fig_spec_mlt_99em}), since
  it matches in relative flux, in absolute flux, and corresponds to the same post-explosion
  time of 22\,d as inferred for SN\,1999em \citep{DH06_SN1999em}. During the recombination phase,
  the agreement is also very good but to match the color we need to take a 13\,d older
  ejecta model  (bottom-left panel of Fig.~\ref{fig_spec_mlt_99em}). The match at
  nebular times is excellent (bottom-right panel of Fig.~\ref{fig_spec_mlt_99em}), which supports
  the expectation that the H-rich envelope properties (e.g., its total mass or original extent)
  have a weaker influence on the nebular-phase properties.

  Our model  m15mlt3 has properties $E_{\rm kin}=$\,1.3\,B, $M_{\rm H-env}=$\,10.2\,\msun and
  $R_{\star}=$\,500\,\rsun. This is in good agreement with the findings of \citet{utrobin_07}.
  \citet{bersten_etal_11} find a somewhat larger progenitor radius and argue for extensive
  \iso{56}Ni mixing into the H-rich envelope in order to match the near-constant plateau luminosity.
  These two works quote ejecta masses but what these truly model
  is the H-rich envelope mass, which in both works is on the order of 10\,\msun.
  We stress again that the plateau light curve sets a meaningful constraint on the H-rich envelope properties
  (including its mass), but provides no information on the He-core mass.
  Determining the total {\it ejecta} mass requires the additional modeling of the nebular spectra to infer
  the oxygen and calcium masses \citep{li_mccray_92,li_mccray_93,jerkstrand_etal_12},
  and helium-core kinematics \citep{DLW10b}.

\subsection{Implications}

  The only simulations in our grid (Table~\ref{tab_progprop}) that show a systematic fading
  in the $U$-band light curve
  are models m15mlt3 and m15z2m3, which both have a progenitor radius of $\sim$\,500\,\rsun.
  All other simulations, characterized by larger progenitor radii between 600-1100\,\rsun,
  exhibit a $U$-band plateau before fading, or even a bump for the largest progenitor stars.
  Our results thus indicate that in order to match the color evolution of SNe II-P, progenitor radii
  on the order of $\sim$\,500\,\rsun\ may be required.

  This seems incompatible with the inferences of RSG stars in the Galaxy and in the
  Magellanic clouds \citep{levesque_etal_05,levesque_etal_10}. Indeed, it is generally thought
  that RSG stars are very extended stars, with radii even as large as 1500\,\rsun. This notion
  is supported by interferometric observations of nearby RSG stars, yielding
  photospheric radii of $\sim$\,950\,\rsun\ for Betelgeuse \citep{haubois_etal_09,neilson_etal.11},
  and as much as 1420\,\rsun\ for VY CMa \citep{wittkowski_etal_12}.
  It is important, however, to realize that the RSG radius we constrain from SN II-P radiation
  modeling is the radius that contains the bulk of the mass. Photometric and spectroscopic
  observations of RSGs reveal instead the location where the photons decouple from the atmosphere,
  which corresponds to an optical-depth at that wavelength of 2/3. Unfortunately, there is ample evidence
  that these two radii do not coincide in RSGs, and that the situation is complicated further by the
  inherent uncertainties associated with stellar evolution/structure and stellar atmosphere calculations.

  As we demonstrate in Section~\ref{sect_mesa}, varying the mixing length parameter can dramatically
  alter the radius of the stellar model, i.e., the location where the Rosseland mean optical depth would be 2/3.
  This property is well known (see, e.g., \citealt{maeder_meynet_87}) and implies that stellar evolutionary models
  predict the RSG luminosity but make no reliable prediction about the surface radius and effective temperature.
  By varying the mixing length parameter, one can produce a huge range of RSG radii, without impacting the RSG luminosity
  or the helium-core properties. Experimenting with \mesa, we have produced 15\,\msun\ models that die as RSGs
  with radii of 2500\,\rsun\ ($\alpha=$\,0.1) down to 350\,\rsun\ ($\alpha=$\,5), all with $\log(L/L_{\odot})=$\,4.8.
  Note that for increasing convective fluxes, the convective motions
  become eventually supersonic, requiring a multi-dimensional and hydrodynamical treatment.
  Such supersonic motions are in fact inferred in RSG atmospheres \citep{josselin_plez_07}.

  The observational side is also problematic. Recently, using both optical and near IR ranges, \citet{davies_etal_13}
  have modeled the spectral energy distribution of RSG stars and found that these objects are much warmer than previously
  inferred, in particular through optical analyses alone \citep{levesque_etal_05}. This revision of RSG effective
  temperatures leads to a reduction of RSG radii by as much as 20--30\,\%. Furthermore,  convection,
  molecule and dust formation, as well as time-dependent mass loss rates complicate considerably the inference
  of a meaningful $\tau=$\,2/3 surface. Its location is function of the effective opacity of the medium, which varies
  with wavelength, and the atmospheric structure (density scale height for example).
  It is thus unclear how that surface, located somewhere within
  the outer $\sim$\,10$^{-5}$\,\msun\  may relate to the radius that contains the bulk of the RSG mass.

  Finally, as already discussed in Section~\ref{sect_mesa}, a fundamental property of RSG H-rich envelopes
is their low binding energy. This implies that little work is needed to modify their structure, and in particular
their extent.  For example, VY CMa is characterized by a huge radius, but it is also known to have an extremely
intense mass loss history \citep{decin_etal_06}. If it were to explode today, it would probably be classified
not as a SN II-P, but more probably as a SN IIn \citep{smith_vycma_09}, or as a SN II-Linear.
Phases of intense mass loss in RSGs may be associated with
a surface phenomenon, e.g., pulsations or some other mass loss mechanism \citep{yoon_cantiello_10}.
But alternatively, the envelope may be reacting to the strengthening of shell burning at the edge or within the helium core,
or to the emergence of acoustic/gravity waves excited by convection deep in the star \citep{quataert_shiode_12}.
In other words, the extreme radii of such RSG stars that we measure today may results from processes that are
ignored in current stellar-evolution codes, which assume a quasi-hydrostatic, 1-D, stellar structure.

  The strong sensitivity of SN II-P radiation to progenitor radius makes it a valuable probe of the pre-SN
  structure. It represents a useful observable to constrain
  RSG ``mass-radii'' (that same quantity coming out of stellar-evolution calculations), potentially more reliable
  than interferometric or spectroscopic/photometric measurements on RSG atmospheres which constrain
  an ambiguous, time-dependent, wavelength dependent, mass-loss rate dependent ``$\tau$-radius''.

\begin{figure}
\epsfig{file=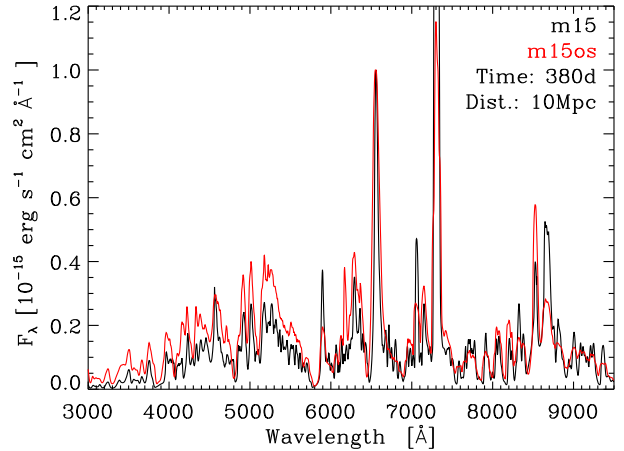,width=8.cm}
\caption{Comparison between synthetic spectra for models m15 and m15os  at 380\,d after explosion.
The m15os flux is normalized to that of model m15 at 6560\,\AA.
\label{fig_os}
}
\end{figure}

\section{Dependency on core overshooting}
\label{sect_os}

  As discussed in Section~\ref{sect_mesa}, increasing the efficiency of core overshooting
  in a 15\,\msun\ star has the primary effect of increasing the helium core size, while the associated
  increase in luminosity and mass loss leads to a reduction of the total star mass at death. In our
  simulations, the original surface chemistry is mildly altered, typically at the few percent level
  for intermediate mass elements.
  We note, however, that in the case of strong core-overshooting, the RSG surface
  chemistry might reflect instead the influence of the CNO cycle operating in the core,
  with Helium enhanced by a few percent,
  Nitrogen increased by a factor of $\sim$\,4, while C and O mass fractions are decreased by a few 10\%.
  Here, the main impact of enhancing core overshooting is to make a star act as if it was more
  massive, and thus bears a similar impact to stellar rotation \citep{maeder_meynet_00}.

  Using radiation hydrodynamics simulations of RSG star explosions,
  \citet{DLW10b} demonstrated that the velocity at the junction between the He core and the
  H envelope is a critical marker of the pre-SN star. They find that in order to reproduce the
  typical observed width of the O\one\,6300\,\AA\ fine-structure line or H$\alpha$, the outer edge of the He core
  or the inner edge of the H-rich envelope in the progenitor must be expelled at velocities on
  the order of $\sim$\,1500\,\kms. In turn, this implies a progenitor star of $\lesssim$\,20\,\msun,
  because higher mass stars would place that junction at higher velocities (for the same
  standard explosion energy). They note that if stellar rotation is allowed for, it pushes this
  limit to lower main-sequence masses. Since overshooting and rotation cause the same
  basic effect on the He-core mass, enhanced overshooting will lower that mass limit
  of SN II-P progenitors to lower values than the estimated $\sim$\,20\,\msun\
  \citep{DLW10b,jerkstrand_etal_12}.

  Inferences about the core are best done at nebular times, when the radiation comes
  from ejecta regions that used to reside in the He-core and at the base of the H-rich envelope.
  In our synthetic spectra for model m15 and m15os (Fig.~\ref{fig_os}), we find that the typical
  half-width-at-half-maximum (HWHM) of H$\alpha$ is 2500\,\kms\ in model m15os and only
  $\sim$\,1000\,\kms\ in model m15.
  These values are close to the ejecta velocity we measure at the junction
  between H-rich and He-rich shells, which are 2500 and 1500\,\kms\ for models m15os and m15,
  respectively. Given the similar ejecta kinetic energy of 1.4 and 1.27\,B for these models,
  the difference stems in part from the contrast in $M_{\rm ejecta}/M_{\rm He \ts core}$  ratio.

  We note that the greater \iso{56}Ni mass synthesized in model m15os probably contributes to making the
  HWHM of H$\alpha$ larger. In both models, the Doppler velocity at maximum absorption in H$\alpha$
  is located within the H-rich ejecta shells, as 3800 and 5300\,\kms, both larger than observed in SN\,1999em.
  Getting the profile absorption of H$\alpha$ ``right" is however a recurrent problem, even at photospheric
  times, because the line is so sensitive to even subtle changes in ionization.

  To summarize, core overshooting increases the uncertainty when inferring the
  mass of a SN II-P progenitor, because, like rotation, it makes the star evolve  as if it was more
  massive but unaffected by overshooting.

\begin{figure}
\epsfig{file=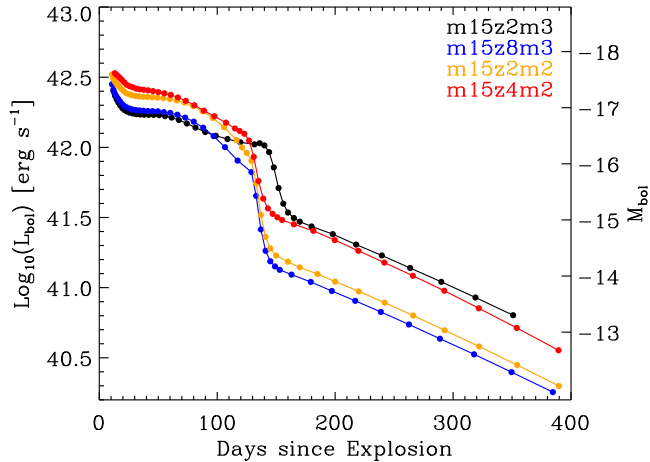,width=8.5cm}
\caption{Bolometric light curves for the m15 series for different progenitor
metallicities, from a tenth to twice the solar value. See Table~\ref{tab_progprop}
for details.
\label{fig_m15_z_lc}
}
\end{figure}

\begin{figure*}
\epsfig{file=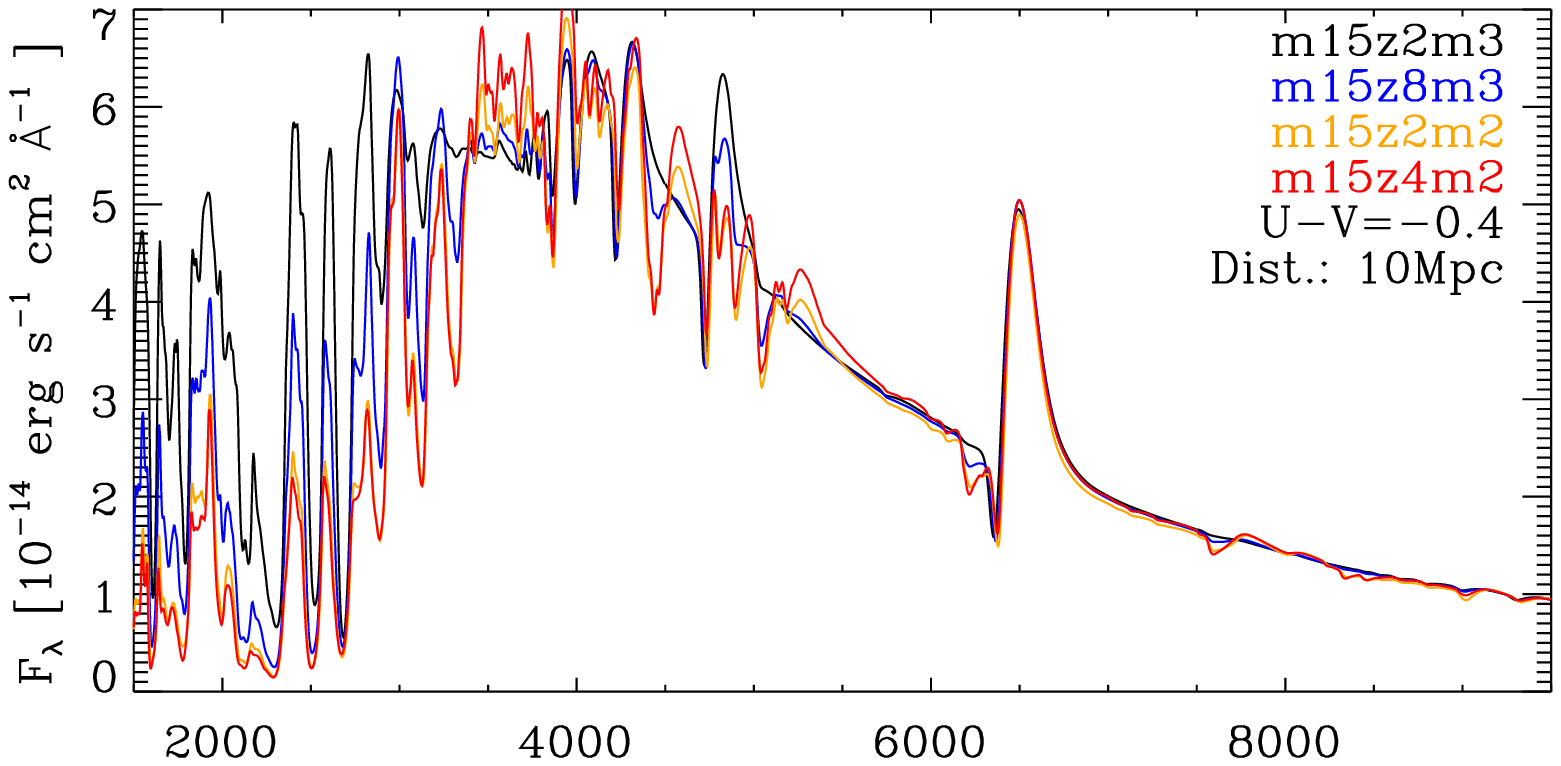,width=15.25cm}
\epsfig{file=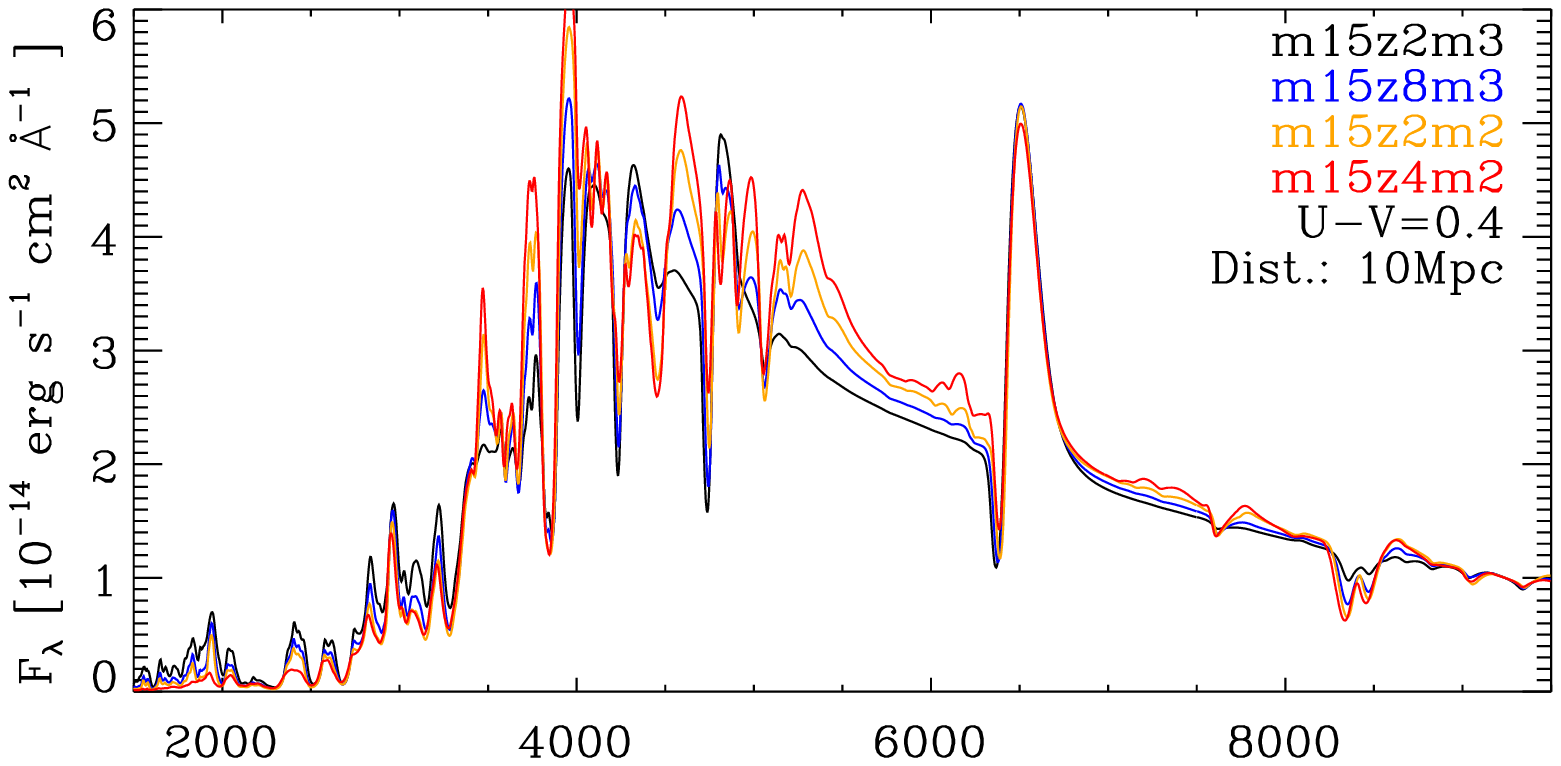,width=15.25cm}
\epsfig{file=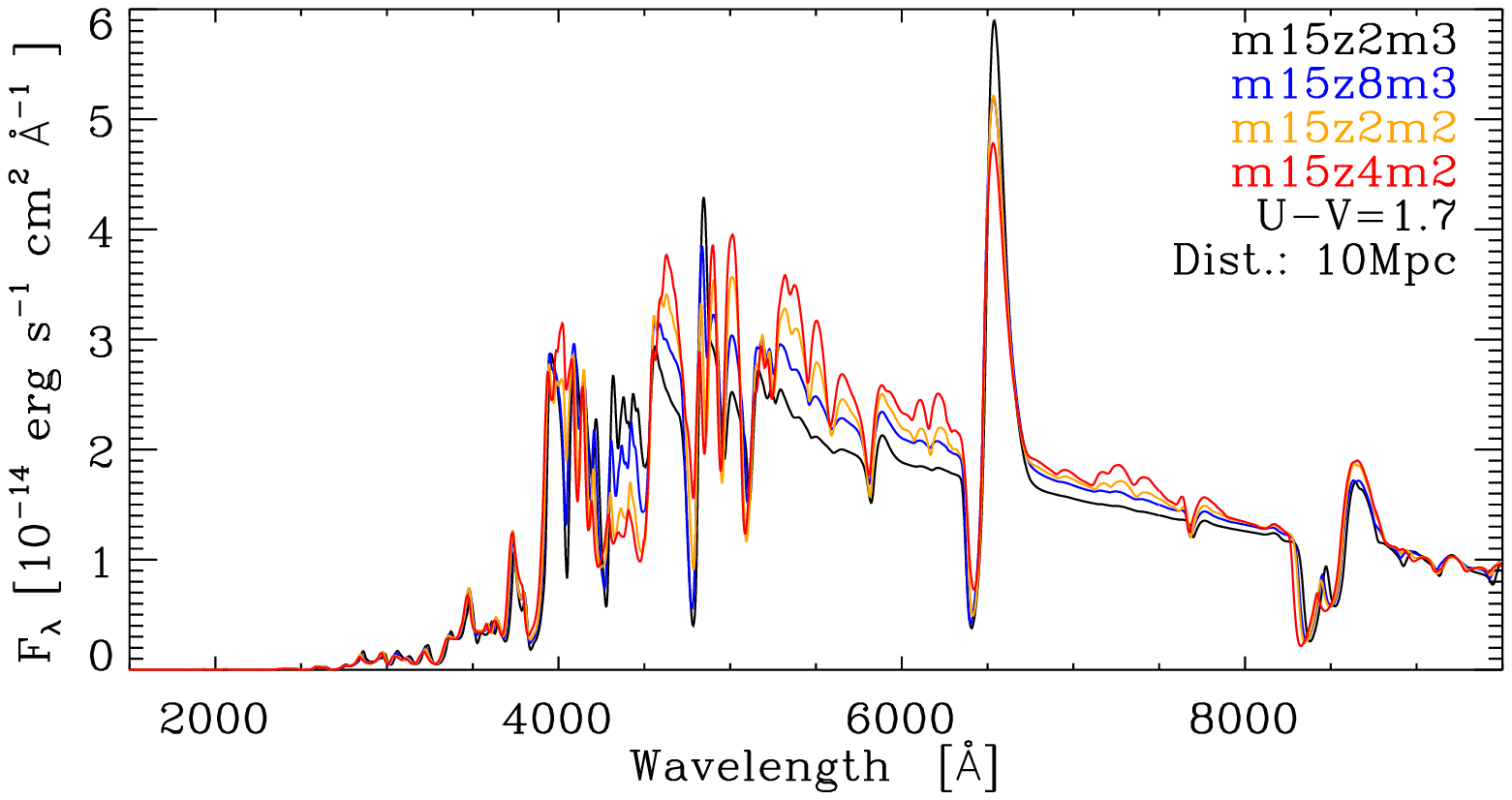,width=15.25cm}
\vspace{-0.2cm}
\caption{
Comparison of synthetic spectra at selected $U-V$ colors for SNe II-P arising from a 15\,\msun\
progenitor evolved with different initial metallicities, from sub-solar (model m15z2m3)
to super-solar (model m15z4m2) --- see Table~\ref{tab_progprop}.
Because the color evolution is different for each model, these correspond to different post-explosion times.
When $U-V=-$\,0.4, the SN ages are 17.1, 19.1, 22.1, and 22.9\,d for models m15z2m3, m15z8m3,
m15z2m2, and m15z4m2.
In the same model order and when $U-V=$\,0.4, the SN ages are 30.2, 30.7, 33.7, and 31.0\,d,
while when $U-V=$\,1.7, the SN ages are 80.2, 66.0, 72.6, and 63.3\,d.
\label{fig_m15_z_spec}
}
\end{figure*}

\begin{figure*}
\epsfig{file=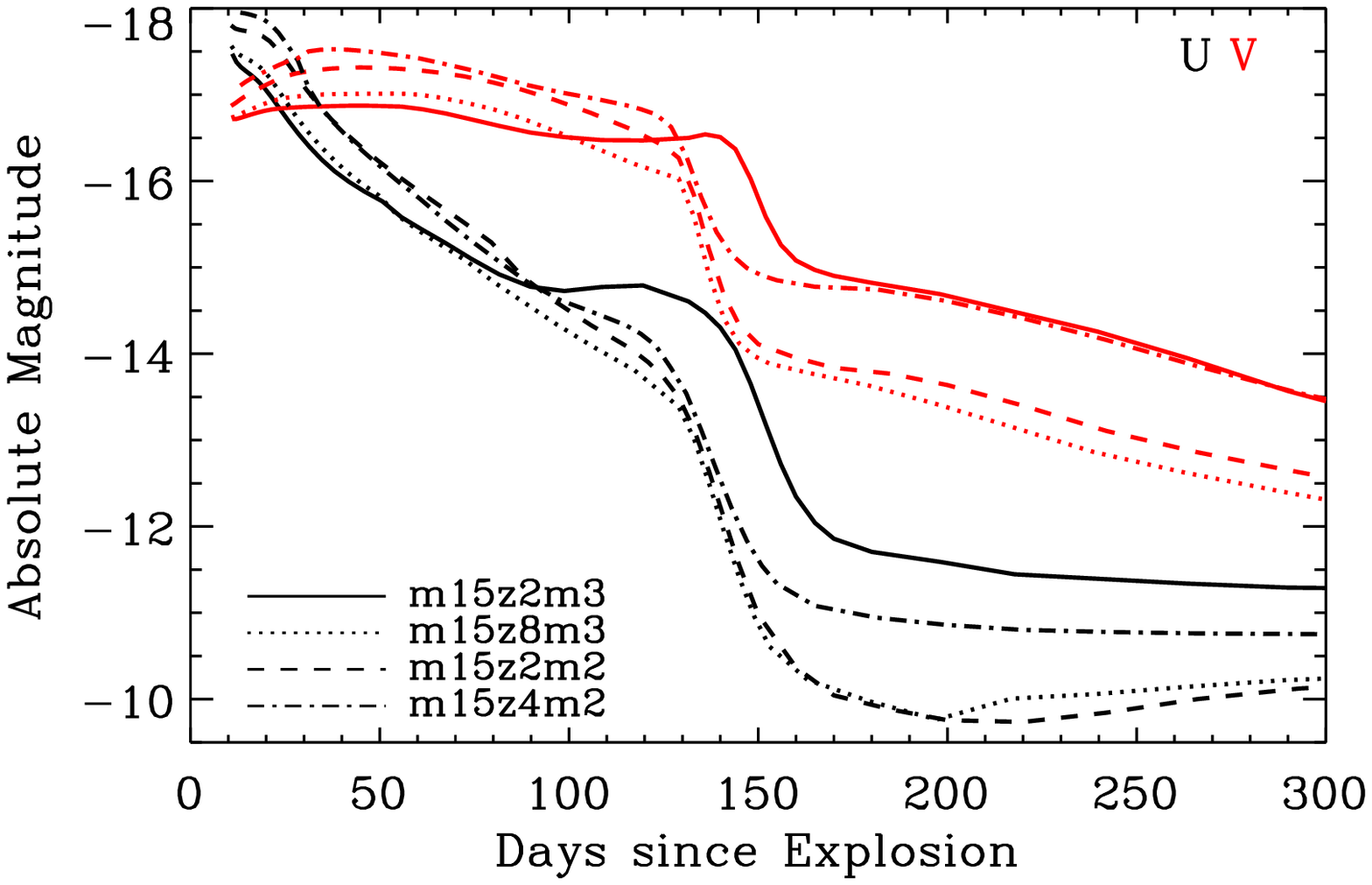,width=8.5cm}
\epsfig{file=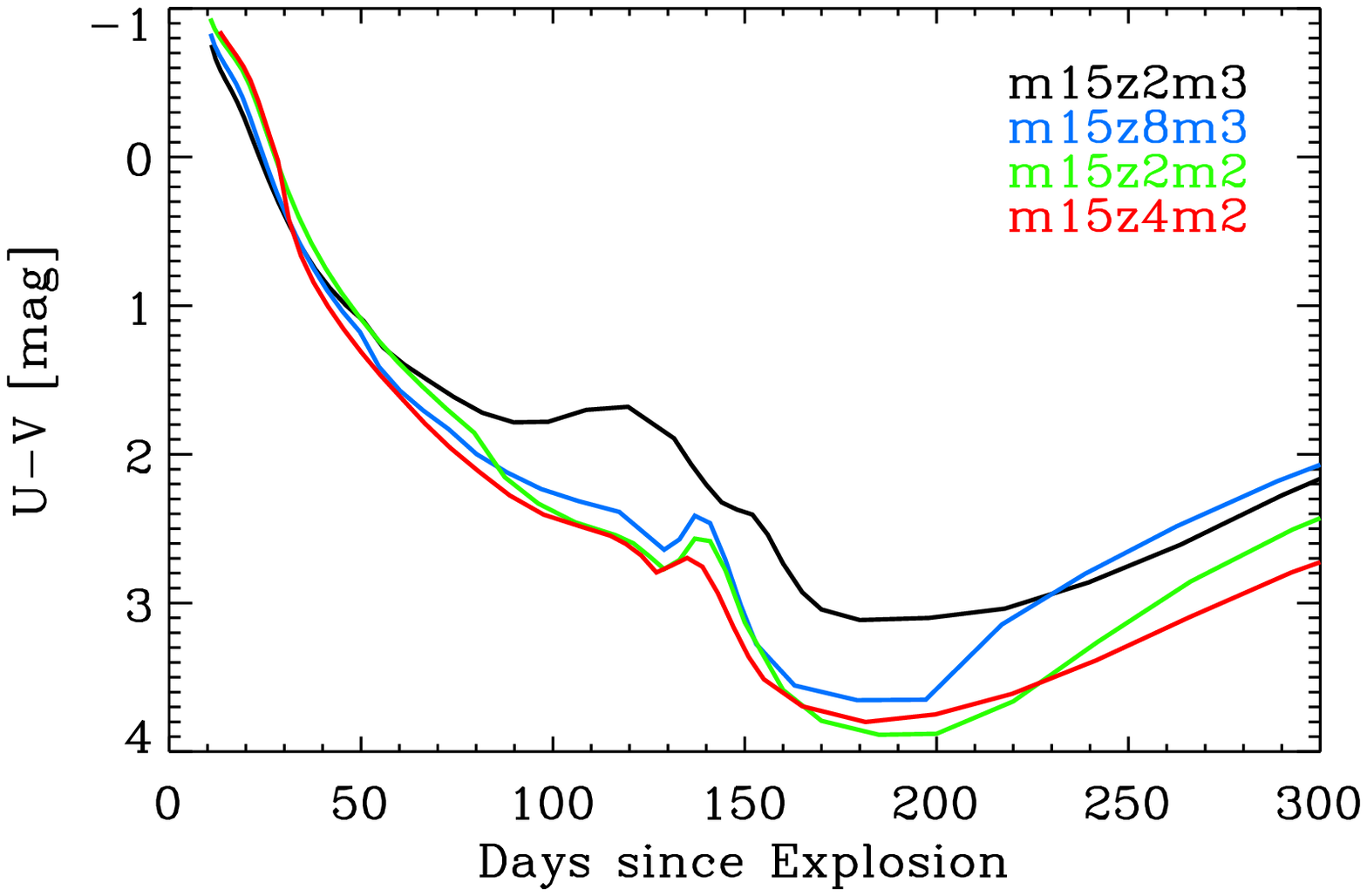,width=8.5cm}
\caption{{\it Left:} $U$ and $V$-band light curves for models m15z2m3, m15z8m3, m15z2m2,
and m15z4m2. {\it Right:} Evolution of the $U-V$ color.
\label{fig_m15_z_mag}
}
\end{figure*}

 \section{Dependency on metallicity}
\label{sect_z}

   The impact of metallicity on SN II-P radiation can be split into two categories.
First, metallicity is a key ingredient affecting pre-SN evolution. It may be a source of
under-abundant elements for nuclear reactions. Its influence on opacity,
in particular through variations in iron abundance, can cause the envelope
to shrink or expand because radiative diffusion in the envelope controls in part
the radius of the star. But more importantly, it can modulate the stellar-wind mass
loss rate and thus result in pre-SN stars of varying masses. Metallicity is critical
here because it provides the necessary species for molecule and dust formation
in cool stars, and ions with optically-thick lines in hotter stars \citep{cak}.
In their respective environments, each opacity source is key for driving a
stellar wind.

Second, metallicity will directly influence SN II-P spectra and colors.
Indeed, metals, even at solar metallicity, play a critical blanketing role
on the photospheric flux, blocking the flux at specific wavelengths or over wide
spectral regions through line overlap. Hence, metallicity has the potential to alter
the color as well as the line-profile signatures of SNe II-P. Through its effect on pre-SN
mass loss and the H-rich envelope mass, varying metallicities should lead to a range
of ejecta masses of SNe II-P in different environments, influencing in particular the length
of the plateau phase.

We expect all these effects to play a role in SN II-P properties but it is unclear today
if we have seen such metallicity effects. The main impediment is the wide-spread use,
only starting to change now, of targeted surveys, which focus on similar types of galaxies and
thus deliver a restricted view of the diversity of SNe in our Universe. Below, we discuss what
distinctive signatures we expect from SNe II-P at different metallicities.

With \mesa, we have run simulations for a 15\,\msun\ star at metallicities of 0.002 (model m15z2m3; one
tenth solar), 0.008 (m15z8m3), 0.02 (m15; our reference model, also called m15z2m2), and 0.04 (m15z4m2).
We find that at lower metallicity, the lower mass-loss rates produce higher mass stars at the time of death,
while the reduced H-rich envelope opacity (dominated by contributions from electron scattering and metal
lines) produces smaller radii (Table~\ref{tab_progprop}). Consequently, the cooling from SN expansion is modulated,
producing a fainter plateau and a faster color evolution for lower-metallicity SNe II-P. The lower envelope masses
at higher metallicity, on the other end, produce a SN II-P with a shorter plateau.
Finally, the varying \iso{56}Ni masses synthesized in the explosion introduce a range of nebular-phase luminosities
(Fig.~\ref{fig_m15_z_lc}).
By themselves, these properties do not set any constraint on the metallicity because they could stem
from different convection efficiencies affecting the progenitor radius, or different mass-loss rates peeling
differentially the H-rich envelope.
Unambiguous metallicity signatures are to be found instead in SN II-P spectra.

For as long as the SN II-P photosphere is in the outer parts of the H envelope (which holds for typically
half the photospheric phase; \citealt{DH11}), the spectrum formation region is located in shells that have
the metallicity at which the star formed, thus unaffected by nuclear burning
in the course of the evolution and explosion, or by mixing following explosion.
If mixing occurs during the pre-SN evolution, ashes from the CNO cycle may pollute the
outer progenitor envelope, but we find that apart from a sizeable impact on nitrogen, other species
are unaffected (e.g., H, He) or only mildly affected (e.g., C and O, at the few tens of percent level).
This is to be contrasted with the potential variation of {\it all} metal abundances by a factor of a few
if the SN is located in sub-solar or super-solar metallicity environments.

 To best reveal the influence of metallicity on the spectra, without being affected by the different cooling rates
 of the associated ejecta, we compare SN II-P simulations m15z2m3, m15z8m3,  m15z2m2 (our reference model),
 and m15z4m2 when they have the same $U-V$ color. This ensures that we are comparing the spectra
 when their photospheric (or color) temperatures are similar. An observer can make a similar comparison
 when the time of explosion is not known --- the reddening must, however, be known.
 In Fig.~\ref{fig_m15_z_spec},
 we show snapshots at three different epochs corresponding to $U-V$ of $-$\,0.4, $+$\,0.4, and
 $+$\,1.7\,mag (the corresponding post-explosion times in each case are given in the figure caption).

\begin{figure}
\epsfig{file=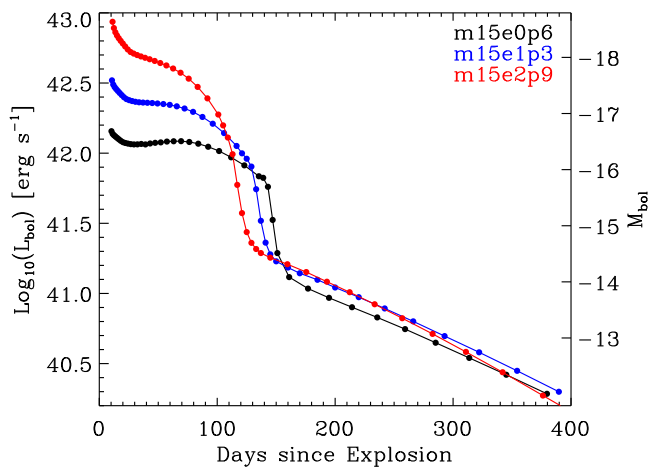,width=8.5cm}
\caption{Bolometric light curves for the reference model m15 but exploded to produce
ejecta kinetic energies of 0.6, 1.3, and 2.9\,B (models m15e0p6, m15e1p3, and m15e2p9, respectively).
\label{fig_m15_ekin_lc}
}
\end{figure}

\begin{figure*}
\epsfig{file=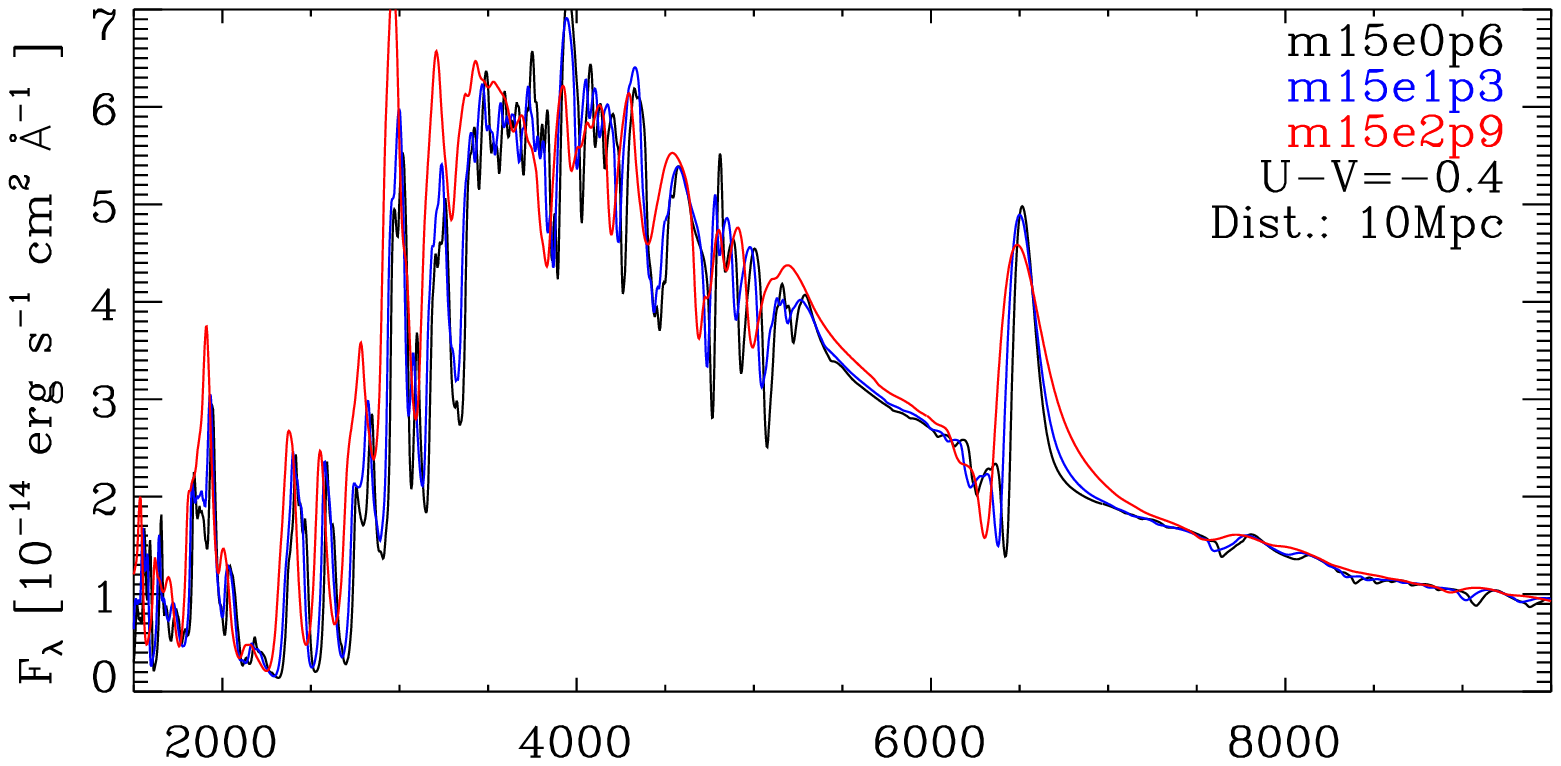,width=15.25cm}
\epsfig{file=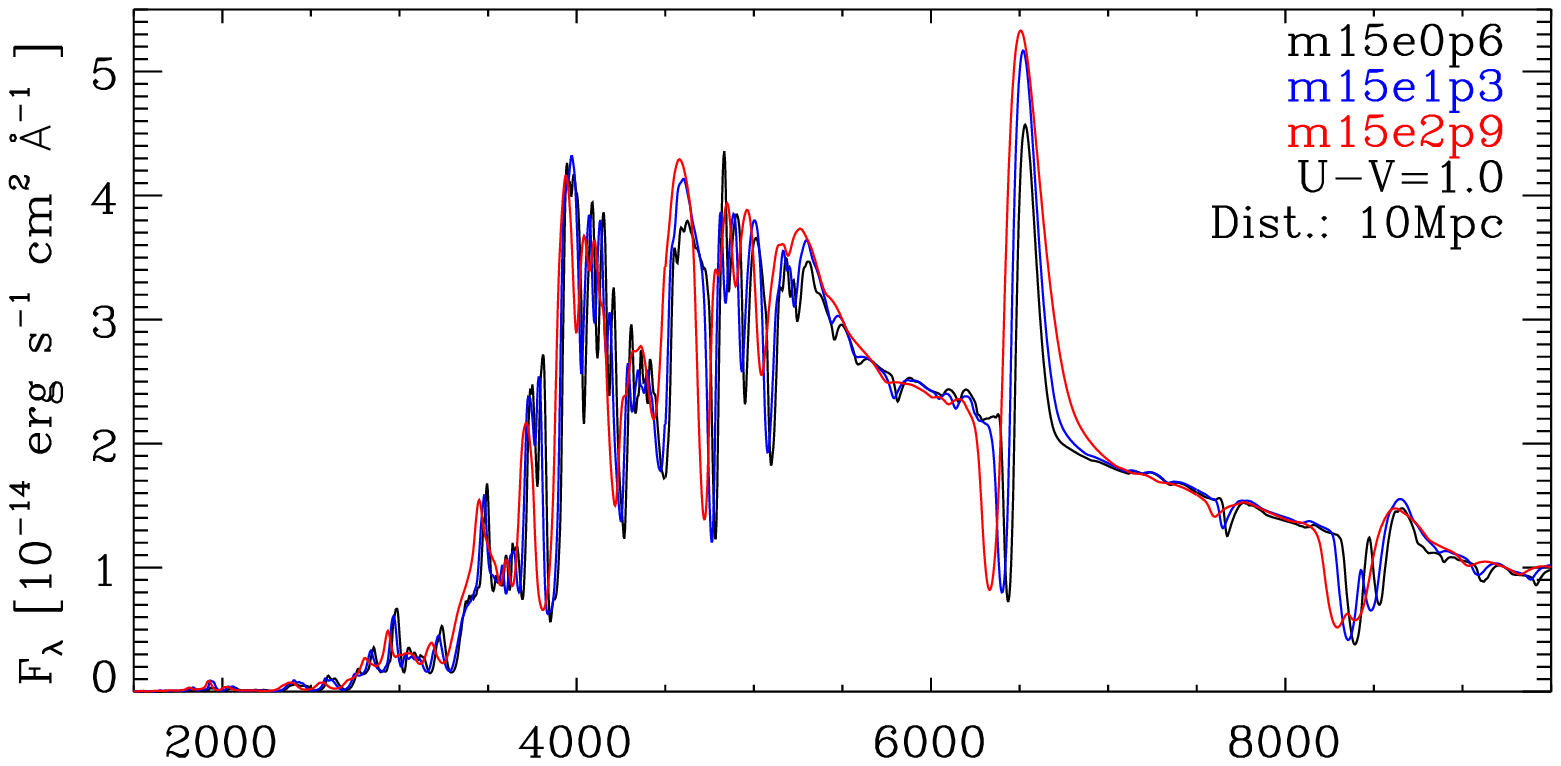,width=15.25cm}
\epsfig{file=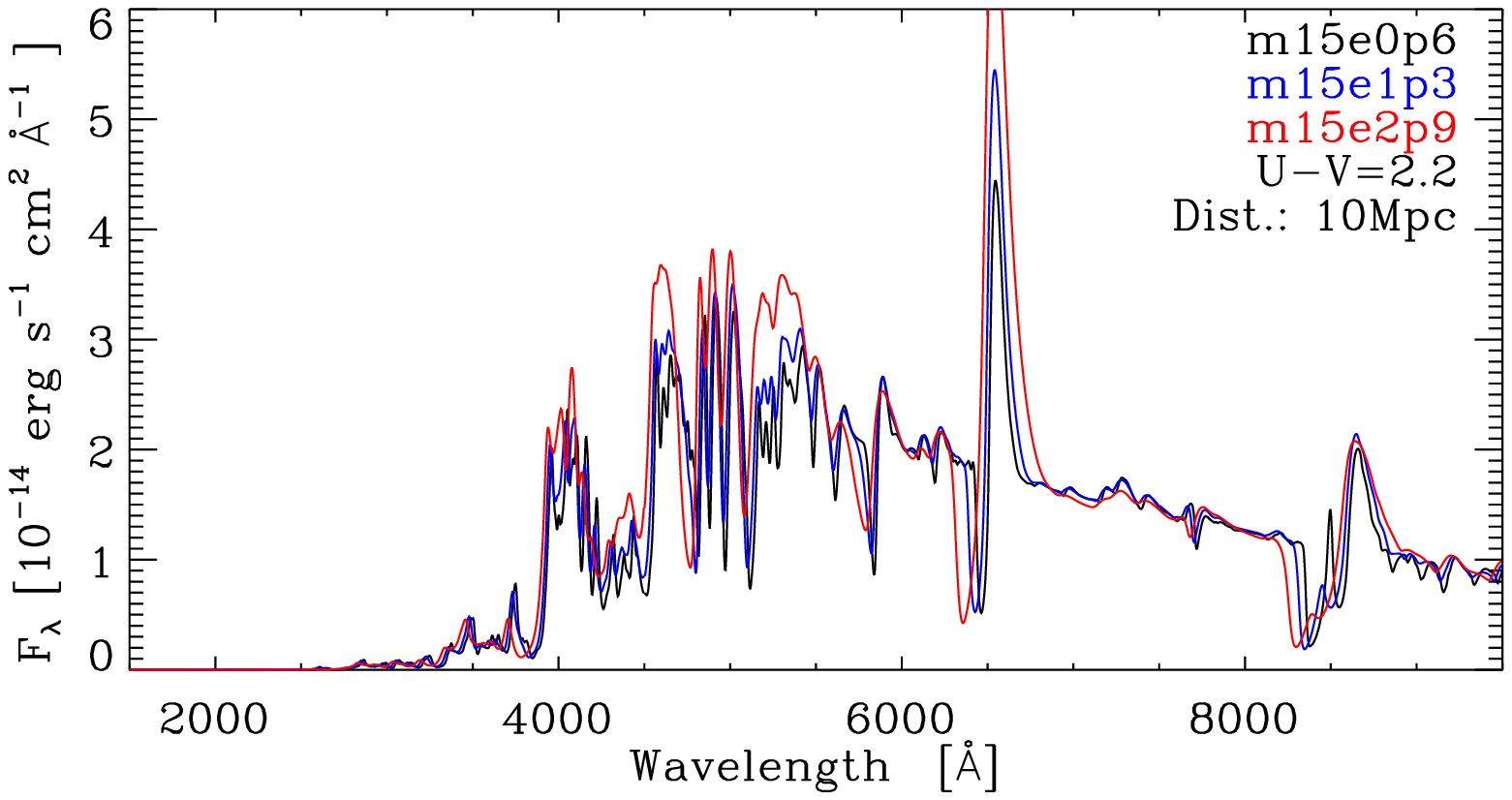,width=15.25cm}
\caption{
Spectral comparison at selected $U-V$ colors  for the reference model m15 but exploded to produce
ejecta kinetic energies of 0.6, 1.3, and 2.9\,B (models m15e0p6, m15e1p3, and m15e2p9, respectively).
The bolometric light curves for these models are presented in Fig.~\ref{fig_m15_ekin_lc}.
Each comparison corresponds to contemporaneous post-explosion times because all models have essentially
the same color at a given post-explosion time.
\label{fig_m15_ekin_spec}
}
\end{figure*}

  At early times ($U-V=-$\,0.4), the influence of metallicity is strong throughout the spectrum. In the UV,
  the flux level is significantly enhanced at lower $Z$, owing to the reduction of line blanketing,
  and in the optical, metal lines appear much weaker. Since all
  species other than H and He are at the environmental metallicity, we observe an effect in lines associated with
  O, Sc, Si, and Fe. As the ejecta recombines ($U-V=$\,0.4), the flux level in the UV becomes insensitive to
  metallicity variations (probably because of a saturation in line absorption and a reduction in gas emissivity
  at UV wavelengths), and the contrast in the optical is as strong as before, but is now also apparent in Ca lines
  (Ca\two\,H\&K as well as the triplet at 8500\,\AA) --- H$\alpha$ hardly
  changes between models at those times. Nearer the end of the plateau phase, the UV flux is identical amongst
  these models. Differences in the optical are caused primarily by Ti\two\ (around 4000-4500\,\AA) and Fe\two\ lines
  (model m15z2m3 shows very weak Fe\two\ lines). Individual lines like Na\one\,D or O\one\,7770\,\AA\ reveal the
  systematic abundance difference between models.

  Comparing the multi-band light curves for models m15z2m3, m15z8m3, m15z2m2, and m15z4m2 is
  difficult because they differ not just in metallicity but also in
  progenitor radii (from $\sim$\,500 to $\sim$\,800\,\rsun; resulting from the different envelope opacity)
  and \iso{56}Ni mass ($\sim$\,0.04 to $\sim$\,0.09\,\msun; inherent to the way the piston-driven ejecta
  was produced). Although lower metallicity models suffer less blanketing in the UV and should
  appear bluer, the effect is compensated by their lower progenitor radii that cause stronger
  cooling through expansion and hence cooler photospheres (Fig.~\ref{fig_m15_z_mag}).
  Identifying a metallicity effect in the color evolution is thus non trivial, in particular compared
  to the ease with which we find metallicity-dependent signatures in the spectra (Fig.~\ref{fig_m15_z_spec}).
  Although detailed modeling specifically tailored for SN\,1999em will be needed, the strength of lines
  from IMEs and IGEs in
  the spectra of SN\,1999em, and the good agreement with the solar-metallicity model m15mlt3 shown in
  Fig.~\ref{fig_spec_mlt_99em} (see also synthetic fits in \citealt{DH06_SN1999em,DH11}) suggest that the
  metallicity of the progenitor star may be solar. Hence, our models do not support a one-fifth-solar metallicity
  for SN\,1999em, as suggested by \citet{baklanov_etal_05}.

  The interesting result from this exploration is that SNe II-P show signs of differing metallicities not just
  in lines of Fe, but also in lines of O, Na, Sc, Ti, or Ca, and can thus provide a much richer assessment
  of metallicity variations than, e.g.,
  nebular-line analyses, which focus primarily on oxygen abundances. Being so luminous, one may
  use SN II-P spectra to determine the SN metallicity, which is critical to infer the
  influence of metallicity on pre-SN evolution. Importantly, we are measuring the metallicity  of the SN rather
  than a mean metallicity of the galaxy or the SN environment.

\begin{figure*}
\epsfig{file=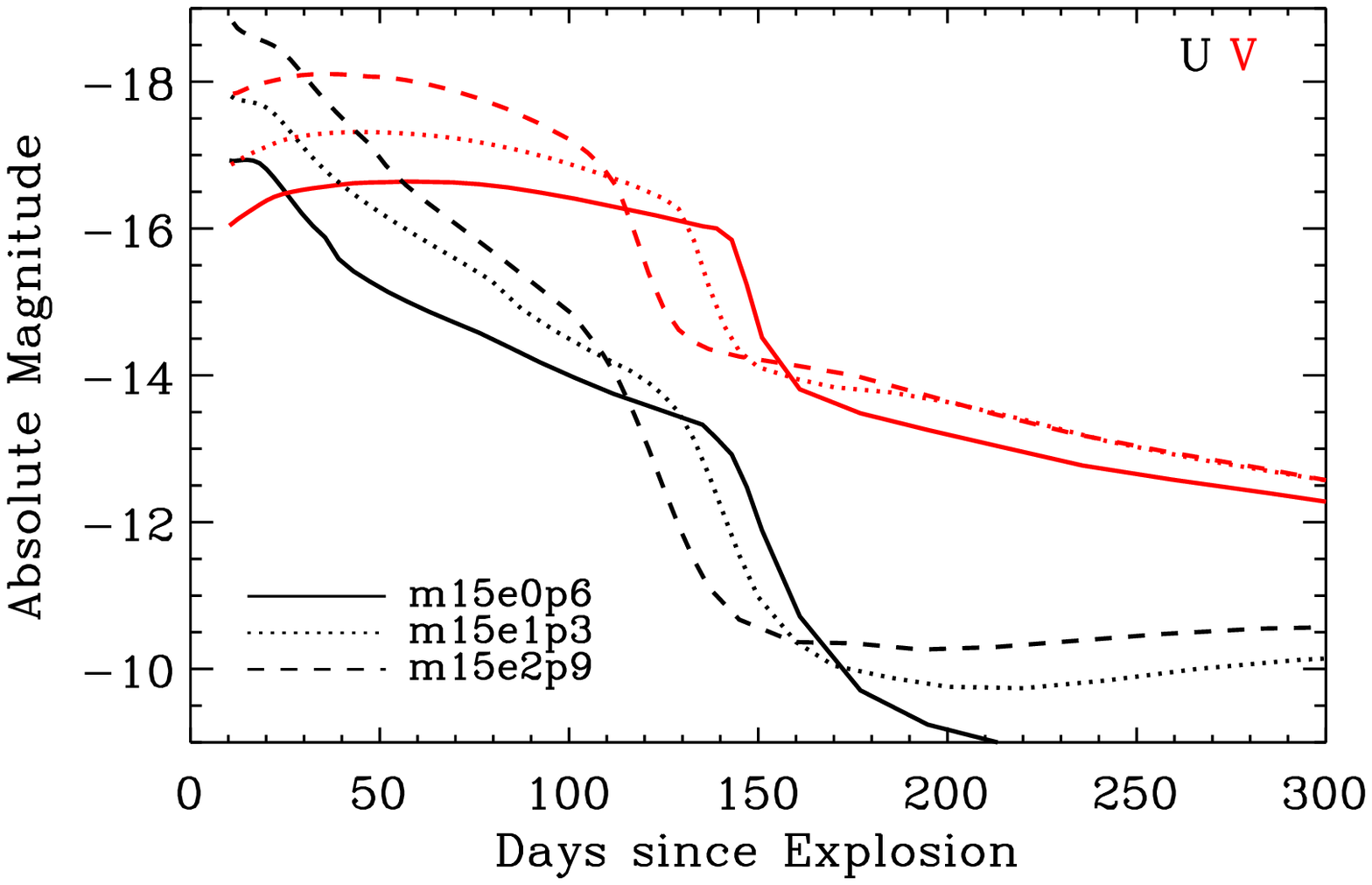,width=8.5cm}
\epsfig{file=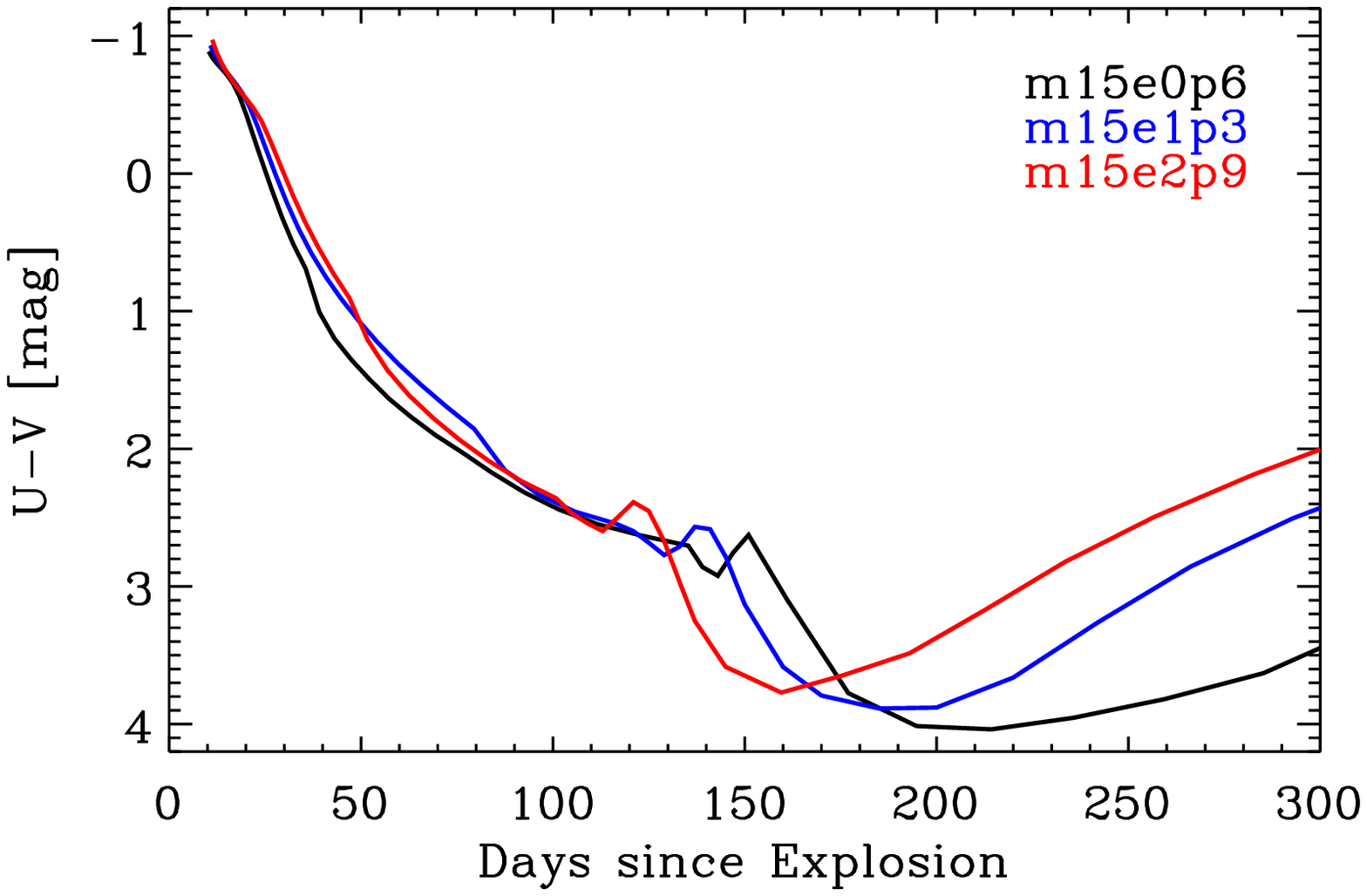,width=8.5cm}
\caption{{\it Left:} $U$ and $V$-band light curves for models m15e0p6, m15e1p3, and
m15e2p9. {\it Right:} Evolution of the $U-V$ color. Notice the strong degeneracy
between models at photospheric epochs.
\label{fig_m15_ekin_mag}
}
\end{figure*}

\section{Dependencies on ejecta kinetic energy}
\label{sect_ekin}

   In this last section, we explore the dependency of SN II-P radiation on explosion energy employing
   our reference m15 model for starting conditions. For clarity, we refer to this reference model as m15e1p3,
   to specify its kinetic energy of 1.27\,B. We ran two additional models exploded to yield an ejecta kinetic
   energy of 0.6\,B (model m15e0p6) and 2.9\,B (model m15e2p9).

   A larger explosion energy yields a brighter and shorter plateau.
  Conversely, a smaller explosion energy yields a fainter and longer plateau.
   Halfway through its photospheric phase, model m15e2p9 is 1.35\,mag brighter in $V$
   than model m15e0p6, which is close to the 1.5\,mag contrast predicted by the correlations
   of \citet{LN85} for the corresponding model parameters.
   Also, more \iso{56}Ni
   is synthesized in higher-energy explosions which translates into a stronger nebular flux
   (Fig.~\ref{fig_m15_ekin_lc}).  Faster expansion
   in the H-rich envelope leads to broader lines during the photospheric phase. Photospheric velocity
   measurements at mid-plateau phase can in fact provide a fair assessment of the ejecta kinetic
   energy \citep{DLW10b}. Interestingly, our models suggest that SN II-P color evolution is independent
   of explosion energy. For example, we show a spectral comparison at three epochs in Fig.~\ref{fig_m15_ekin_spec},
   corresponding to $U-V=-$\,0.4, 1, and 2\,mag, and find that each correspond to the same post-explosion
   date to within a few days at most. This degenerate color evolution is shown in Fig.~\ref{fig_m15_ekin_mag}.
   While there is an obvious shift in absolute magnitude reflecting the different radii of the photosphere
   for each model (left panel), the $U-V$ color is the same at a given post-explosion date
   for all models during the photospheric phase.
   This property is somewhat coincidental. The energy deposited by
   the shock corresponds to a representative expansion rate (considering a given ejecta mass). And,
   the greater the shock-deposited energy, the greater the cooling from the subsequent expansion.
   Higher energy explosions start with a larger ejecta temperature, but that energy is degraded more strongly
   the faster the ejecta expands (for an adiabatic evolution, we have $T \propto 1/R$).

\begin{figure*}
\epsfig{file=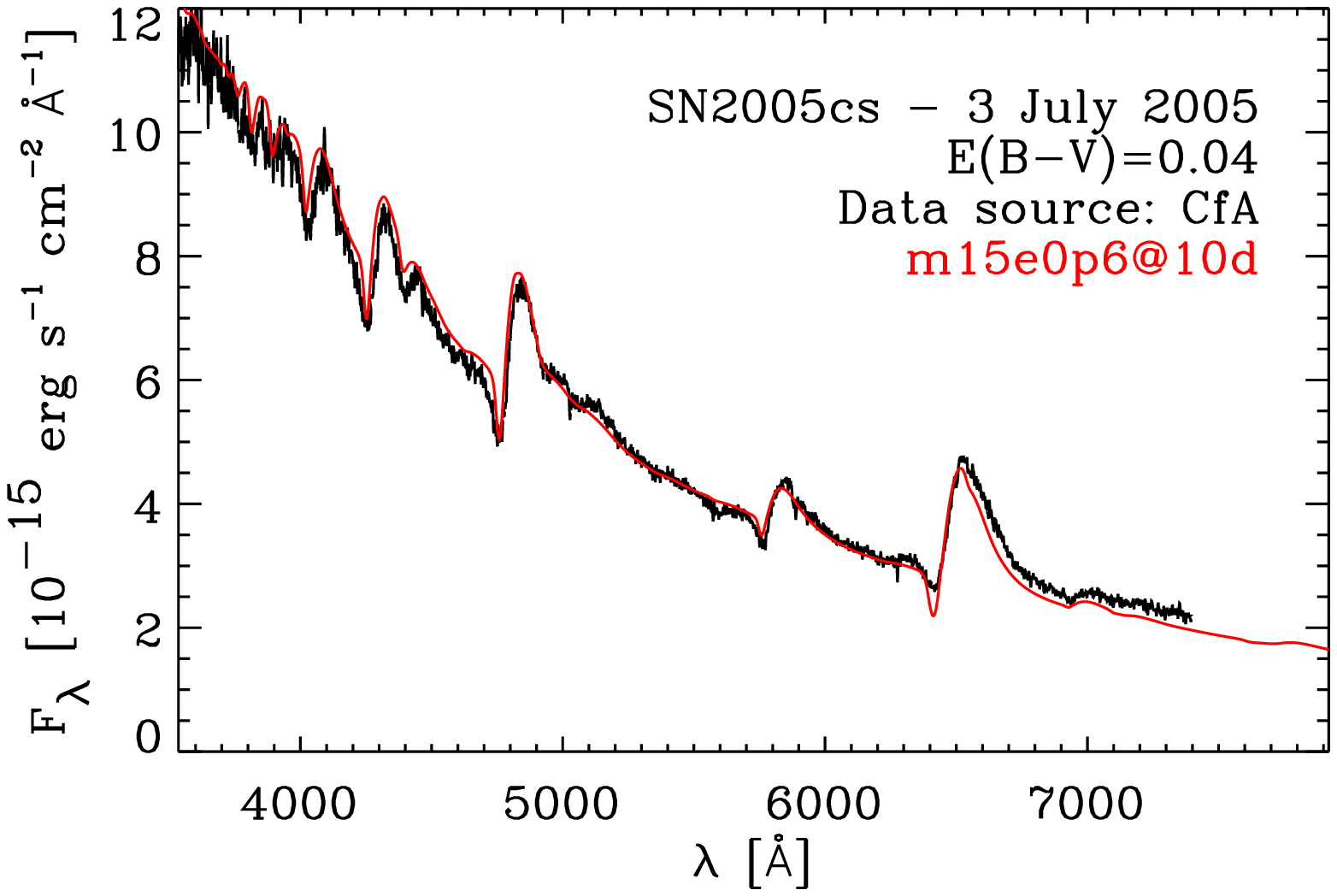,width=8.5cm}
\epsfig{file=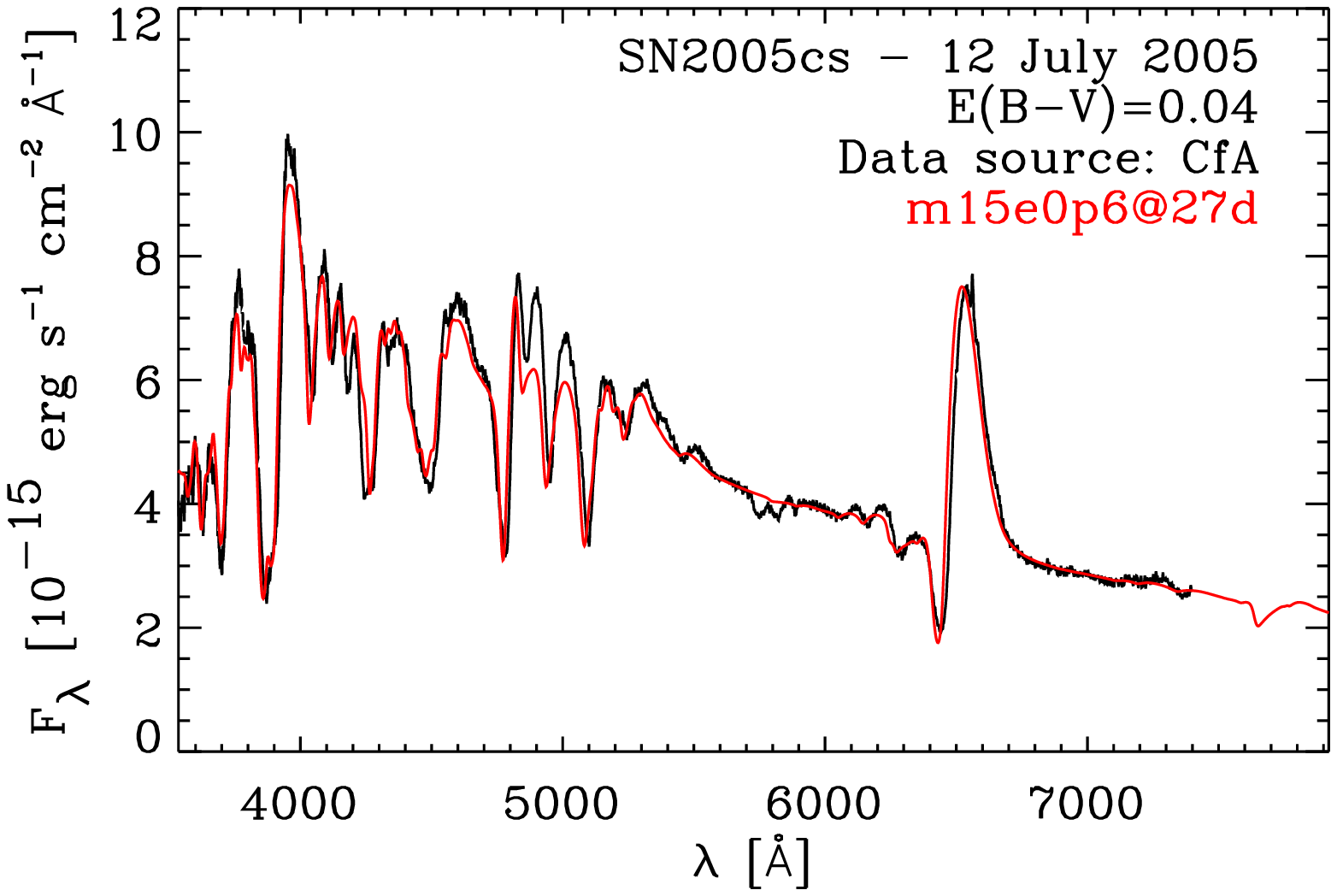,width=8.5cm}
\epsfig{file=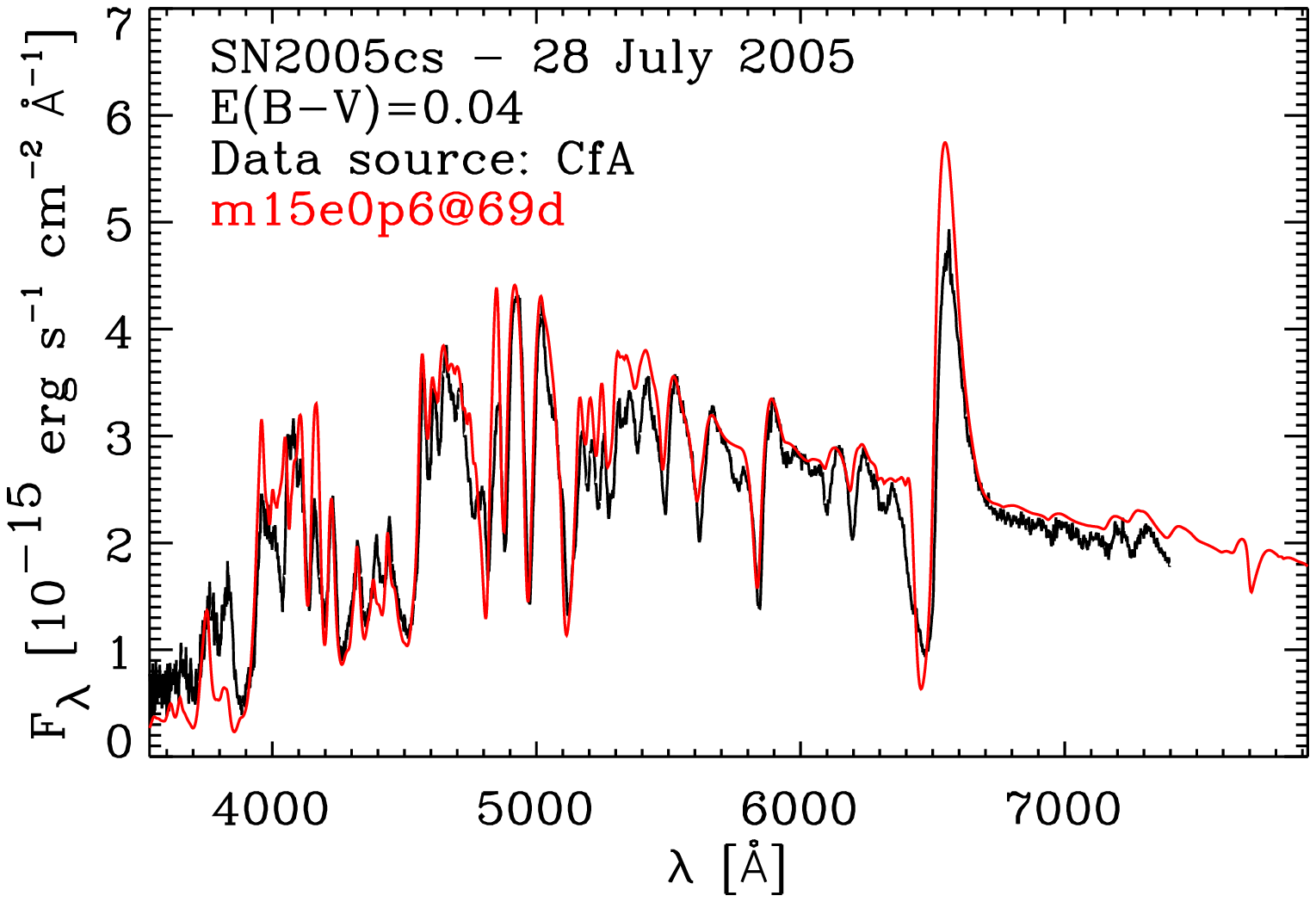,width=8.5cm}
\epsfig{file=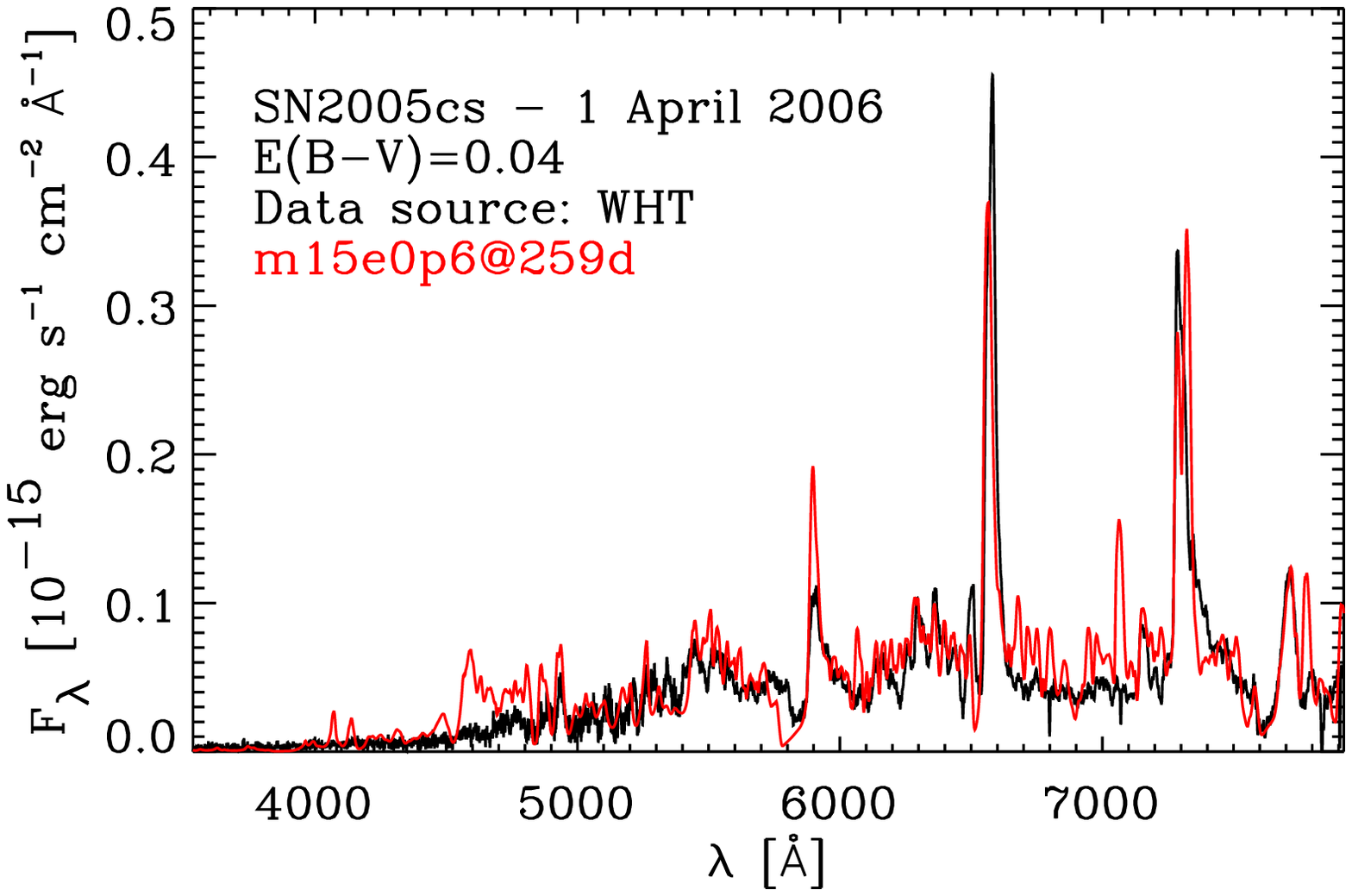,width=8.5cm}
\caption{Spectral comparison between SN\,2005cs \citep{dessart_etal_08} and
the under-energetic model m15e0p6 (the synthetic spectra are first reddened with $E(B-V)=0.04$\,mag
and later normalized to observations at 6000\,\AA). In this illustration, we are merely interested in
comparing the relative agreement between model and observations when the colors match ---
the absolute flux level, which may be offset due to an inadequate explosion energy, progenitor
radius, or \iso{56}Ni mass, is not of primary interest here.
This agreement suggests we are modeling satisfactorily the radiative transfer
in the photospheric layers.
\label{fig_m15_05cs}
}
\end{figure*}

As a last illustration, we show in Fig.~\ref{fig_m15_05cs} the correspondence between model
m15e0p6 and the under-luminous
SN\,2005cs \citep{PST06_SN2005cs,brown_etal_07,dessart_etal_08}. It is not the
scope of the present work to provide a detailed analysis of that SN. However,
we see that our model m15e0p6 provides a satisfactory match to the
observations during the photospheric phase (line widths and strenths, colors).
At nebular times, the comparison is less satisfactory. Our model is too
luminous (our ejecta model has initially 0.046\,\msun\ of \iso{56}Ni when
$\sim$\,0.008\,\msun\ was inferred by \citealt{utrobin_chugai_08})
and shows broader lines than observed for Na\one\,D or H$\alpha$. As before,
He\one\,7065\,\AA\ is too strong. However, the slower expansion rate of this
ejecta shows clearly the P-Cygni profile associated with the K\one\ resonance
line at 7664--7698\,\AA.

Hence, from the restricted exploration performed in this paper, the diversity in color
evolution of SNe II-P is found to stem not from different explosion energies but primarily
from different progenitor radii.

\section{Summary and outlook}
\label{sect_disc}

   We have presented a set of simulations for the evolution of a massive star, the explosion
   that follows core collapse, and the evolution of the radiation of the resulting supernova.
   Our goal is to connect the SN radiation properties with physical processes affecting
   stellar evolution and the explosion, and build a consistent picture encompassing all relevant
   aspects. This is important because determining the pre-SN star mass provides only a lower limit
   on the main-sequence star mass, and no information on the processes that make the two differ.
   Determining the H-rich envelope mass from the bolometric light curve places no constraint
   on the Helium core mass, which is in fact a much more reliable tracer of the main-sequence mass
   \citep{DLW10b}. A consistent picture of SN explosions must attempt to address not just the
   progenitor mass, but also the role of rotation, convection/overshooting, metallicity etc. on the
   evolution of a star until core collapse, and thus tackling these issues from first principles
   is a sensible goal.

   We have reported on two distinct efforts. First, we revisited the simulations of \citet{DH11}
   to identify the origin of the discrepancies with the observations of the standard SN II-P 1999em.
   This has proven successful with our upgraded simulations including larger model atoms
   for Fe\one\ and Fe\two, together with the treatment of non-thermal processes and non-local
   energy deposition. However, a color problem remains (model is too blue for too long),
   which all attempts at improving the quality of the radiative transfer have failed to resolve fully.

   The second effort has been to explore what variation in progenitor and explosion properties
   could cure this color problem. Using a grid of simulations for  a 15\,\msun\ main sequence
   star calculated with \mesa, we study the dependency of pre-SN star properties on
   mixing length parameter, rotation, overshooting, and metallicity. Variations in these parameters
   alter global quantities (e.g., the progenitor radius, the H-rich envelope mass,
   the ratio of H-rich envelope (or ejecta) mass to the helium-core mass), which
   influence significantly the synthetic SN II-P light curves, although the same
   main-sequence mass and the same ejecta kinetic energy characterize these simulations.
   This confirms the difficulty of associating a given SN II-P with a main-sequence progenitor star.

   We have also compared the results of \mesa\ with those obtained by \kepler. With a suitable (and sensible)
   choice of parameters,  both codes produce a similar pre-SN star from a 15\,\msun\ star on
   the main sequence. The resulting explosions for these \kepler\ and \mesa\ pre-SN star models
   yield comparable LCs and spectra. The specific case we describe highlights the degeneracy of SN II-P radiation
   --- strong variations in mixing and chemical composition in sub-dominant species are associated with only
   modest variations in observables.

   In the \mesa\ exploration, the most interesting result is obtained with the 15\,\msun\ models evolved using different
   mixing length parameters. All else being the same (opacities, mass, metallicity etc), more efficient convection
   yields more compact RSG stars. This change does not affect other stellar properties at death, such as total
   mass or helium core mass. We find that the reduction of the progenitor radius from 800\,\rsun\ to 500\,\rsun\
   cures nearly
   completely the color problem identified in all our SN II-P simulations.  Independent of our work and diagnostics,
   recent radiative-transfer calculations of RSG atmospheres propose a systematic reduction of RSG radii
    \citep{davies_etal_13}.
       One may argue that very extended RSG stars exist. However, these RSGs often exhibit intense
   mass loss episodes, as in the case of VY CMa. If it were to explode, rather than producing
   a type II-P,  the resulting SN would probably be a Type IIn, or a II-L.

   The effect of metallicity is interesting. It is visible in the pre-SN object through the influence on the final mass
   and radius. In the SN II-P spectra, variations in metallicity modulate the magnitude of line blanketing, which
   affects the entire UV range. In the optical, the effect is visible through strong individual line features for numerous
   metal species, including O, Na, Ca, Ti, and Fe. Hence, the outer layers of SN II-P ejecta, unaffected by
   steady/explosive burning or mixing,  represent a great potential for metallicity studies.
   SN II-P observations could be used to infer the metallicity at the SN site, which is important to understand the
   progenitor evolution, but also as a tool to infer metallicities in the nearby and distant Universe.
  This may prove particularly useful in the coming years with JWST and extremely large telescopes.

   The influence of kinetic energy corroborates previous findings, more energetic explosions yielding a brighter
   and shorter plateau while the larger explosion energy produces broader spectral lines at all times.
   We find that the SN color evolution is independent of explosion energy, i.e. the SN exhibits the same color
   at a given post-explosion date irrespective of explosion energy. This suggests that, at a given metallicity,
   the main factor affecting SN II-P colors is the progenitor radius. We note that SNe II-P will systematically evolve from
   high to low (recombination) temperatures and will therefore step through the same spectral
   morphology (in the UV or in the optical).
   It is the {\it rate} at which their color evolves that will differ between SNe II-P, i.e., being faster in more
   compact RSGs than in more extended ones.

  Since SN II-P light curves are only weakly dependent on the helium core mass, they cannot be used
  to infer the total ejecta mass. Instead, they constrain the H-rich envelope mass.
  In hydrodynamical simulations of SNe II-P, this distinction is often missing because, instead
  of using physically-consistent stellar-evolution models, a simple power law or a polytrope is adopted
  for the progenitor density structure  \citep{LN83,utrobin_07,bersten_etal_11}.
   Such approximated models tend to have a different surface scale height.
   Furthermore, the Lagrangian mass
   for the edge of the helium core is arbitrarily positioned. In stellar evolutionary calculations, this
   edge (which corresponds to the base of the H-rich envelope), is not arbitrarily located, but is instead
   a very stiff function of main-sequence mass (influenced as well by rotation or core-overshooting).
   The RSG progenitor envelope thus reveals a
    highly-bound helium core and a loosely-bound hydrogen-rich envelope.
    The mass contained in the helium core,
   which can be large for higher mass progenitors and the corresponding SN ejecta, influences nebular spectral
   properties but not the light curve.

   These issues are in part illustrated by our models computed with rotation and overshooting.
   Both effects tend to make a given main-sequence star evolve as if it was more massive.
   For a 15\,\msun\ progenitor, rotation and/or overshooting produce a star at death that
   has a bigger helium core, has a higher luminosity, is more extended (bigger radius), has suffered strong mass
   loss (lighter H-rich envelope). In such stars, the helium core contributes a greater fraction of the total
   ejecta mass, at the expense of the H-rich envelope mass. However, only the
   latter is constrained from light-curve modeling.  Much of the mass discrepancy for SN II-P progenitors
   today stems from the ambiguity of what is meant by ejecta/progenitor mass. Inferences based
   on pre-explosion images are fundamentally tied to the star luminosity and thus to the (helium) core mass.
   In contrast, in the often-adopted formalism of \citet{LN83}, the ``mass" corresponds to the material
   affecting the plateau light curve, and that mass is the H-rich envelope mass in the progenitor star.

   In the future, we will experiment with \mesa, \v1d, and \cmfgen\ to produce ejecta models
   whose computed SN light curves and spectral evolution, at both photospheric and nebular times,
   are compatible with well observed SNe II-P.

\section*{Acknowledgments}

LD acknowledges financial support from the European Community through
an International Re-integration Grant, under grant number
PIRG04-GA-2008-239184, and from ``Agence Nationale de la Recherche"
grant ANR-2011-Blanc-SIMI-5-6-007-01.  DJH acknowledges support from
STScI theory grants HST-AR-11756.01.A and HST-AR-12640.01, and NASA
theory grant NNX10AC80G.  This work was granted access to the HPC
resources of CINES under the allocation 2011--c2011046608 and
c2012046608 made by GENCI (Grand Equipement National de Calcul
Intensif).  A subset of the computations were also performed at
Caltech's Center for Advanced Computing Research on the cluster Zwicky
funded through NSF grant no. PHY-0960291 and the Sherman Fairchild
Foundation.

\appendix

\section{} Illustration of line features at photospheric and nebular times in a SN II-P
\section{} Multi-band light curves for the SN II-P models discussed in the main body of the paper.


\begin{thebibliography}{77}
\expandafter\ifx\csname natexlab\endcsname\relax\def\natexlab#1{#1}\fi

\bibitem[{{Arcavi} {et~al.}(2010){Arcavi}, {Gal-Yam}, {Kasliwal}, {Quimby},
  {Ofek}, {Kulkarni}, {Nugent}, {Cenko}, {Bloom}, {Sullivan}, {Howell},
  {Poznanski}, {Filippenko}, {Law}, {Hook}, {J{\"o}nsson}, {Blake}, {Cooke},
  {Dekany}, {Rahmer}, {Hale}, {Smith}, {Zolkower}, {Velur}, {Walters},
  {Henning}, {Bui}, {McKenna}, \& {Jacobsen}}]{arcavi_etal_10}
{Arcavi}, I., {Gal-Yam}, A., {Kasliwal}, M.~M., {Quimby}, R.~M., {Ofek}, E.~O.,
  {Kulkarni}, S.~R., {Nugent}, P.~E., {Cenko}, S.~B., {Bloom}, J.~S.,
  {Sullivan}, M., {Howell}, D.~A., {Poznanski}, D., {Filippenko}, A.~V., {Law},
  N., {Hook}, I., {J{\"o}nsson}, J., {Blake}, S., {Cooke}, J., {Dekany}, R.,
  {Rahmer}, G., {Hale}, D., {Smith}, R., {Zolkower}, J., {Velur}, V.,
  {Walters}, R., {Henning}, J., {Bui}, K., {McKenna}, D., \& {Jacobsen}, J.
  2010, \apj, 721, 777

\bibitem[{{Baklanov} {et~al.}(2005){Baklanov}, {Blinnikov}, \&
  {Pavlyuk}}]{baklanov_etal_05}
{Baklanov}, P.~V., {Blinnikov}, S.~I., \& {Pavlyuk}, N.~N. 2005, Astronomy
  Letters, 31, 429

\bibitem[{{Bersten} {et~al.}(2011){Bersten}, {Benvenuto}, \&
  {Hamuy}}]{bersten_etal_11}
{Bersten}, M.~C., {Benvenuto}, O., \& {Hamuy}, M. 2011, \apj, 729, 61

\bibitem[{{Bersten} \& {Hamuy}(2009)}]{bersten_hamuy_09}
{Bersten}, M.~C. \& {Hamuy}, M. 2009, \apj, 701, 200

\bibitem[{{Blinnikov} {et~al.}(1998){Blinnikov}, {Eastman}, {Bartunov},
  {Popolitov}, \& {Woosley}}]{blinnikov_etal_98}
{Blinnikov}, S.~I., {Eastman}, R., {Bartunov}, O.~S., {Popolitov}, V.~A., \&
  {Woosley}, S.~E. 1998, \apj, 496, 454

\bibitem[{{Brott} {et~al.}(2011){Brott}, {de Mink}, {Cantiello}, {Langer}, {de
  Koter}, {Evans}, {Hunter}, {Trundle}, \& {Vink}}]{brott_etal_11}
{Brott}, I., {de Mink}, S.~E., {Cantiello}, M., {Langer}, N., {de Koter}, A.,
  {Evans}, C.~J., {Hunter}, I., {Trundle}, C., \& {Vink}, J.~S. 2011, \aap,
  530, A115

\bibitem[{{Brown} {et~al.}(2007){Brown}, {Dessart}, {Holland}, {Immler},
  {Landsman}, {Blondin}, {Blustin}, {Breeveld}, {Dewangan}, {Gehrels},
  {Hutchins}, {Kirshner}, {Mason}, {Mazzali}, {Milne}, {Modjaz}, \&
  {Roming}}]{brown_etal_07}
{Brown}, P.~J., {Dessart}, L., {Holland}, S.~T., {Immler}, S., {Landsman}, W.,
  {Blondin}, S., {Blustin}, A.~J., {Breeveld}, A., {Dewangan}, G.~C.,
  {Gehrels}, N., {Hutchins}, R.~B., {Kirshner}, R.~P., {Mason}, K.~O.,
  {Mazzali}, P.~A., {Milne}, P., {Modjaz}, M., \& {Roming}, P.~W.~A. 2007,
  \apj, 659, 1488

\bibitem[{{Brown} {et~al.}(2009){Brown}, {Holland}, {Immler}, {Milne},
  {Roming}, {Gehrels}, {Nousek}, {Panagia}, {Still}, \& {Vanden
  Berk}}]{brown_etal_09}
{Brown}, P.~J., {Holland}, S.~T., {Immler}, S., {Milne}, P., {Roming},
  P.~W.~A., {Gehrels}, N., {Nousek}, J., {Panagia}, N., {Still}, M., \& {Vanden
  Berk}, D. 2009, \aj, 137, 4517

\bibitem[{{Cardelli} {et~al.}(1988){Cardelli}, {Clayton}, \&
  {Mathis}}]{CCM88_ISE}
{Cardelli}, J.~A., {Clayton}, G.~C., \& {Mathis}, J.~S. 1988, \apjl, 329, L33

\bibitem[{{Castor} {et~al.}(1975){Castor}, {Abbott}, \& {Klein}}]{cak}
{Castor}, J.~I., {Abbott}, D.~C., \& {Klein}, R.~I. 1975, \apj, 195, 157

\bibitem[{{Chornock} {et~al.}(2010){Chornock}, {Filippenko}, {Li}, \&
  {Silverman}}]{chornock_etal_10a}
{Chornock}, R., {Filippenko}, A.~V., {Li}, W., \& {Silverman}, J.~M. 2010,
  \apj, 713, 1363

\bibitem[{{Davies} {et~al.}(2013){Davies}, {Kudritzki}, {Plez}, {Trager},
  {Lan{\c c}on}, {Gazak}, {Bergemann}, {Evans}, \&
  {Chiavassa}}]{davies_etal_13}
{Davies}, B., {Kudritzki}, R.-P., {Plez}, B., {Trager}, S., {Lan{\c c}on}, A.,
  {Gazak}, Z., {Bergemann}, M., {Evans}, C., \& {Chiavassa}, A. 2013, \apj,
  767, 3

\bibitem[{{Decin} {et~al.}(2006){Decin}, {Hony}, {de Koter}, {Justtanont},
  {Tielens}, \& {Waters}}]{decin_etal_06}
{Decin}, L., {Hony}, S., {de Koter}, A., {Justtanont}, K., {Tielens},
  A.~G.~G.~M., \& {Waters}, L.~B.~F.~M. 2006, \aap, 456, 549

\bibitem[{{Dessart} {et~al.}(2008){Dessart}, {Blondin}, {Brown}, {Hicken},
  {Hillier}, {Holland}, {Immler}, {Kirshner}, {Milne}, {Modjaz}, \&
  {Roming}}]{dessart_etal_08}
{Dessart}, L., {Blondin}, S., {Brown}, P.~J., {Hicken}, M., {Hillier}, D.~J.,
  {Holland}, S.~T., {Immler}, S., {Kirshner}, R.~P., {Milne}, P., {Modjaz}, M.,
  \& {Roming}, P.~W.~A. 2008, \apj, 675, 644

\bibitem[{{Dessart} \& {Hillier}(2005)}]{DH05_qs_SN}
{Dessart}, L. \& {Hillier}, D.~J. 2005, \aap, 437, 667

\bibitem[{{Dessart} \& {Hillier}(2006)}]{DH06_SN1999em}
---. 2006, \aap, 447, 691

\bibitem[{{Dessart} \& {Hillier}(2008{\natexlab{a}})}]{DH08}
---. 2008{\natexlab{a}}, \mnras, 383, 57

\bibitem[{{Dessart} \& {Hillier}(2008{\natexlab{b}})}]{DH08_time}
---. 2008{\natexlab{b}}, \mnras, 383, 57

\bibitem[{{Dessart} \& {Hillier}(2010)}]{DH10}
---. 2010, \mnras, 405, 2141

\bibitem[{{Dessart} \& {Hillier}(2011)}]{DH11}
---. 2011, \mnras, 410, 1739

\bibitem[{{Dessart} {et~al.}(2012){Dessart}, {Hillier}, {Li}, \&
  {Woosley}}]{dessart_etal_12}
{Dessart}, L., {Hillier}, D.~J., {Li}, C., \& {Woosley}, S. 2012, \mnras, 424,
  2139

\bibitem[{{Dessart} {et~al.}(2010{\natexlab{a}}){Dessart}, {Livne}, \&
  {Waldman}}]{DLW10b}
{Dessart}, L., {Livne}, E., \& {Waldman}, R. 2010{\natexlab{a}}, \mnras, 408,
  827

\bibitem[{{Dessart} {et~al.}(2010{\natexlab{b}}){Dessart}, {Livne}, \&
  {Waldman}}]{DLW10a}
---. 2010{\natexlab{b}}, \mnras, 405, 2113

\bibitem[{{Dessart} {et~al.}(2013){Dessart}, {Waldman}, {Livne}, {Hillier}, \&
  {Blondin}}]{dessart_etal_13}
{Dessart}, L., {Waldman}, R., {Livne}, E., {Hillier}, D.~J., \& {Blondin}, S.
  2013, \mnras, 428, 3227

\bibitem[{{Eastman} {et~al.}(1994){Eastman}, {Woosley}, {Weaver}, \&
  {Pinto}}]{eastman_etal_94}
{Eastman}, R.~G., {Woosley}, S.~E., {Weaver}, T.~A., \& {Pinto}, P.~A. 1994,
  \apj, 430, 300

\bibitem[{{Falk} \& {Arnett}(1977)}]{falk_arnett_77}
{Falk}, S.~W. \& {Arnett}, W.~D. 1977, \apjs, 33, 515

\bibitem[{{Gezari} {et~al.}(2008){Gezari}, {Dessart}, {Basa}, {Martin},
  {Neill}, {Woosley}, {Hillier}, {Bazin}, {Forster}, {Friedman}, {Le Du},
  {Mazure}, {Morrissey}, {Neff}, {Schiminovich}, \& {Wyder}}]{gezari_etal_08}
{Gezari}, S., {Dessart}, L., {Basa}, S., {Martin}, D.~C., {Neill}, J.~D.,
  {Woosley}, S.~E., {Hillier}, D.~J., {Bazin}, G., {Forster}, K., {Friedman},
  P.~G., {Le Du}, J., {Mazure}, A., {Morrissey}, P., {Neff}, S.~G.,
  {Schiminovich}, D., \& {Wyder}, T.~K. 2008, \apjl, 683, L131

\bibitem[{{Grassberg} {et~al.}(1971){Grassberg}, {Imshennik}, \&
  {Nadyozhin}}]{grassberg_etal_71}
{Grassberg}, E.~K., {Imshennik}, V.~S., \& {Nadyozhin}, D.~K. 1971, \apss, 10,
  28

\bibitem[{{Grevesse} \& {Sauval}(1998)}]{GS98}
{Grevesse}, N. \& {Sauval}, A.~J. 1998, \ssr, 85, 161

\bibitem[{{Hamuy} {et~al.}(2001){Hamuy}, {Pinto}, {Maza}, {Suntzeff},
  {Phillips}, {Eastman}, {Smith}, {Corbally}, {Burstein}, {Li}, {Ivanov},
  {Moro-Martin}, {Strolger}, {de Souza}, {dos Anjos}, {Green}, {Pickering},
  {Gonz{\'a}lez}, {Antezana}, {Wischnjewsky}, {Galaz}, {Roth}, {Persson}, \&
  {Schommer}}]{hamuy_etal_01}
{Hamuy}, M., {Pinto}, P.~A., {Maza}, J., {Suntzeff}, N.~B., {Phillips}, M.~M.,
  {Eastman}, R.~G., {Smith}, R.~C., {Corbally}, C.~J., {Burstein}, D., {Li},
  Y., {Ivanov}, V., {Moro-Martin}, A., {Strolger}, L.~G., {de Souza}, R.~E.,
  {dos Anjos}, S., {Green}, E.~M., {Pickering}, T.~E., {Gonz{\'a}lez}, L.,
  {Antezana}, R., {Wischnjewsky}, M., {Galaz}, G., {Roth}, M., {Persson},
  S.~E., \& {Schommer}, R.~A. 2001, \apj, 558, 615

\bibitem[{{Haubois} {et~al.}(2009){Haubois}, {Perrin}, {Lacour}, {Verhoelst},
  {Meimon}, {Mugnier}, {Thi{\'e}baut}, {Berger}, {Ridgway}, {Monnier},
  {Millan-Gabet}, \& {Traub}}]{haubois_etal_09}
{Haubois}, X., {Perrin}, G., {Lacour}, S., {Verhoelst}, T., {Meimon}, S.,
  {Mugnier}, L., {Thi{\'e}baut}, E., {Berger}, J.~P., {Ridgway}, S.~T.,
  {Monnier}, J.~D., {Millan-Gabet}, R., \& {Traub}, W. 2009, \aap, 508, 923

\bibitem[{{Hillier} \& {Dessart}(2012)}]{HD12}
{Hillier}, D.~J. \& {Dessart}, L. 2012, \mnras, 424, 252

\bibitem[{{Hillier} \& {Miller}(1998)}]{HM98_lb}
{Hillier}, D.~J. \& {Miller}, D.~L. 1998, \apj, 496, 407

\bibitem[{{Hoeflich}(1988)}]{hoeflich_88}
{Hoeflich}, P. 1988, Proceedings of the Astronomical Society of Australia, 7,
  434

\bibitem[{{Jerkstrand} {et~al.}(2012){Jerkstrand}, {Fransson}, {Maguire},
  {Smartt}, {Ergon}, \& {Spyromilio}}]{jerkstrand_etal_12}
{Jerkstrand}, A., {Fransson}, C., {Maguire}, K., {Smartt}, S., {Ergon}, M., \&
  {Spyromilio}, J. 2012, \aap, 546, A28

\bibitem[{{Josselin} \& {Plez}(2007)}]{josselin_plez_07}
{Josselin}, E. \& {Plez}, B. 2007, \aap, 469, 671

\bibitem[{{Kasen} {et~al.}(2008){Kasen}, {Thomas}, {R{\"o}pke}, \&
  {Woosley}}]{kasen_etal_08}
{Kasen}, D., {Thomas}, R.~C., {R{\"o}pke}, F., \& {Woosley}, S.~E. 2008,
  Journal of Physics Conference Series, 125, 012007

\bibitem[{{Kasen} \& {Woosley}(2009)}]{KW09}
{Kasen}, D. \& {Woosley}, S.~E. 2009, \apj, 703, 2205

\bibitem[{{Kozma} \& {Fransson}(1992)}]{KF92}
{Kozma}, C. \& {Fransson}, C. 1992, \apj, 390, 602

\bibitem[{{Kozma} \& {Fransson}(1998{\natexlab{a}})}]{KF98a}
---. 1998{\natexlab{a}}, \apj, 496, 946

\bibitem[{{Kozma} \& {Fransson}(1998{\natexlab{b}})}]{KF98b}
---. 1998{\natexlab{b}}, \apj, 497, 431

\bibitem[{{Leonard} {et~al.}(2002){Leonard}, {Filippenko}, {Gates}, {Li},
  {Eastman}, {Barth}, {Bus}, {Chornock}, {Coil}, {Frink}, {Grady}, {Harris},
  {Malkan}, {Matheson}, {Quirrenbach}, \& {Treffers}}]{leonard_etal_02a}
{Leonard}, D.~C., {Filippenko}, A.~V., {Gates}, E.~L., {Li}, W., {Eastman},
  R.~G., {Barth}, A.~J., {Bus}, S.~J., {Chornock}, R., {Coil}, A.~L., {Frink},
  S., {Grady}, C.~A., {Harris}, A.~W., {Malkan}, M.~A., {Matheson}, T.,
  {Quirrenbach}, A., \& {Treffers}, R.~R. 2002, \pasp, 114, 35

\bibitem[{{Levesque} {et~al.}(2010){Levesque}, {Berger}, {Kewley}, \&
  {Bagley}}]{levesque_etal_10}
{Levesque}, E.~M., {Berger}, E., {Kewley}, L.~J., \& {Bagley}, M.~M. 2010, \aj,
  139, 694

\bibitem[{{Levesque} {et~al.}(2005){Levesque}, {Massey}, {Olsen}, {Plez},
  {Josselin}, {Maeder}, \& {Meynet}}]{levesque_etal_05}
{Levesque}, E.~M., {Massey}, P., {Olsen}, K.~A.~G., {Plez}, B., {Josselin}, E.,
  {Maeder}, A., \& {Meynet}, G. 2005, \apj, 628, 973

\bibitem[{{Li} {et~al.}(2012){Li}, {Hillier}, \& {Dessart}}]{li_etal_12}
{Li}, C., {Hillier}, D.~J., \& {Dessart}, L. 2012, \mnras, 426, 1671

\bibitem[{{Li} \& {McCray}(1992)}]{li_mccray_92}
{Li}, H. \& {McCray}, R. 1992, \apj, 387, 309

\bibitem[{{Li} \& {McCray}(1993)}]{li_mccray_93}
---. 1993, \apj, 405, 730

\bibitem[{{Litvinova} \& {Nadezhin}(1983)}]{LN83}
{Litvinova}, I.~I. \& {Nadezhin}, D.~K. 1983, \apss, 89, 89

\bibitem[{{Litvinova} \& {Nadezhin}(1985)}]{LN85}
{Litvinova}, I.~Y. \& {Nadezhin}, D.~K. 1985, Soviet Astronomy Letters, 11, 145

\bibitem[{{Livne}(1993)}]{livne_93}
{Livne}, E. 1993, \apj, 412, 634

\bibitem[{{Lucy}(1991)}]{lucy_91}
{Lucy}, L.~B. 1991, \apj, 383, 308

\bibitem[{{Maeder} \& {Meynet}(1987)}]{maeder_meynet_87}
{Maeder}, A. \& {Meynet}, G. 1987, \aap, 182, 243

\bibitem[{{Maeder} \& {Meynet}(2000)}]{maeder_meynet_00}
---. 2000, \aap, 361, 159

\bibitem[{{Maguire} {et~al.}(2012){Maguire}, {Jerkstrand}, {Smartt},
  {Fransson}, {Pastorello}, {Benetti}, {Valenti}, {Bufano}, \&
  {Leloudas}}]{maguire_etal_12}
{Maguire}, K., {Jerkstrand}, A., {Smartt}, S.~J., {Fransson}, C., {Pastorello},
  A., {Benetti}, S., {Valenti}, S., {Bufano}, F., \& {Leloudas}, G. 2012,
  \mnras, 420, 3451

\bibitem[{{Meakin} \& {Arnett}(2007)}]{meakin_arnett_07}
{Meakin}, C.~A. \& {Arnett}, D. 2007, \apj, 667, 448

\bibitem[{{Neilson} {et~al.}(2011){Neilson}, {Lester}, \&
  {Haubois}}]{neilson_etal.11}
{Neilson}, H.~R., {Lester}, J.~B., \& {Haubois}, X. 2011, in Astronomical
  Society of the Pacific Conference Series, Vol. 451, Astronomical Society of
  the Pacific Conference Series, ed. S.~{Qain}, K.~{Leung}, L.~{Zhu}, \&
  S.~{Kwok}, 117

\bibitem[{{Pastorello} {et~al.}(2006){Pastorello}, {Sauer}, {Taubenberger},
  {Mazzali}, {Nomoto}, {Kawabata}, {Benetti}, {Elias-Rosa}, {Harutyunyan},
  {Navasardyan}, {Zampieri}, {Iijima}, {Botticella}, {Di Rico}, {Del Principe},
  {Dolci}, {Gagliardi}, {Ragni}, \& {Valentini}}]{PST06_SN2005cs}
{Pastorello}, A., {Sauer}, D., {Taubenberger}, S., {Mazzali}, P.~A., {Nomoto},
  K., {Kawabata}, K.~S., {Benetti}, S., {Elias-Rosa}, N., {Harutyunyan}, A.,
  {Navasardyan}, H., {Zampieri}, L., {Iijima}, T., {Botticella}, M.~T., {Di
  Rico}, G., {Del Principe}, M., {Dolci}, M., {Gagliardi}, S., {Ragni}, M., \&
  {Valentini}, G. 2006, \mnras, 370, 1752

\bibitem[{{Paxton} {et~al.}(2011){Paxton}, {Bildsten}, {Dotter}, {Herwig},
  {Lesaffre}, \& {Timmes}}]{paxton_etal_11}
{Paxton}, B., {Bildsten}, L., {Dotter}, A., {Herwig}, F., {Lesaffre}, P., \&
  {Timmes}, F. 2011, \apjs, 192, 3

\bibitem[{{Paxton} {et~al.}(2013){Paxton}, {Cantiello}, {Arras}, {Bildsten},
  {Brown}, {Dotter}, {Mankovich}, {Montgomery}, {Stello}, {Timmes}, \&
  {Townsend}}]{paxton_etal_13}
{Paxton}, B., {Cantiello}, M., {Arras}, P., {Bildsten}, L., {Brown}, E.~F.,
  {Dotter}, A., {Mankovich}, C., {Montgomery}, M.~H., {Stello}, D., {Timmes},
  F.~X., \& {Townsend}, R. 2013, ArXiv:1301.0319

\bibitem[{{Popov}(1993)}]{popov_93}
{Popov}, D.~V. 1993, \apj, 414, 712

\bibitem[{{Pumo} \& {Zampieri}(2011)}]{pumo_zampieri_11}
{Pumo}, M.~L. \& {Zampieri}, L. 2011, \apj, 741, 41

\bibitem[{{Quataert} \& {Shiode}(2012)}]{quataert_shiode_12}
{Quataert}, E. \& {Shiode}, J. 2012, \mnras, 423, L92

\bibitem[{{Quimby} {et~al.}(2007){Quimby}, {Wheeler}, {H{\"o}flich}, {Akerlof},
  {Brown}, \& {Rykoff}}]{quimby_etal_07}
{Quimby}, R.~M., {Wheeler}, J.~C., {H{\"o}flich}, P., {Akerlof}, C.~W.,
  {Brown}, P.~J., \& {Rykoff}, E.~S. 2007, \apj, 666, 1093

\bibitem[{{Smartt}(2009)}]{smartt_09}
{Smartt}, S.~J. 2009, \araa, 47, 63

\bibitem[{{Smith} {et~al.}(2009){Smith}, {Hinkle}, \& {Ryde}}]{smith_vycma_09}
{Smith}, N., {Hinkle}, K.~H., \& {Ryde}, N. 2009, \aj, 137, 3558

\bibitem[{{Smith} {et~al.}(2011){Smith}, {Li}, {Filippenko}, \&
  {Chornock}}]{smith_etal_11}
{Smith}, N., {Li}, W., {Filippenko}, A.~V., \& {Chornock}, R. 2011, \mnras,
  412, 1522

\bibitem[{{Timmes}(1999)}]{timmes_99}
{Timmes}, F.~X. 1999, \apjs, 124, 241

\bibitem[{{Utrobin}(2007)}]{utrobin_07}
{Utrobin}, V.~P. 2007, \aap, 461, 233

\bibitem[{{Utrobin} \& {Chugai}(2005)}]{UC05}
{Utrobin}, V.~P. \& {Chugai}, N.~N. 2005, \aap, 441, 271

\bibitem[{{Utrobin} \& {Chugai}(2008)}]{utrobin_chugai_08}
---. 2008, \aap, 491, 507

\bibitem[{{Utrobin} \& {Chugai}(2009)}]{utrobin_chugai_09}
---. 2009, \aap, 506, 829

\bibitem[{{Weaver} {et~al.}(1978){Weaver}, {Zimmerman}, \&
  {Woosley}}]{weaver_etal_78}
{Weaver}, T.~A., {Zimmerman}, G.~B., \& {Woosley}, S.~E. 1978, \apj, 225, 1021

\bibitem[{{Wittkowski} {et~al.}(2012){Wittkowski}, {Hauschildt},
  {Arroyo-Torres}, \& {Marcaide}}]{wittkowski_etal_12}
{Wittkowski}, M., {Hauschildt}, P.~H., {Arroyo-Torres}, B., \& {Marcaide},
  J.~M. 2012, \aap, 540, L12

\bibitem[{{Woosley} \& {Heger}(2007)}]{WH07}
{Woosley}, S.~E. \& {Heger}, A. 2007, \physrep, 442, 269

\bibitem[{{Woosley} {et~al.}(2002){Woosley}, {Heger}, \& {Weaver}}]{WHW02}
{Woosley}, S.~E., {Heger}, A., \& {Weaver}, T.~A. 2002, Reviews of Modern
  Physics, 74, 1015

\bibitem[{{Yoon} \& {Cantiello}(2010)}]{yoon_cantiello_10}
{Yoon}, S.-C. \& {Cantiello}, M. 2010, \apjl, 717, L62

\bibitem[{{Young}(2004)}]{Y04}
{Young}, T.~R. 2004, \apj, 617, 1233

\end{thebibliography}

\label{lastpage}

\end{document}